\documentclass[aps,pra,reprint,superscriptaddress,longbibliography]{revtex4-1}

\usepackage{amsmath,amsthm,amsfonts,amssymb,revsymb,bm}
\usepackage{url}
\usepackage{colonequals}
\usepackage{color}
\usepackage{times}
\usepackage{multirow}
\usepackage{graphicx}

\usepackage[colorlinks={true},pdftex]{hyperref}  
\hypersetup{citecolor={blue}, filecolor={blue}, linkcolor={blue}, urlcolor={blue},
  colorlinks=true, pdfauthor={Constantin Brif, et al.},
  pdftitle={Optimizing squeezing in a coherent quantum feedback network}}

\begin{document}

\title{Optimizing squeezing in a coherent quantum feedback network of optical parametric oscillators}

\author{Constantin Brif}
\affiliation{Sandia National Laboratories, Livermore, CA 94550, USA}
\author{Mohan Sarovar}
\affiliation{Sandia National Laboratories, Livermore, CA 94550, USA}
\author{Daniel B. S. Soh}
\affiliation{Sandia National Laboratories, Livermore, CA 94550, USA}
\affiliation{Ginzton Laboratory, Stanford University, Stanford, CA 94305, USA}
\author{David R. Farley}
\affiliation{Sandia National Laboratories, Livermore, CA 94550, USA}
\author{Scott E. Bisson}
\affiliation{Sandia National Laboratories, Livermore, CA 94550, USA}

\date{\today}

\begin{abstract}
Advances in the emerging field of coherent quantum feedback control (CQFC) have led to the development of new capabilities in the areas of quantum control and quantum engineering, with a particular impact on the theory and applications of quantum optical networks. We consider a CQFC network consisting of two coupled optical parametric oscillators (OPOs) and study the squeezing spectrum of its output field. The performance of this network as a squeezed-light source with desired spectral characteristics is optimized by searching over the space of model parameters with experimentally motivated bounds. We use the QNET package to model the network's dynamics and the PyGMO package of global optimization algorithms to maximize the degree of squeezing at a selected sideband frequency or the average degree of squeezing over a selected bandwidth. The use of global search methods is critical for identifying the best possible performance of the CQFC network, especially for squeezing at higher-frequency sidebands and higher bandwidths. The results demonstrate that the CQFC network of two coupled OPOs makes it possible to vary the squeezing spectrum, effectively utilize the available pump power, and overall significantly outperform a single OPO. Additionally, the Hessian eigenvalue analysis shows that the squeezing generation performance of the optimally operated CQFC network is robust to small variations of phase parameters.
\end{abstract}

\maketitle

\section{Introduction}
\label{sec:intro}

Feedback control is ubiquitous in classical engineering. However, its extension to the quantum realm has been challenging due to the unique character of the quantum measurement, which requires coupling of the observed quantum system to a classical measurement apparatus. Consequently, measurement-based quantum control has to deal with the fundamental effect of stochastic measurement back action on the quantum system, along with the need to amplify quantum signals up to macroscopic levels and high latency of classical controllers in comparison to typical quantum dynamic time scales~\cite{Brif.NJP.12.075008.2010, Wiseman.Milburn.book.2014}. An alternative approach that has attracted significant interest in the last decade is coherent quantum feedback control (CQFC)~\cite{Zhang.James.CSB.57.2200.2012, Gough.PTRSA.370.5241.2012, Combes.arXiv.1611.00375.2016}, which considers networks where the quantum system of interest (called the \emph{plant}) is controlled via coupling (either direct or, more often, through intermediate quantum fields) to an auxiliary quantum system (called the \emph{controller}). CQFC schemes utilize coherent quantum signals circulating between the plant and controller, thus avoiding the need for signal amplification and associated excess noise. Also, both plant and controller can evolve on the same time scale, which eliminates the latency issues. Due to these advantages, CQFC makes it possible to engineer quantum networks with new and unique characteristics~\cite{Gough.PTRSA.370.5241.2012, Combes.arXiv.1611.00375.2016, Jacobs.NJP.16.073036.2014, Yamamoto.PRX.4.041029.2014}.

The theoretical foundation of CQFC is a powerful framework based on input-output theory, which is used for modeling networks of open quantum systems connected by electromagnetic fields~\cite{Hudson.CMP.93.301.1984, Gardiner.Collett.PRA.31.3761.1985, Gardiner.PRL.70.2269.1993, Wiseman.Milburn.PRA.49.4110.1994} (see also~\cite{Zhang.James.CSB.57.2200.2012, Gough.PTRSA.370.5241.2012, Combes.arXiv.1611.00375.2016} for reviews). Moreover, recent developments, including the SLH formalism~\cite{Gough.James.IEEE-TAC.54.2530.2007, Gough.James.CMP.287.1109.2009, Gough.PRA.81.023804.2010}, the quantum hardware description language (QHDL)~\cite{Tezak.PTRSA.370.5270.2012}, and the QNET software package~\cite{QNET.url}, have added important capabilities for, respectively, modular analysis, specification, and simulation of such quantum optical networks. Together, the existing theoretical tools enable efficient and automated design and modeling of CQFC networks.

Proposed and experimentally demonstrated applications of CQFC include the development of autonomous devices for preparation, manipulation, and stabilization  of quantum states~\cite{Kerckhoff.PRL.105.040502.2010, Kerckhoff.NJP.13.055022.2011, Hamerly.PRL.109.173602.2012, Hamerly.PRA.87.013815.2013, Zhang.IEEE-TAC.57.1997.2012, Liu.JPB.48.105501.2015}, disturbance rejection by a dynamic compensator~\cite{Mabuchi.PRA.78.032323.2008}, linear-optics implementation of a modular quantum memory~\cite{Nurdin.Gough.QIC.15.1017.2015}, generation of optical squeezing~\cite{Gough.Wildfeuer.PRA.80.042107.2009, Iida.IEEE-TAC.57.2045.2012, Crisafulli.OE.21.18371.2013, Nemet.Parkins.PRA.94.023809.2016}, generation of quantum entanglement between optical field modes~\cite{Yan.PRA.84.062304.2011, Zhou.SciRep.5.11132.2015, Shi.Nurdin.QIP.14.337.2015, Shi.Nurdin.arXiv.1502.01070.2015, Shi.Nurdin.QIC.15.1141.2015, Shi.Nurdin.arXiv.1508.04584.2015}, coherent estimation of open quantum systems~\cite{Miao.PRA.92.012115.2015, Roy.arXiv.1502.03729.2016}, and ultra-low-power optical processing elements for optical switching~\cite{Mabuchi.PRA.80.045802.2009, Mabuchi.APL.98.193109.2011, Santori.PRAppl.1.054005.2014} and analog computing~\cite{Pavlichin.Mabuchi.NJP.16.105017.2014, Tezak.Mabuchi.EPJ-QT.2.10.2015}. In addition to tabletop bulk-optics implementations, CQFC networks have been also implemented using integrated silicon photonics~\cite{Sarovar.EPJ-QT.3.14.2016} and superconducting microwave devices~\cite{Kerckhoff.PRL.109.153602.2012, Kerckhoff.PRX.3.021013.2013}.

Squeezed states of light~\cite{Collett.Gardiner.PRA.30.1386.1984, Collett.Walls.PRA.32.2887.1985, Wu.JOSAB.4.1465.1987, Lvovsky.chapter.2015} have found numerous applications in quantum metrology and quantum information sciences, including interferometric detection of gravitational waves~\cite{Caves.PRD.23.1693.1981, Grote.PRL.110.181101.2013}, continuous-variable quantum key distribution (CV-QKD)~\cite{Furrer.PRL.109.100502.2012, Furrer.PRA.90.042325.2014, Madsen.NatCommun.3.1083.2012, Jacobsen.arXiv.1408.4566.2014, Eberle.NJP.15.053049.2013, Gehring.NatCommun.6.8795.2015}, generation of Gaussian entanglement~\cite{Eberle.NJP.15.053049.2013, Gehring.NatCommun.6.8795.2015, Ast.OL.41.5094.2016}, and quantum computing with continuous-variable cluster states~\cite{Yukawa.PRA.78.012301.2008, Gu.PRA.79.062318.2009, Weedbrook.RMP.84.621.2012, Menicucci.PRL.112.120504.2014}. Different applications require squeezed states with different properties. For example, detectable gravitational waves are expected to have frequencies in the range from $10$~Hz to $10$~kHz, and, consequently, quadrature squeezed states used to increase the measurement sensitivity in interferometric detectors should have a high degree of squeezing at sideband frequencies in this range. On the other hand, in CV-QKD the secure key rate is proportional to the bandwidth of squeezing, and hence it would be useful to generate states with squeezing bandwidth extending to $100$~MHz or even higher. It would be also of interest to extend the maximum of squeezing to high sideband frequencies.

In recent years, there have been remarkable advances in the generation of squeezed states~\cite{Takeno.OE.15.4321.2007, Vahlbruch.PRL.97.011101.2006, Eberle.PRL.104.251102.2010, Mehmet.OE.19.25763.2011, Khalaidovski.CQG.29.075001.2012, Mehmet.PRA.81.013814.2010, Ast.OL.37.2367.2012, Ast.OE.21.13572.2013, Baune.OE.23.16035.2015, Yan.PRA.85.040305.2012, Kaiser.Optica.3.362.2016, Dutt.PRAppl.3.044005.2015, Dutt.OL.41.223.2016}, however, achieving significant control over the squeezing spectrum still remains an ongoing effort. In 2009, Gough and Wildfeuer~\cite{Gough.Wildfeuer.PRA.80.042107.2009} proposed to enhance squeezing in the output field of a degenerate optical parametric oscillator (OPO) by incorporating the OPO into a CQFC network, where a part of the output beam is split off and then fed back into the OPO. Iida et al.~\cite{Iida.IEEE-TAC.57.2045.2012} reported an experimental demonstration of this scheme, while N\'emet and Parkins~\cite{Nemet.Parkins.PRA.94.023809.2016} proposed to modify it by including a time delay into the feedback loop. Another significant modification of this scheme was proposed and experimentally demonstrated by Crisafulli~et~al.~\cite{Crisafulli.OE.21.18371.2013}, who included a second OPO to act as the controller, with the plant OPO and the controller OPO coupled by two fields propagating between them in opposite directions. Due to the presence of quantum-limited gains in both arms of the feedback loop, this CQFC network has a very rich dynamics. In particular, by tuning the network's parameters it is possible to significantly vary the squeezing spectrum of its output field, for example, shift the maximum of squeezing from the resonance to a high-frequency sideband~\cite{Crisafulli.OE.21.18371.2013}.

The full range of performance of the CQFC network of two coupled OPOs as a squeezed-light source, however, still remains to be explored. In this paper, we study the limits of the network's performance by performing two types of optimizations: (1) maximizing the degree of squeezing at a chosen sideband frequency and (2) maximizing the average degree of squeezing over a chosen bandwidth; in both cases, the searches are executed over the space of network parameters with experimentally motivated bounds. To maximize the chances of finding a globally optimal solution, we use the PyGMO package of global optimization algorithms~\cite{pygmo.url} and employ a hybrid strategy which executes in parallel eight searches (using seven different global algorithms). Before each optimization is completed, the searches are repeated multiple times, and intermediate solutions are exchanged between them after each repetition. This strategy enabled us to discover that the CQFC network, when optimally operated, is capable of achieving a remarkably high degree of squeezing at sideband frequencies and bandwidths as high as $100$~MHz, with a very effective utilization of the available pump power. We also find that the obtained optimal solutions are quite robust to small variations of phase parameters.

\section{Background}
\label{sec:back}

The derivations in this section largely follow those in Refs.~\cite{Gough.Wildfeuer.PRA.80.042107.2009, Crisafulli.OE.21.18371.2013}, with some additional details and modifications.

\subsection{Input-output model of a quantum optical network}
\label{sec:IO-model}

Consider a network of coupled linear and bilinear optical elements such as mirrors, beam-splitters, phase-shifters, lasers, and degenerate OPOs. The quantum theory of such a network considers quantized cavity field modes which are coupled through cavity mirrors to external (input and output) quantum fields~\cite{Gardiner.Collett.PRA.31.3761.1985, Gardiner.PRL.70.2269.1993, Wiseman.Milburn.PRA.49.4110.1994}. Let $n$ be the number of the network's input ports (equal to the number of output ports) and $m$ be the number of cavities (in this model, we assume that each cavity supports one internal field mode). Let $\mathbf{a}$, $\mathbf{a}_{\mathrm{in}}$, and $\mathbf{a}_{\mathrm{out}}$ denote vectors of boson annihilation operators for, respectively, the cavity modes, the input fields, and the output fields:
\begin{equation}
\label{eq:mode-vectors}
\mathbf{a} = \begin{bmatrix} a_1 \\ \vdots \\ a_m \end{bmatrix} , \quad
\mathbf{a}_{\mathrm{in}} = \begin{bmatrix} a_{\mathrm{in},1} \\ \vdots \\ a_{\mathrm{in},n} \end{bmatrix} , \quad
\mathbf{a}_{\mathrm{out}} = \begin{bmatrix} a_{\mathrm{out},1} \\ \vdots \\ a_{\mathrm{out},n} \end{bmatrix} .
\end{equation}

Assuming that all input fields are in the vacuum state, the network is fully described by the $(\mathbf{S}, \mathbf{L}, H)$ model (also called the SLH model)~\cite{Gough.James.IEEE-TAC.54.2530.2007, Gough.James.CMP.287.1109.2009, Gough.PRA.81.023804.2010}, which includes the $n \times n$ matrix $\mathbf{S}$ that describes the scattering of external fields, the $n$-dimensional vector $\mathbf{L}$ that describes the coupling of cavity modes and external fields, and the Hamiltonian $H$ that describes the intracavity dynamics. For the model considered here, elements $\{S_{i j}\}$ of $\mathbf{S}$ are c-numbers, while $H$ and elements $\{L_i\}$ of $\mathbf{L}$ are operators on the combined Hilbert space of all cavity modes in the network. The Heisenberg equations of motion (also known as quantum Langevin equations) for the cavity mode operators $\{a_{\ell}(t)\}$ are ($\hbar = 1$)
\begin{equation}
\label{eq:HEOM-1}
\frac{d a_{\ell}}{d t} = -i [a_{\ell},H] + \mathcal{L}_L [a_{\ell}] + \Gamma_l ,
\quad \ell = 1, \ldots, m .
\end{equation}
Here, $\mathcal{L}_L$ is the Lindblad superoperator:
\begin{equation}
\label{eq:Lindblad-superoperator}
\mathcal{L}_L [a_{\ell}] = \sum_{i = 1}^n \left( L_i^{\dag} a_{\ell} L_i 
- \frac{1}{2} L_i^{\dag} L_i a_{\ell} - \frac{1}{2} a_{\ell} L_i^{\dag} L_i \right) ,
\end{equation}
and $\Gamma_l$ is the noise operator:
\begin{equation}
\label{eq:noise-operator}
\Gamma_l =  \mathbf{a}_{\mathrm{in}}^\dag \mathbf{S}^\dag [a_{\ell},\mathbf{L}] 
+ [\mathbf{L}^\dag , a_{\ell}] \mathbf{S} \mathbf{a}_{\mathrm{in}} ,
\end{equation}
where $\mathbf{a}_{\mathrm{in}}^\dag = [a_{\mathrm{in},1}^\dag , \ldots , a_{\mathrm{in},n}^\dag]$ and $\mathbf{L}^\dag = [L_1^\dag , \ldots , L_n^\dag]$ are row vectors of respective Hermitian conjugate operators. The generalized boundary condition for the network is
\begin{equation}
\label{eq:boundary-condition}
\mathbf{a}_{\mathrm{out}} = \mathbf{S} \mathbf{a}_{\mathrm{in}} + \mathbf{L} .
\end{equation}

For the type of networks that we consider, elements of $\mathbf{L}$ are linear in annihilation operators of the cavity modes, i.e., 
\begin{equation}
\label{eq:L-vector}
\mathbf{L} = \mathbf{K} \mathbf{a} ,
\end{equation}
where $\mathbf{K}$ is an $n \times m$ complex matrix with elements $\{K_{i \ell} = [ L_i, a_{\ell}^\dag ]\}$,
and the Hamiltonian has the bilinear form:
\begin{equation}
\label{eq:Ham-2}
H = \mathbf{a}^\dag \mathbf{\Omega} \mathbf{a} 
+ {\textstyle\frac{i}{2}} \mathbf{a}^\dag \mathbf{W} \mathbf{a}^\ddag
- {\textstyle\frac{i}{2}} \mathbf{a}^{\mathsf{T}} \mathbf{W}^\dag \mathbf{a} ,
\end{equation}
where $\mathbf{a}^\dag = [a_1^\dag , \ldots , a_m^\dag]$ and $\mathbf{a}^\ddag = \mathbf{a}^{\dag \mathsf{T}}$ are, respectively, row and column vectors of boson creation operators for the cavity modes, $\mathbf{\Omega}$ is an $m \times m$ Hermitian matrix, and $\mathbf{W}$ is an $m \times m$ complex matrix. With such $\mathbf{L}$ and $H$, the Heisenberg equations of motion~\eqref{eq:HEOM-1} take the form:
\begin{equation}
\label{eq:HEOM-2}
\frac{d \mathbf{a}}{d t} = \mathbf{V} \mathbf{a} + \mathbf{W} \mathbf{a}^\ddag 
+ \mathbf{Y} \mathbf{a}_{\mathrm{in}} ,
\end{equation}
where $\mathbf{V} = - \frac{1}{2} \mathbf{K}^\dag \mathbf{K} - i \mathbf{\Omega}$ is an $m \times m$ complex matrix and $\mathbf{Y} = -\mathbf{K}^\dag \mathbf{S}$ is an $m \times n$ complex matrix.

To obtain the transfer-matrix function from input to output fields, we seek the solution of Eq.~\eqref{eq:HEOM-2} in the frequency domain. Using the Fourier transform, we define:
\begin{subequations}
\label{eq:FT}
\begin{align}
& b(t) = \frac{1}{\sqrt{2\pi}} \int_{-\infty}^{\infty} d \omega \, b(\omega) e^{-i \omega t} , \\
& b^\dag(t) = \frac{1}{\sqrt{2\pi}} \int_{-\infty}^{\infty} d \omega \, b^\dag(-\omega) e^{-i \omega t} ,
\end{align}
\end{subequations}
where $b(t)$ stands for any element of $\mathbf{a}(t)$, $\mathbf{a}_{\mathrm{in}}(t)$, and $\mathbf{a}_{\mathrm{out}}(t)$. The field operators are in the interaction frame, and therefore $\omega$ is the sideband frequency (relative to the carrier frequency). We also use the double-length column vectors of the form:
\begin{equation}
\label{eq:b-breve}
\breve{\mathbf{b}}(\omega) = \begin{bmatrix} \mathbf{b}(\omega) \\ \mathbf{b}^\ddag (-\omega) 
\end{bmatrix} ,
\end{equation}
where $\mathbf{b}(\omega)$ stands for either of $\mathbf{a}(\omega)$, $\mathbf{a}_{\mathrm{in}}(\omega)$, and $\mathbf{a}_{\mathrm{out}}(\omega)$. With this notation, Eq.~\eqref{eq:HEOM-2} together with its Hermitian conjugate can be transformed into one matrix equation and solved for $\breve{\mathbf{a}}(\omega)$ in the frequency domain:
\begin{equation}
\label{eq:HEOM-FD}
\breve{\mathbf{a}}(\omega) = (\breve{\mathbf{A}} + i\omega \mathbf{I}_{2 m})^{-1} 
\breve{\mathbf{K}}^\dag \breve{\mathbf{S}}\, \breve{\mathbf{a}}_{\mathrm{in}}(\omega) .
\end{equation}
Here, $\mathbf{I}_{2 m}$ is the $2 m \times 2 m$ identity matrix, $\breve{\mathbf{A}} = \Delta(\mathbf{V},\mathbf{W})$, $\breve{\mathbf{K}} = \Delta(\mathbf{K},\mathbf{0})$, $\breve{\mathbf{S}} = \Delta(\mathbf{S},\mathbf{0})$, and we use the notation:
$\Delta(\mathbf{A},\mathbf{B}) = \begin{bmatrix} \mathbf{A} & \mathbf{B} \\ \mathbf{B}^\ast & \mathbf{A}^\ast
\end{bmatrix}$.
Analogously, the boundary condition of Eq.~\eqref{eq:boundary-condition} together with its Hermitian conjugate can be transformed into one matrix equation in the frequency domain:
\begin{equation}
\label{eq:boundary-condition-FD}
\breve{\mathbf{a}}_{\mathrm{out}}(\omega) = \breve{\mathbf{S}} \breve{\mathbf{a}}_{\mathrm{in}}(\omega) 
+ \breve{\mathbf{K}} \breve{\mathbf{a}}(\omega) .
\end{equation}
In Eqs.~\eqref{eq:HEOM-FD} and \eqref{eq:boundary-condition-FD}, $\breve{\mathbf{a}}(\omega)$ is a $2 m$-dimensional vector, $\breve{\mathbf{a}}_{\mathrm{in}}(\omega)$ and $\breve{\mathbf{a}}_{\mathrm{out}}(\omega)$ are $2 n$-dimensional vectors, $\breve{\mathbf{A}}$ is a $2 m \times 2 m$ matrix, $\breve{\mathbf{K}}$ is a $2 n \times 2 m$ matrix, and $\breve{\mathbf{S}}$ is a $2 n \times 2 n$ matrix. By substituting Eq.~\eqref{eq:HEOM-FD} into Eq.~\eqref{eq:boundary-condition-FD}, one obtains the quantum input-output relations in the matrix form:
\begin{equation}
\label{eq:IO-FD}
\breve{\mathbf{a}}_{\mathrm{out}}(\omega) = \breve{\mathbf{Z}}(\omega) \breve{\mathbf{a}}_{\mathrm{in}}(\omega) ,
\end{equation}
where
\begin{equation}
\label{eq:TF-1}
\breve{\mathbf{Z}}(\omega) = \left[ \mathbf{I}_{2 n} 
+ \breve{\mathbf{K}} (\breve{\mathbf{A}} + i\omega \mathbf{I}_{2 m})^{-1} \breve{\mathbf{K}}^\dag \right] 
\breve{\mathbf{S}} 
\end{equation}
is the network's transfer-matrix function. The $2 n \times 2n$ matrix $\breve{\mathbf{Z}}(\omega)$ can be decomposed into the block form:
\begin{equation}
\label{eq:TF-2}
\breve{\mathbf{Z}}(\omega) = \begin{bmatrix} \mathbf{Z}^- (\omega) & \mathbf{Z}^+ (\omega) \\ 
{\mathbf{Z}^+ (-\omega)}^\ast & {\mathbf{Z}^- (-\omega)}^\ast \end{bmatrix} ,
\end{equation}
where $\mathbf{Z}^- (\omega)$ and $\mathbf{Z}^+ (\omega)$ are $n \times n$ matrices. Correspondingly, input-output relations of Eq.~\eqref{eq:IO-FD} can be expressed for each of the output fields ($i = 1,\ldots,n$) as:
\begin{subequations}
\label{eq:IO-FD-1}
\begin{align}
a_{\mathrm{out},i}(\omega) = \sum_{j=1}^n \big[ & Z_{i j}^-(\omega) a_{\mathrm{in},j}(\omega) \nonumber \\
& + Z_{i j}^+(\omega) a_{\mathrm{in},j}^\dag(-\omega) \big], \\
a_{\mathrm{out},i}^\dag(-\omega) = \sum_{j=1}^n \big[ & {Z_{i j}^+(-\omega)}^\ast a_{\mathrm{in},j}(\omega) \nonumber \\
& + {Z_{i j}^-(-\omega)}^\ast a_{\mathrm{in},j}^\dag(-\omega) \big].
\end{align}
\end{subequations}

\subsection{Squeezing spectrum}
\label{sec:SS}

Consider the quadrature of the $i$th output field in time and frequency domains:
\begin{subequations}
\label{eq:X-quadrature}
\begin{align}
& X_i (t,\theta) = a_{\mathrm{out},i}(t) e^{-i \theta} + a_{\mathrm{out},i}^\dag(t) e^{i \theta} , \\
& X_i (\omega,\theta) = a_{\mathrm{out},i}(\omega) e^{-i \theta} + a_{\mathrm{out},i}^\dag(-\omega) e^{i \theta} ,
\end{align}
\end{subequations}
where $\theta$ is the homodyne phase. The power spectral density of the quadrature's quantum noise (commonly referred to as the \emph{squeezing spectrum}) is~\cite{Collett.Gardiner.PRA.30.1386.1984, Collett.Walls.PRA.32.2887.1985}:
\begin{equation}
\label{eq:spectrum-1}
\mathcal{P}_i(\omega,\theta) = 1 + \int_{-\infty}^{\infty} d \omega' 
\langle {:\! X_i (\omega,\theta) , X_i (\omega',\theta)\! :} \rangle ,
\end{equation}
where $:\,:$ denotes the normal ordering of boson operators and $\langle x , y \rangle = \langle x y \rangle - \langle x \rangle \langle y \rangle$. Since all input fields are in the vacuum state, $\langle X_i (\omega,\theta) \rangle = \langle X_i (\omega',\theta) \rangle = 0$, and one obtains:
\begin{align}
\mathcal{P}_i(\omega,\theta) = & \: 1 + \mathcal{N}_i(\omega) + \mathcal{N}_i(-\omega) \nonumber \\
& + \mathcal{M}_i(\omega) e^{-2 i \theta} + {\mathcal{M}_i(\omega)}^\ast e^{2 i \theta} ,
\label{eq:spectrum-2}
\end{align}
where
\begin{subequations}
\label{eq:NM-spectrum-1}
\begin{align}
& \mathcal{N}_i(\omega) = \int_{-\infty}^{\infty} d \omega' 
\langle a_{\mathrm{out},i}^\dag(-\omega') a_{\mathrm{out},i}(\omega) \rangle , \\
& \mathcal{M}_i(\omega) = \int_{-\infty}^{\infty} d \omega' 
\langle a_{\mathrm{out},i}(\omega) a_{\mathrm{out},i}(\omega') \rangle .
\end{align}
\end{subequations}
By substituting Eqs.~\eqref{eq:IO-FD-1} into Eqs.~\eqref{eq:NM-spectrum-1} and evaluating expectation values for vacuum input fields, one obtains:
\begin{subequations}
\label{eq:NM-spectrum-2}
\begin{align}
& \mathcal{N}_i(\omega) = \sum_{j=1}^n \left| Z_{i j}^+(\omega) \right|^2 , \\
& \mathcal{M}_i(\omega) = \sum_{j=1}^n Z_{i j}^-(\omega) Z_{i j}^+(-\omega) .
\end{align}
\end{subequations}

In this work, we are only concerned with squeezing properties of the field at one of the output ports. We will designate this port as corresponding to $i = 1$ and denote the squeezing spectrum of this output field as $\mathcal{P}(\omega,\theta) = \mathcal{P}_1(\omega,\theta)$. In squeezing generation, the figure of merit is the quantum noise change relative to the vacuum level, measured in decibels, and since $\mathcal{P}_{\mathrm{vac}}(\omega,\theta) = 1$, the corresponding spectral quantity is
\begin{equation}
\label{eq:Q}
\mathcal{Q}(\omega,\theta) = 10 \log_{10} \mathcal{P}(\omega,\theta) .
\end{equation}
Negative values of $\mathcal{Q}$ correspond to quantum noise reduction below the vacuum level (i.e., squeezing of the quadrature uncertainty). The maximum degree of squeezing corresponds to the minimum value of $\mathcal{Q}$. The maximum and minimum of $\mathcal{P}(\omega,\theta)$ as a function of $\theta$,
\begin{equation}
\label{eq:Ppm}
\mathcal{P}^+(\omega) = \max_{\theta} \mathcal{P}(\omega,\theta) , \quad
\mathcal{P}^-(\omega) = \min_{\theta} \mathcal{P}(\omega,\theta) ,
\end{equation}
are power spectral densities of the quantum noise in anti-squeezed and squeezed quadrature, respectively. Analogously to Eq.~\eqref{eq:Q}, logarithmic spectral measures of anti-squeezing and squeezing for the two quadratures are defined as $\mathcal{Q}^{\pm}(\omega) = 10 \log_{10} \mathcal{P}^{\pm}(\omega)$, respectively.
Expressing $\mathcal{M}(\omega)$ as $\mathcal{M}(\omega) = |\mathcal{M}(\omega)| e^{i\theta_{\mathcal{M}}(\omega)}$ and using Eq.~\eqref{eq:spectrum-2}, it is easy to find (we omit the subscript $i = 1$ for simplicity):
\begin{equation}
\mathcal{P}^{\pm}(\omega) = 1 + \mathcal{N}(\omega) + \mathcal{N}(-\omega) 
\pm 2 |\mathcal{M}(\omega)| ,
\end{equation}
with anti-squeezed and squeezed quadrature corresponding to $\theta = \theta_{\mathcal{M}}(\omega)/2$ and $\theta = [\theta_{\mathcal{M}}(\omega) - \pi]/2$, respectively. Note that, in general, these optimum values of the homodyne phase $\theta$ depend on the sideband frequency $\omega$, so, for example, if the goal is to maximize the degree of squeezing at a particular sideband frequency $\omega_{\mathrm{opt}}$, then the optimum phase value $\theta_{\mathrm{opt}} = [\theta_{\mathcal{M}}(\omega_{\mathrm{opt}}) - \pi]/2$ should be selected accordingly.

\section{Squeezing from a single OPO}
\label{sec:OPO-1}

\begin{figure}[htbp]
\centering
\includegraphics[width=1.0\columnwidth]{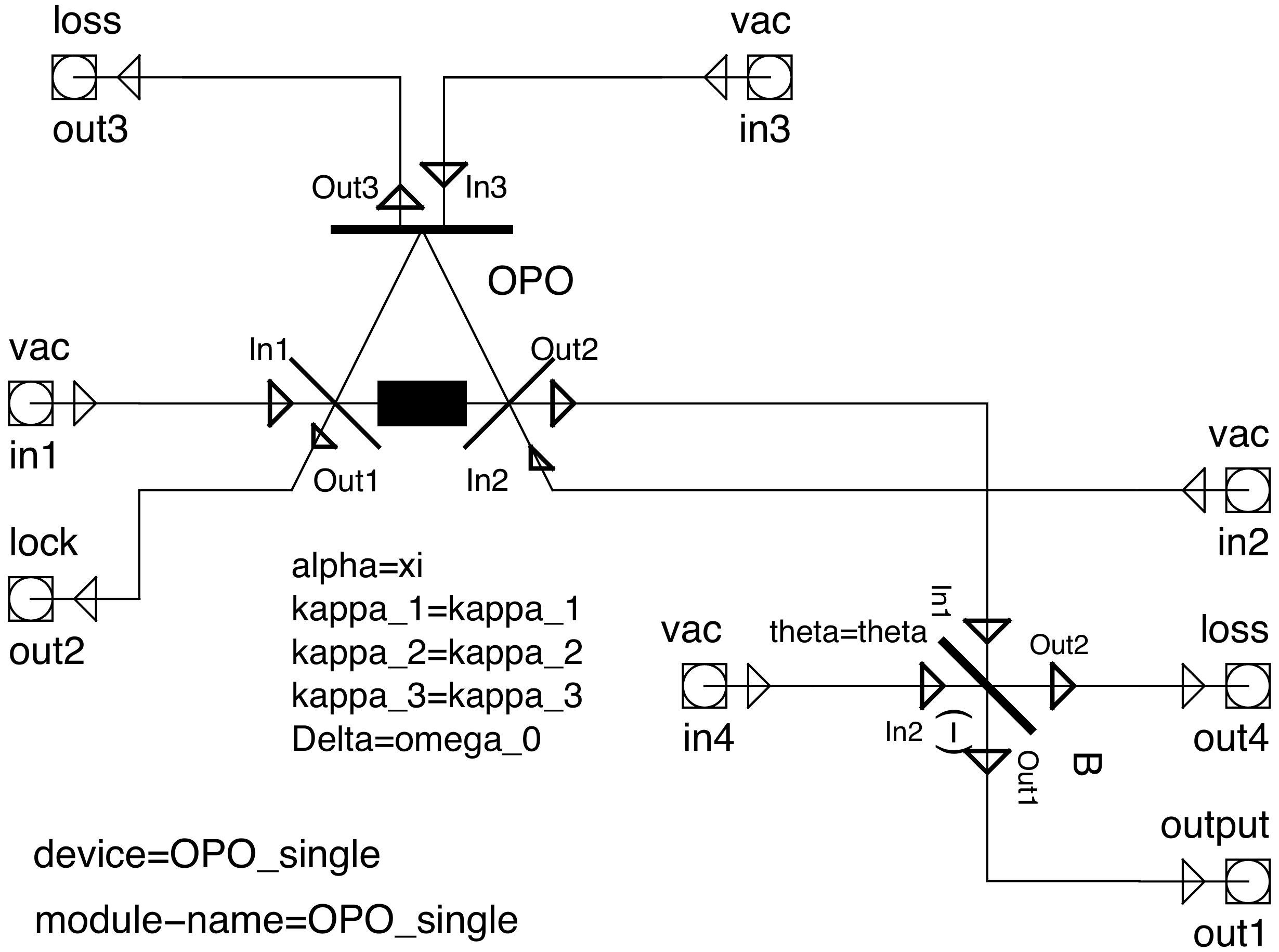}
\caption{A schematic depiction of the single OPO network.}
\label{fig:OPO_single_scheme}
\end{figure}

A network that produces squeezed light by means of a single degenerate OPO~\cite{Wu.JOSAB.4.1465.1987} is schematically shown in Fig.~\ref{fig:OPO_single_scheme}. The OPO consists of a nonlinear crystal enclosed in a Fabry-P\'erot cavity. The pump field for the OPO is assumed to be classical and not shown in the scheme. Each partially transparent mirror in the network (including cavity mirrors and a beamsplitter) has two input ports and two output ports. A vacuum field  enters into each input port. The OPO cavity has a fictitious third mirror to model intracavity losses (mainly due to absorption in the crystal as well as scattering and Fresnel reflection at the crystal's facets). The beamsplitter B models losses in the output transmission line (e.g., due to coupling into a fiber) and inefficiencies in the homodyne detector (not shown) used to measure the squeezing spectrum of the output field. Taking into account all optical elements, the network is modeled as having four input ports, four output ports, and one cavity mode ($n = 4$, $m = 1$).

\begin{table}[htbp]
\caption{\label{tab:params-1}Parameters of the single OPO network.}
\begin{ruledtabular}
\begin{tabular}{lll}
Parameter & Type & Description \\ \hline
$\kappa_1$ & Positive & Leakage rate for the left cavity mirror \\
$\kappa_2$ & Positive & Leakage rate for the right cavity mirror \\
$\kappa_3$ & Positive & Leakage rate for intracavity losses \\
$\omega_0$ & Real     & Frequency detuning of the cavity \\
$\xi$      & Complex  & Pump amplitude of the OPO \\
$\theta_{\mathrm{B}}$ & Real     & Rotation angle of the beamsplitter \\
\end{tabular}
\end{ruledtabular}
\end{table}

Parameters of the single OPO network are described in Table~\ref{tab:params-1}. With $\xi = |\xi| e^{i \theta_{\xi}}$, there is a total of seven real parameters. Note that we use angular frequencies throughout this paper. For each cavity mirror, the leakage rate is 
\begin{equation}
\label{eq:kappa-T-relation}
\kappa_i = \frac{c T_i}{2 l_{\mathrm{eff}}} , \quad i = 1,2,3,
\end{equation}
where $T_i$ is the power transmittance of the $i$th mirror, $c$ is the speed of light, and $l_{\mathrm{eff}}$ is the effective cavity length (taking into account the length and refractive index of the crystal). To simplify the notation, we also use alternative parameters:
\begin{equation}
\gamma = \kappa_1 + \kappa_2 + \kappa_3 , \\
\end{equation}
to denote the total leakage rate (including losses) from the cavity, and
\begin{equation}
t_{\mathrm{B}} = \cos(\theta_{\mathrm{B}}) , \ \ \ r_{\mathrm{B}} = \sin(\theta_{\mathrm{B}}) ,
\end{equation}
to denote, respectively, the transmittivity and reflectivity of the beamsplitter.

The QNET package~\cite{QNET.url} is used to derive the $(\mathbf{S}, \mathbf{L}, H)$ model of the network, and the resulting components of the model are
\begin{eqnarray*}
& \mathbf{S} = \begin{bmatrix} 
0 & t_{\mathrm{B}} & 0 & - r_{\mathrm{B}} \\ 
1 & 0 & 0 & 0 \\ 
0 & 0 & 1 & 0 \\ 
0 & r_{\mathrm{B}} & 0 & t_{\mathrm{B}} 
\end{bmatrix} , \quad
\mathbf{L} = \begin{bmatrix} 
\sqrt{\kappa_2} t_{\mathrm{B}} a \\  
\sqrt{\kappa_1} a \\  
\sqrt{\kappa_3} a \\ 
\sqrt{\kappa_2} r_{\mathrm{B}} a
\end{bmatrix}, & \\
& H = \omega_0 a^\dag a + {\textstyle\frac{i}{2}} \xi a^{\dag 2} - {\textstyle\frac{i}{2}} \xi^\ast a^2 , &
\end{eqnarray*}
where $a$ is the annihilation operator of the cavity field mode. Using the formalism of Sec.~\ref{sec:IO-model}, we obtain: $\mathbf{\Omega} = \omega_0$, $\mathbf{W} = \xi$, 
\begin{align*}
& \mathbf{K} = \left[ \sqrt{\kappa_2} t_{\mathrm{B}}, \sqrt{\kappa_1}, \sqrt{\kappa_3}, \sqrt{\kappa_2} r_{\mathrm{B}} \right]^{\mathsf{T}}, \\
& \mathbf{V} = -\eta , \quad
\mathbf{Y} = -\left[\sqrt{\kappa_1}, \sqrt{\kappa_2}, \sqrt{\kappa_3}, 0 \right], \\
& \breve{\mathbf{A}} = \begin{bmatrix}
- \eta & \xi \\
\xi^\ast & - \eta^\ast
\end{bmatrix}, \\
& (\breve{\mathbf{A}} + i\omega \mathbf{I}_2)^{-1} = \displaystyle
-\frac{1}{\lambda(\omega)} \begin{bmatrix}
\eta^\ast - i\omega & \xi \\
\xi^\ast & \eta - i\omega
\end{bmatrix}, 
\end{align*}
where we defined auxiliary parameters:
$$
\eta = {\textstyle\frac{1}{2}} \gamma + i \omega_0 , \quad
\lambda(\omega) = (\eta^\ast - i\omega)  (\eta - i\omega) - |\xi|^2 .
$$
These results make it straightforward to analytically compute the transfer-matrix function $\breve{\mathbf{Z}}(\omega)$ of Eq.~\eqref{eq:TF-1}. Since we are only interested in squeezing properties of the field at the output port~1, it is sufficient to use only the respective rows of matrices $\mathbf{Z}^- (\omega)$ and $\mathbf{Z}^+ (\omega)$, i.e.,
\begin{subequations}
\label{eq:IO-OPO-1}
\begin{align}
& \mathbf{Z}_1^-(\omega) = \frac{\sqrt{\kappa_2} t_{\mathrm{B}} (\eta^\ast - i\omega)
}{\lambda(\omega)} \mathbf{Y} 
+ \left[ 0, t_{\mathrm{B}}, 0, - r_{\mathrm{B}} \right] , \\
& \mathbf{Z}_1^+(\omega) = \frac{\sqrt{\kappa_2} t_{\mathrm{B}} \xi}{\lambda(\omega)} \mathbf{Y} .
\end{align}
\end{subequations}
By substituting elements of $\mathbf{Z}_1^-(\omega)$ and $\mathbf{Z}_1^+(\omega)$ into Eqs.~\eqref{eq:NM-spectrum-2}, we obtain:
\begin{subequations}
\label{eq:NM-spectrum-OPO-1}
\begin{align}
& \mathcal{N}_1(\omega) = \frac{\gamma \kappa_2 T_{\mathrm{B}} |\xi|^2}{|\lambda(\omega)|^2}, \\
& \mathcal{M}_1(\omega) = \frac{\gamma (\eta^\ast - i\omega) - \lambda(\omega)}{|\lambda(\omega)|^2} 
\kappa_2 T_{\mathrm{B}} \xi ,
\end{align}
\end{subequations}
where $T_{\mathrm{B}} = t_{\mathrm{B}}^2$ is the power transmittance of the beam splitter. Using Eq.~\eqref{eq:spectrum-2}, the resulting squeezing spectrum is
\begin{equation}
\label{eq:PSD-OPO-1}
\mathcal{P}(\omega,\theta) = 1 + 2 \kappa_2 T_{\mathrm{B}} |\xi| \frac{\gamma |\xi| 
+ \mu(\omega) \cos\varphi + \gamma \omega_0 \sin\varphi}{|\lambda(\omega)|^2} ,
\end{equation}
where $\mu(\omega) = \frac{1}{4}\gamma^2 + |\xi|^2 + \omega^2 - \omega_0^2$ and $\varphi = \theta_{\xi} - 2 \theta$. The spectra for anti-squeezed and squeezed quadrature are obtained as the maximum and minimum (cf.~Eq.~\eqref{eq:Ppm}) of $\mathcal{P}(\omega,\theta)$ in Eq.~\eqref{eq:PSD-OPO-1} for $\varphi = \tan^{-1} [\gamma \omega_0 / \mu(\omega)]$ and $\varphi = \tan^{-1} [\gamma \omega_0 / \mu(\omega)] + \pi$, respectively, and are given by
\begin{equation}
\label{eq:S-pm-1}
\mathcal{P}^{\pm}(\omega) = 1 \pm 2 \kappa_2 T_{\mathrm{B}} |\xi| \frac{\sqrt{\mu^2(\omega) 
+ \gamma^2 \omega_0^2} \pm \gamma |\xi|}{|\lambda(\omega)|^2} .
\end{equation}
In order to compare the theoretical spectra with experimental data, it is common to express the pump amplitude as
\begin{equation}
\label{eq:x-scaled}
|\xi| = {\textstyle\frac{1}{2}} \gamma x , \quad x = \sqrt{P/P_{\mathrm{th}}} ,
\end{equation}
where $P$ is the OPO pump power and $P_{\mathrm{th}}$ is its threshold value. Analogously to the scaled pump amplitude $x = 2 |\xi|/\gamma$, it is convenient to use scaled frequencies $\Omega = 2 \omega/\gamma$ and $\Omega_0 = 2 \omega_0/\gamma$. With this notation, Eq.~\eqref{eq:S-pm-1} takes the form:
\begin{equation}
\label{eq:S-pm-2}
\mathcal{P}^{\pm}(\omega) = 1 \pm 4 T_{\mathrm{B}} \rho x
\frac{\sqrt{(1 + y^2)^2 + 4 \Omega_0^2} \pm 2 x}{(1 - y^2)^2 + 4 \Omega^2} ,
\end{equation}
where $\rho = \kappa_2/\gamma = T_2/(T_1 + T_2 + L)$ is the escape efficiency of the cavity, $L = T_3$ denotes the intracavity power loss, and $y^2 = x^2 + \Omega^2 - \Omega_0^2$.

In the case of zero detuning, $\omega_0 = 0$, the squeezing spectrum of Eq.~\eqref{eq:PSD-OPO-1} becomes
\begin{equation}
\label{eq:PSD-OPO-1-0}
\mathcal{P}(\omega,\theta) = 1 + 2 \kappa_2 T_{\mathrm{B}} |\xi| \frac{\gamma |\xi| 
+ (\frac{1}{4}\gamma^2 + |\xi|^2 + \omega^2) \cos\varphi}{(\frac{1}{4}\gamma^2 - |\xi|^2 - \omega^2)^2
+ \gamma^2 \omega^2} .
\end{equation}
The corresponding spectra for anti-squeezed and squeezed quadrature are obtained for $\varphi = 0$ and $\varphi = \pi$, respectively. They can be expressed by taking $\Omega_0 = 0$ in Eq.~\eqref{eq:S-pm-2}, which reproduces the familiar result~\cite{Collett.Walls.PRA.32.2887.1985, Wu.JOSAB.4.1465.1987}:
\begin{equation}
\label{eq:S-pm-2-0}
\mathcal{P}^{\pm}(\omega) = 1 \pm T_{\mathrm{B}} \rho \frac{4 x}{(1 \mp x)^2 + \Omega^2} .
\end{equation}

The spectra of Eq.~\eqref{eq:S-pm-2-0} have Lorentzian shapes with maximum (for anti-squeezing) and minimum (for squeezing) at the resonance (zero sideband frequency), and with the degree of squeezing rapidly decreasing as the sideband frequency increases. For applications such as CV-QKD, it would be valuable to significantly extend the squeezing bandwidth. It would be also of interest to achieve a maximum degree of squeezing (i.e., a minimum value of $\mathcal{P}^{-}$) at a high-frequency sideband. Therefore, we investigate whether such modifications of the squeezing spectrum are possible by using a nonzero value of the cavity's frequency detuning. 

\begin{figure}[t]
\centering
\includegraphics[width=1.0\columnwidth]{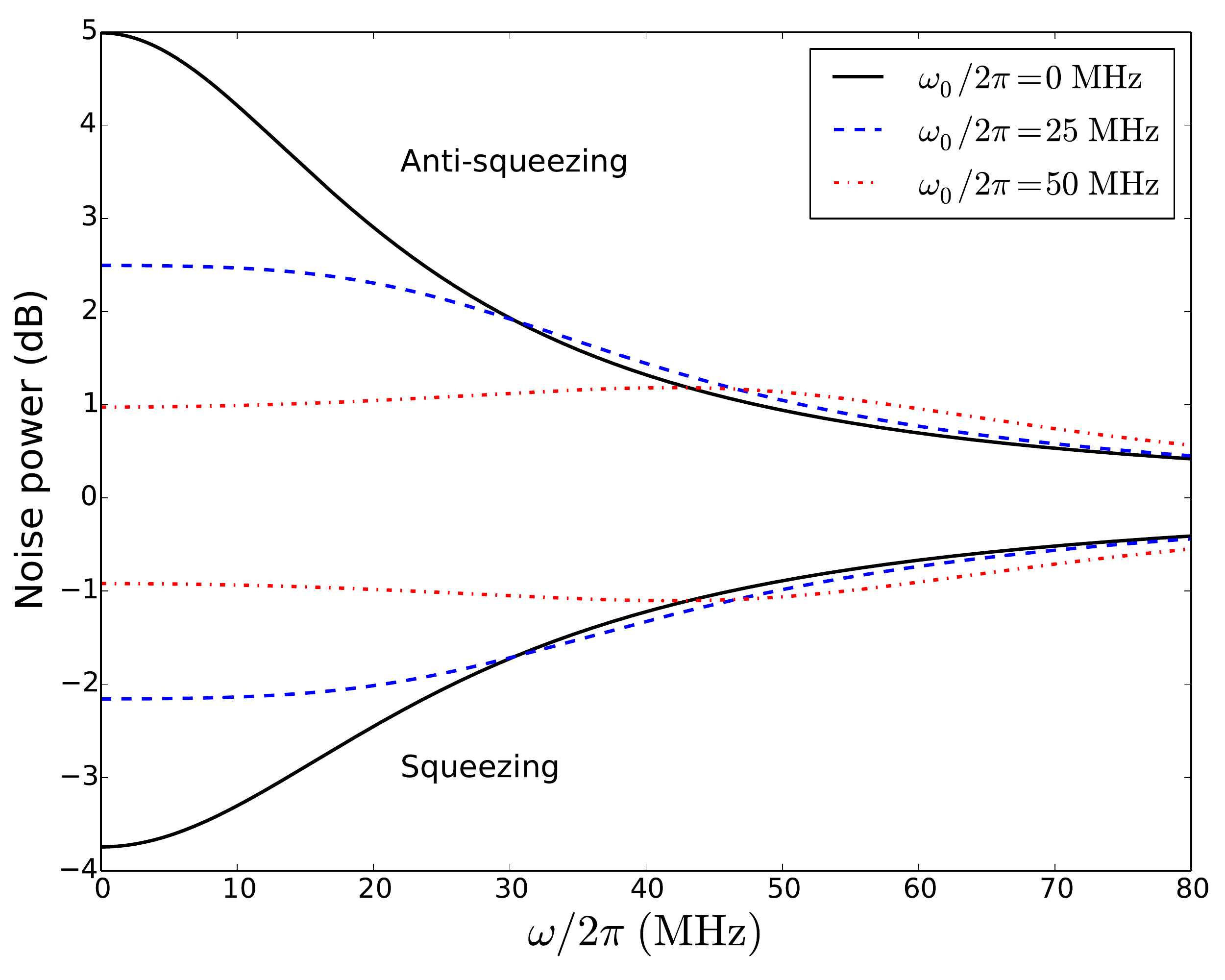}
\caption{Squeezing spectra of the output light field from a single OPO network with different values of the cavity's frequency detuning $\omega_0/2\pi$ (given in the legend). Logarithmic power spectral densities of the quantum noise in anti-squeezed and squeezed quadrature, $\mathcal{Q}^{\pm}(\omega) = 10 \log_{10} \mathcal{P}^{\pm}(\omega)$, are shown versus the sideband frequency $\omega/2\pi$ for $\mathcal{P}^{\pm}(\omega)$ of Eq.~\eqref{eq:S-pm-2}. The values of network parameters are listed in the text.}
\label{fig:OPO_single_spectrum}
\end{figure}

Consider a single OPO with a set of experimentally motivated parameters: pump power $P = 1.5$~W, pump wavelength $\lambda_p = 775$~nm, and signal wavelength $\lambda_s = 1550$~nm; an MgO:PPLN crystal with length $l_c = 20$~mm, refractive index (at $\lambda_s$) $n_s = 2.1$, and effective nonlinear coefficient $d_{\mathrm{eff}} = 14$~pm/V; a Fabry-P\'erot cavity with effective length $l_{\mathrm{eff}} = 87$~mm, left mirror reflectance $R_1 = 0.98$ ($T_1 = 0.02$, $\kappa_1 / 2\pi \approx 5.484$~MHz), right mirror reflectance $R_2 = 0.85$ ($T_2 = 0.15$, $\kappa_2 / 2\pi \approx 41.132$~MHz), intracavity loss $L = 0.02$ ($\kappa_3 / 2\pi \approx 5.484$~MHz), and total leakage rate $\gamma/2\pi \approx 52.1$~MHz; output transmission line loss $L_{\mathrm{tl}} = R_{\mathrm{B}} = 0$ ($T_{\mathrm{B}} = 1$). These parameters correspond to OPO's threshold power $P_{\mathrm{th}} \approx 14.86$~W and scaled pump amplitude $x = \sqrt{P/P_{\mathrm{th}}} \approx 0.318$. Using these parameters, we compute the squeezing spectra $\mathcal{P}^{\pm}(\omega)$ of Eq.~\eqref{eq:S-pm-2} for three detuning values: $\omega_0/2\pi = \{0, 25, 50\}$~MHz. The resulting logarithmic spectra $\mathcal{Q}^{\pm}(\omega) = 10 \log_{10} \mathcal{P}^{\pm}(\omega)$ for anti-squeezed and squeezed quadrature are shown in Fig.~\ref{fig:OPO_single_spectrum}. These results indicate that, while the use of nonzero detuning can increase the degree of squeezing at higher-frequency sidebands as compared to the case of $\omega_0 = 0$, this increase is very small. Also, no improvement in the squeezing bandwidth (quantified as the average degree of squeezing over a selected bandwidth) is achieved through the use of nonzero detuning. These observations motivate us to explore the use of the CQFC network with two coupled OPOs as a light source with the potential to generate a widely tunable squeezing spectrum.

\begin{figure*}[htbp]
\centering
\includegraphics[width=1.8\columnwidth]{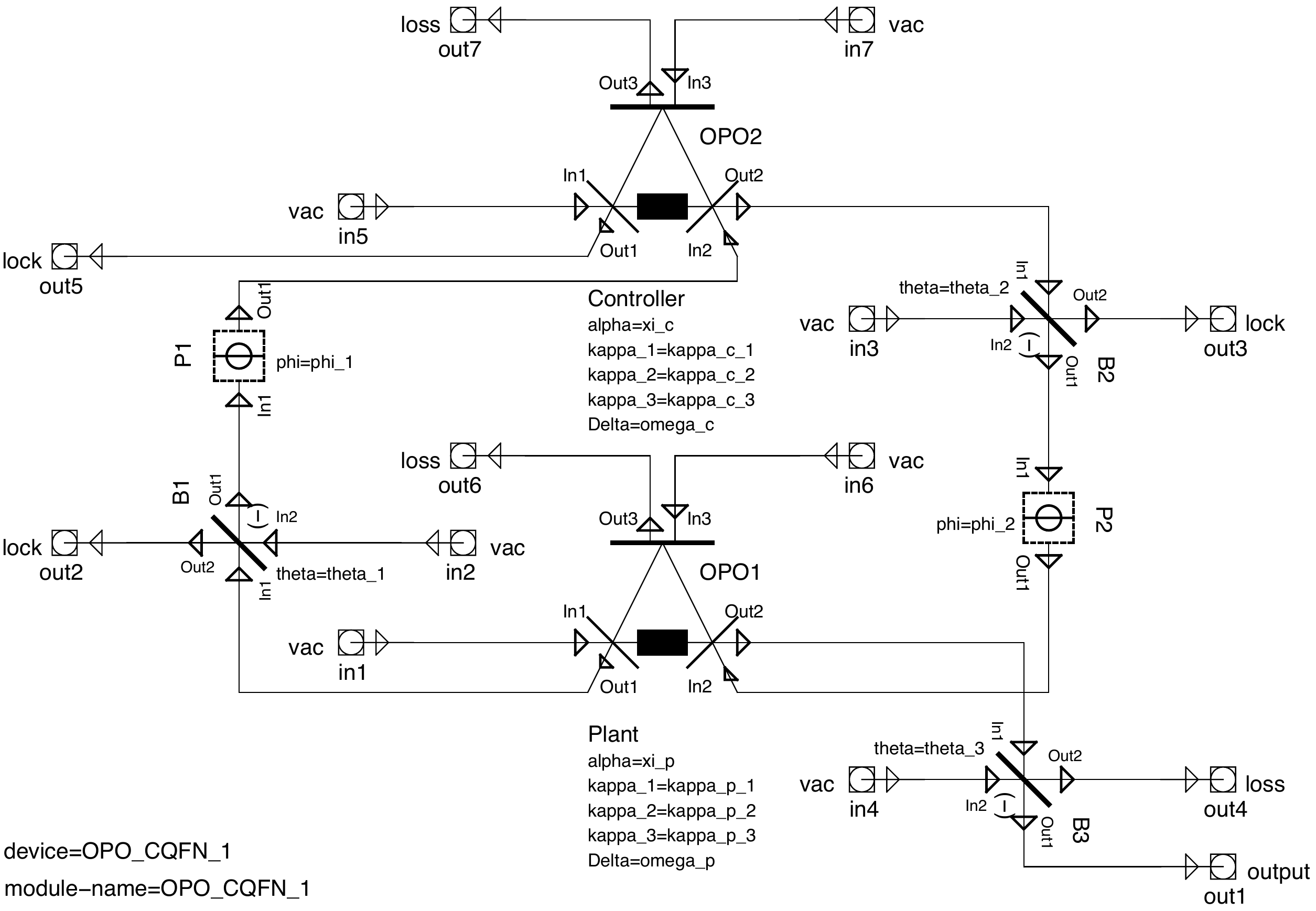}
\caption{A schematic depiction of the CQFC network of two coupled OPOs.}
\label{fig:OPO_CQFN_scheme}
\end{figure*}

\begin{table*}[htbp]
\caption{\label{tab:params-2}Parameters of the CQFC network of two coupled OPOs.}
\begin{ruledtabular}
\begin{tabular}{lll}
Parameter & Type & Description \\ \hline
$\kappa_{\mathrm{p} 1}$ & Positive & Leakage rate for the left mirror of the plant OPO cavity \\
$\kappa_{\mathrm{p} 2}$ & Positive & Leakage rate for the right mirror of the plant OPO cavity \\
$\kappa_{\mathrm{p} 3}$ & Positive & Leakage rate for losses in the plant OPO cavity \\
$\omega_{\mathrm{p}}$   & Real     & Frequency detuning of the plant OPO cavity \\
$\xi_{\mathrm{p}}$      & Complex  & Pump amplitude of the plant OPO \\ \hline
$\kappa_{\mathrm{c} 1}$ & Positive & Leakage rate for the left mirror of the controller OPO cavity \\
$\kappa_{\mathrm{c} 2}$ & Positive & Leakage rate for the right mirror of the controller OPO cavity \\
$\kappa_{\mathrm{c} 3}$ & Positive & Leakage rate for losses in the controller OPO cavity \\
$\omega_{\mathrm{c}}$   & Real     & Frequency detuning of the controller OPO cavity \\
$\xi_{\mathrm{c}}$      & Complex  & Pump amplitude of the controller OPO \\ \hline
$\phi_1$              & Real     & Phase shift of the first phase shifter \\
$\phi_2$              & Real     & Phase shift of the second phase shifter \\
$\theta_1$            & Real     & Rotation angle of the first beamsplitter \\
$\theta_2$            & Real     & Rotation angle of the second beamsplitter \\
$\theta_3$            & Real     & Rotation angle of the third beamsplitter \\
\end{tabular}
\end{ruledtabular}
\end{table*}

\section{Squeezing from a network of two coupled OPOs}
\label{sec:OPO-2}

The CQFC network that includes two coupled degenerate OPOs~\cite{Crisafulli.OE.21.18371.2013} is schematically shown in Fig.~\ref{fig:OPO_CQFN_scheme}. Each OPO consists of a nonlinear crystal enclosed in a Fabry-P\'erot cavity. Pump fields for both OPOs are assumed to be classical and not shown in the scheme. From the control theory perspective, OPO1 is considered to be the \emph{plant} and OPO2 the (quantum) \emph{controller}. Each partially transparent mirror in the network (including cavity mirrors and beamsplitters) has two input ports and two output ports. A vacuum field enters into each input port, except for two input ports of cavity mirrors used for the feedback loop between the plant and controller. Each OPO cavity has a fictitious third mirror to model intracavity losses. Beamsplitters B1 and B2 represent the light diverted to lock the cavities as well as losses in optical transmission lines between the OPO cavities. Beamsplitter B3 represents losses in the output transmission line (e.g., due to coupling into a fiber) and inefficiencies in the homodyne detector (not shown) used to measure the squeezing spectrum of the output field. Phase shifters P1 and P2 are inserted into transmission lines between the OPOs to manipulate the interference underlying the CQFC control. Taking into account the feedback loop between the plant and controller, the network is modeled as having seven input ports, seven output ports, and two cavity modes ($n = 7$, $m = 2$).

Parameters of the network of two coupled OPOs are listed in Table~\ref{tab:params-2}. With $\xi_{\mathrm{p}} = |\xi_{\mathrm{p}}| e^{i \theta_{\mathrm{p}}}$ and $\xi_{\mathrm{c}} = |\xi_{\mathrm{c}}| e^{i \theta_{\mathrm{c}}}$, there is a total of 17 real parameters. The relationship between leakage rate and power transmittance of a cavity mirror is given, similarly to Eq.~\eqref{eq:kappa-T-relation}, by
\begin{equation}
\label{eq:kappa-T-relation-2}
\kappa_{\mathrm{p} i} = \frac{c T_{\mathrm{p} i}}{2 l_{\mathrm{p,eff}}} , \quad
\kappa_{\mathrm{c} i} = \frac{c T_{\mathrm{c} i}}{2 l_{\mathrm{c,eff}}} , \quad i = 1,2,3,
\end{equation}
where $T_{\mathrm{p} i}$ ($T_{\mathrm{c} i}$) is the power transmittance of the $i$th mirror and $l_{\mathrm{p,eff}}$ ($l_{\mathrm{c,eff}}$) is the effective cavity length for the plant (controller). To simplify the notation, we also use alternative parameters:
\begin{equation}
\gamma_{\mathrm{p}} = \kappa_{\mathrm{p} 1} + \kappa_{\mathrm{p} 2} + \kappa_{\mathrm{p} 3} , \quad
\gamma_{\mathrm{c}} = \kappa_{\mathrm{c} 1} + \kappa_{\mathrm{c} 2} + \kappa_{\mathrm{c} 3}
\end{equation}
to denote the total leakage rate (including losses) from, respectively, the plant and controller cavities,
\begin{equation}
t_i = \cos(\theta_i) , \ \ \ r_i = \sin(\theta_i) , \ \ \ i = 1,2,3
\end{equation}
to denote, respectively, the transmittivity and reflectivity of each beamsplitter, and 
\begin{equation}
\phi = \phi_1 + \phi_2
\end{equation}
to denote the total phase shift for the feedback roundtrip path. Similarly to Eq.~\eqref{eq:x-scaled}, we also define the scaled pump amplitudes $x_{\mathrm{p}}$ and $x_{\mathrm{c}}$ for the plant and controller OPOs, respectively:
\begin{equation}
\label{eq:x-scaled-2}
x_{\mathrm{p}} = \frac{2 |\xi_{\mathrm{p}}| }{ \gamma_{\mathrm{p}} } = \sqrt{ \frac{P_{\mathrm{p}} }{ P_{\mathrm{p,th}} }} , \quad
x_{\mathrm{c}} = \frac {2 |\xi_{\mathrm{c}}| }{ \gamma_{\mathrm{c}} } = \sqrt{ \frac{P_{\mathrm{c}} }{ P_{\mathrm{c,th}} }} ,
\end{equation}
where $P_{\mathrm{p}}$ ($P_{\mathrm{c}}$) is the OPO pump power and $P_{\mathrm{p,th}}$ ($P_{\mathrm{c,th}}$) is its threshold value for the plant (controller).

The QNET package~\cite{QNET.url} is used to derive the $(\mathbf{S}, \mathbf{L}, H)$ model of the network, and the resulting components of the model are
\begin{equation}
\mathbf{S} = \begin{bmatrix} 
t_1 t_2 t_3 e^{i \phi} & - r_1 t_2 t_3 e^{i \phi} & - r_2 t_3 e^{i \phi_2} & - r_3 & 0 & 0 & 0 \\
r_1 & t_1 & 0 & 0 & 0 & 0 & 0 \\
t_1 r_2 e^{i \phi_1} & - r_1 r_2 e^{i \phi_1} & t_2 & 0 & 0 & 0 & 0 \\
t_1 t_2 r_3 e^{i \phi} & - r_1 t_2 r_3 e^{i \phi} & - r_2 r_3 e^{i \phi_2} & t_3 & 0 & 0 & 0 \\
0 & 0 & 0 & 0 & 1 & 0 & 0 \\
0 & 0 & 0 & 0 & 0 & 1 & 0 \\
0 & 0 & 0 & 0 & 0 & 0 & 1
\end{bmatrix},
\end{equation}
\begin{equation}
\label{eq:L-2OPOs}
\mathbf{L} = \begin{bmatrix} 
t_3 \left(\sqrt{\kappa_{\mathrm{p} 1}} t_1 t_2 e^{i \phi} 
+ \sqrt{\kappa_{\mathrm{p} 2}}\right) a_{\mathrm{p}} 
+ \sqrt{\kappa_{\mathrm{c} 2}} t_2 t_3 e^{i \phi_2} a_{\mathrm{c}} \\  
\sqrt{\kappa_{\mathrm{p} 1}} r_1 a_{\mathrm{p}} \\
\sqrt{\kappa_{\mathrm{p} 1}} t_1 r_2 e^{i \phi_1} a_{\mathrm{p}} 
+ \sqrt{\kappa_{\mathrm{c} 2}} r_2 a_{\mathrm{c}} \\
r_3 \left(\sqrt{\kappa_{\mathrm{p} 1}} t_1 t_2 e^{i \phi} 
+ \sqrt{\kappa_{\mathrm{p} 2}}\right) a_{\mathrm{p}} 
+ \sqrt{\kappa_{\mathrm{c} 2}} t_2 r_3 e^{i \phi_2} a_{\mathrm{c}} \\
\sqrt{\kappa_{\mathrm{c} 1}} a_{\mathrm{c}} \\
\sqrt{\kappa_{\mathrm{p} 3}} a_{\mathrm{p}} \\
\sqrt{\kappa_{\mathrm{c} 3}} a_{\mathrm{c}}
\end{bmatrix},
\end{equation}
\begin{align}
H = & \left(\omega_{\mathrm{p}} + \mathrm{Im}\,\nu \right) a_{\mathrm{p}}^{\dag} a_{\mathrm{p}}
+ \omega_{\mathrm{c}} a_{\mathrm{c}}^{\dag} a_{\mathrm{c}} 
+ \left( {\textstyle\frac{i}{2}} \nu_{12} a_{\mathrm{p}}^{\dag} a_{\mathrm{c}} + \text{H.c.} \right) \nonumber \\
& + \left[ {\textstyle\frac{i}{2}} \left( 
\xi_{\mathrm{p}} a_{\mathrm{p}}^{\dag 2} + \xi_{\mathrm{c}} a_{\mathrm{c}}^{\dag 2} \right) 
+ \text{H.c.} \right] ,
\label{eq:Ham-fb}
\end{align}
where $a_{\mathrm{p}}$ and $a_{\mathrm{c}}$ denote, respectively, the annihilation operators of the plant's and controller's cavity field modes, and we defined auxiliary parameters:
\begin{eqnarray*}
& \nu_1 = \sqrt{\kappa_{\mathrm{c} 2} \kappa_{\mathrm{p} 1}} t_1 e^{i \phi_1} , \quad
\nu_2 = \sqrt{\kappa_{\mathrm{c} 2} \kappa_{\mathrm{p} 2}} t_2 e^{i \phi_2}, & \\
& \nu_{12} = \nu_1^{\ast} - \nu_2 , \quad
\nu = \sqrt{\kappa_{\mathrm{p} 1} \kappa_{\mathrm{p} 2}} t_1 t_2 e^{i \phi}. & 
\end{eqnarray*}
By comparing Eq.~\eqref{eq:Ham-fb} to the corresponding Hamiltonian without feedback:
\begin{equation}
\label{eq:Ham-nf}
H_{\text{nf}} = \omega_{\mathrm{p}} a_{\mathrm{p}}^{\dag} a_{\mathrm{p}}
+ \omega_{\mathrm{c}} a_{\mathrm{c}}^{\dag} a_{\mathrm{c}}
+ \left[ {\textstyle\frac{i}{2}} \left(
\xi_{\mathrm{p}} a_{\mathrm{p}}^{\dag 2} + \xi_{\mathrm{c}} a_{\mathrm{c}}^{\dag 2} \right) 
+ \text{H.c.} \right] ,
\end{equation}
we observe that two main effects induced by feedback are (1) the appearance of an effective interaction between the plant's and controller's cavity modes, governed by the term $\frac{i}{2} \nu_{12} a_{\mathrm{p}}^{\dag} a_{\mathrm{c}} + \text{H.c.}$, and (2) the modification of the plant detuning by $\mathrm{Im}\,\nu$ which is proportional to $\sin\phi$.

Using the formalism of Sec.~\ref{sec:IO-model}, we obtain:
\begin{align*}
& \mathbf{\Omega} = \begin{bmatrix} 
\omega_{\mathrm{p}} + \mathrm{Im}\,\nu & \frac{i}{2} \nu_{12} \\
-\frac{i}{2} \nu_{12}^\ast & \omega_{\mathrm{c}}
\end{bmatrix}, \quad
\mathbf{W} = \begin{bmatrix}
\xi_{\mathrm{p}} & 0 \\
0 & \xi_{\mathrm{c}}
\end{bmatrix}, \\
& \mathbf{K} = \begin{bmatrix} 
t_3 \left(\sqrt{\kappa_{\mathrm{p} 1}} t_1 t_2 e^{i \phi} + \sqrt{\kappa_{\mathrm{p} 2}}\right) & 
\sqrt{\kappa_{\mathrm{c} 2}} t_2 t_3 e^{i \phi_2} \\  
\sqrt{\kappa_{\mathrm{p} 1}} r_1 & 0 \\
\sqrt{\kappa_{\mathrm{p} 1}} t_1 r_2 e^{i \phi_1} & 
\sqrt{\kappa_{\mathrm{c} 2}} r_2 \\
r_3 \left(\sqrt{\kappa_{\mathrm{p} 1}} t_1 t_2 e^{i \phi} + \sqrt{\kappa_{\mathrm{p} 2}}\right) & 
\sqrt{\kappa_{\mathrm{c} 2}} t_2 r_3 e^{i \phi_2} \\
0 & \sqrt{\kappa_{\mathrm{c} 1}} \\
\sqrt{\kappa_{\mathrm{p} 3}} & 0 \\
0 & \sqrt{\kappa_{\mathrm{c} 3}}
\end{bmatrix}, \\
& \mathbf{V} = - \begin{bmatrix}
\eta_{\mathrm{p}} & \nu_2 \\
\nu_1 & \eta_{\mathrm{c}}
\end{bmatrix}, \\
& \mathbf{Y} = \begin{bmatrix}
- \sqrt{\kappa_{\mathrm{p} 1}} - \sqrt{\kappa_{\mathrm{p} 2}} t_1 t_2 e^{i \phi} & 
- \sqrt{\kappa_{\mathrm{c} 2}} t_1 e^{i \phi_1} \\
\sqrt{\kappa_{\mathrm{p} 2}} r_1 t_2 e^{i \phi} & \sqrt{\kappa_{\mathrm{c} 2}} r_1 e^{i \phi_1} \\
\sqrt{\kappa_{\mathrm{p} 2}} r_2 e^{i \phi_2} & 0 \\
0 & 0 \\
0 & - \sqrt{\kappa_{\mathrm{c} 1}} \\
- \sqrt{\kappa_{\mathrm{p} 3}} & 0 \\
0 & - \sqrt{\kappa_{\mathrm{c} 3}} 
\end{bmatrix}^{\mathsf{T}}, \\
& \breve{\mathbf{A}} = \begin{bmatrix}
-\eta_{\mathrm{p}} & -\nu_2 & \xi_{\mathrm{p}} & 0 \\
-\nu_1 & -\eta_{\mathrm{c}} & 0 & \xi_{\mathrm{c}} \\
\xi_{\mathrm{p}}^\ast & 0 & -\eta_{\mathrm{p}}^\ast & -\nu_2^\ast \\
0 & \xi_{\mathrm{c}}^\ast & -\nu_1^\ast & -\eta_{\mathrm{c}}^\ast
\end{bmatrix},
\end{align*}
where we used additional auxiliary parameters:
$$
\eta_{\mathrm{p}} = {\textstyle\frac{1}{2}} \gamma_{\mathrm{p}} + i \omega_{\mathrm{p}} + \nu , \quad
\eta_{\mathrm{c}} = {\textstyle\frac{1}{2}} \gamma_{\mathrm{c}} + i \omega_{\mathrm{c}} .
$$
It is possible to analytically invert the matrix $\breve{\mathbf{A}} + i\omega \mathbf{I}_4$ in order to obtain the transfer-matrix function $\breve{\mathbf{Z}}(\omega)$ and squeezing spectrum $\mathcal{P}(\omega,\theta)$ in analytic form. However, the resulting expressions are too complicated and visually uninformative to be shown here. For practical purposes, it is more efficient to numerically evaluate $\breve{\mathbf{Z}}(\omega)$ and $\mathcal{P}(\omega,\theta)$ for any given set of parameter values.

\section{Squeezing optimization procedure}
\label{sec:optim}

\subsection{Objective function}

In order to quantitatively investigate the tunability of the squeezing spectrum in the CQFC network of two coupled OPOs, we numerically optimize the degree of squeezing at various sideband frequencies. Specifically, we minimize the objective function of the form:
\begin{equation}
\label{eq:J}
J = \mathcal{P}^{-}(\omega_{\mathrm{opt}}) + g \mathcal{P}^{-}(\omega_{\mathrm{opt}}) \mathcal{P}^{+}(\omega_{\mathrm{opt}}) ,
\end{equation}
where $\omega_{\mathrm{opt}}$ is the selected sideband frequency. The first term in Eq.~\eqref{eq:J} is the minimum of the squeezing spectrum at $\omega_{\mathrm{opt}}$, while the second term is the uncertainty product times the weight parameter $g$. This second term is included in order to eliminate  solutions with a very large uncertainty of the anti-squeezed quadrature. In all optimization results shown below, the weight parameter is $g = 0.001$. With such a small value of $g$, the difference between the values of $J$ and $\mathcal{P}^{-}$ is always insignificant, and therefore, for the sake of simplicity, we refer to the problem of minimizing $J$ as \emph{squeezing optimization}. 

All solutions encountered during a search are checked to satisfy the Routh--Hurwitz stability criterion~\cite{Bishop.Dorf.chapter.2000}, i.e., that all eigenvalues of the matrix $\breve{\mathbf{A}}$ in Eq.~\eqref{eq:TF-1} have negative real parts. Any unstable solution is eliminated from the consideration by assigning to it a very large objective value ($J = 10^6$).

\subsection{Optimization variables}

For a given $\omega_{\mathrm{opt}}$, the objective $J$ is a function of the network parameters --- seven real parameters for the single OPO network:
\begin{equation}
\{ T_1, T_2, L, \omega_0, x, \theta_{\xi}, L_{\mathrm{tl}} \} ,
\end{equation}
and 17 real parameters for the CQFC network of two coupled OPOs:
\begin{eqnarray}
\left\{ T_{\mathrm{p} 1} , T_{\mathrm{p} 2}, L_{\mathrm{p}}, \omega_{\mathrm{p}}, x_{\mathrm{p}}, \theta_{\mathrm{p}},  T_{\mathrm{c} 1} , T_{\mathrm{c} 2}, L_{\mathrm{c}}, \omega_{\mathrm{c}},  x_{\mathrm{c}}, \theta_{\mathrm{c}}, \right. \nonumber \\ 
\left.  \phi_1, \phi_2, L_1, L_2, L_3 \right\} .
\end{eqnarray}
Recall that, for the single OPO network, $L = T_3$ is the intracavity power loss and $L_{\mathrm{tl}} = R_{\mathrm{B}}$ is the power loss in the output transmission line. Similarly, for the CQFC network of two coupled OPOs, $L_{\mathrm{p}} = T_{\mathrm{p} 3}$ and $L_{\mathrm{c}} = T_{\mathrm{c} 3}$ are the intracavity power losses for the plant and controller OPOs, respectively, and $L_i = r_i^2$ $(i = 1,2,3)$ are power losses in the transmission lines. In cases where the two intracavity loss values are equal, we denote $L_{\mathrm{in}} = L_{\mathrm{p}} = L_{\mathrm{c}}$, and where the three transmission line loss values are equal, we denote $L_{\mathrm{out}} = L_1 = L_2 = L_3$.

Numerical simulations demonstrate that an increase in any of the losses always leads to a deterioration of squeezing, and therefore if a loss parameter can vary in a specified interval $[L_{\mathrm{l}}, L_{\mathrm{u}}]$, an optimization will always converge to the lower bound $L_{\mathrm{l}}$. Therefore, it makes sense to to exclude the loss parameters from the optimization variables, i.e., to execute each optimization with all loss parameters having pre-assigned fixed values (of course, these values can vary from one optimization run to another to explore various experimentally relevant regimes). Consequently, there remain five optimization variables for the single OPO network:
\begin{equation}
\{ T_1, T_2, \omega_0, x, \theta_{\xi} \} ,
\end{equation}
and 12 optimization variables for the CQFC network of two coupled OPOs:
\begin{equation}
\{ T_{\mathrm{p} 1} , T_{\mathrm{p} 2}, \omega_{\mathrm{p}}, x_{\mathrm{p}}, \theta_{\mathrm{p}},  T_{\mathrm{c} 1} , T_{\mathrm{c} 2}, \omega_{\mathrm{c}},  
x_{\mathrm{c}}, \theta_{\mathrm{c}}, \phi_1, \phi_2 \} .
\end{equation}

Each optimization variable $z$ can vary in an interval $[z_{\mathrm{l}}, z_{\mathrm{u}}]$ (where $z_{\mathrm{l}}$ is the lower bound and $z_{\mathrm{u}}$ is the upper bound). The bound intervals are 
\begin{itemize}
\item $[0, 2 \pi]$ for all phase variables ($\theta_{\xi}$, $\theta_{\mathrm{p}}$, $\theta_{\mathrm{c}}$, $\phi_1$, $\phi_2$);
\item $[-\omega_{\mathrm{u}}, \omega_{\mathrm{u}}]$ for all cavity detuning frequencies ($\omega_0$, $\omega_{\mathrm{p}}$, $\omega_{\mathrm{c}}$);
\item $[0, T_{\mathrm{u}}]$ for all power transmittances of actual cavity mirrors ($T_1$, $T_2$, $T_{\mathrm{p} 1}$, $T_{\mathrm{p} 2}$, $T_{\mathrm{c} 1}$, $T_{\mathrm{c} 2}$);
\item $[0, x_{\mathrm{u}}]$ for all scaled pump amplitudes ($x$, $x_{\mathrm{p}}$, $x_{\mathrm{c}}$). 
\end{itemize}
The values of upper bounds $\omega_{\mathrm{u}}$, $T_{\mathrm{u}}$ and $x_{\mathrm{u}}$ are specified (along with the values of losses) for each optimization run.

In all optimizations, the fixed physical parameters are selected the same for all OPOs: pump wavelength $\lambda_p = 775$~nm, signal wavelength $\lambda_s = 1550$~nm; an MgO:PPLN crystal with length $l_c = 20$~mm, refractive index (at the signal wavelength) $n_s = 2.1$, and effective nonlinear coefficient $d_{\mathrm{eff}} = 14$~pm/V; a Fabry-P\'erot cavity with effective length $l_{\mathrm{eff}} = 87$~mm. These values are characteristic for a typical tabletop experiment with bulk-optics components.

\begin{table*}[htbp]
\caption{\label{tab:alg-comparison}Performance of different algorithms for squeezing optimization in the CQFC network of two coupled OPOs. The table shows the best degree of squeezing, $\mathcal{Q}^{-}(\omega_{\mathrm{opt}}) = 10 \log_{10} \mathcal{P}^{-}(\omega_{\mathrm{opt}})$ (in dB), found using various algorithms, for $L_{\mathrm{in}} = 0.01$, $L_{\mathrm{out}} = 0.05$, $\omega_{\mathrm{u}}/2\pi = 100.0$~MHz, $x_{\mathrm{u}} = 0.3$, $T_{\mathrm{u}} = 0.9$, and five different $\omega_{\mathrm{opt}}$ values: $\omega_{\mathrm{opt}}/2\pi = \{5, 25, 50, 100, 200\}$~MHz. Optimizations for each individual algorithm execute four parallel searches with the population sizes of $N_{\mathrm{pop}} = 30$, and the evolutions are repeated $N_{\mathrm{ev}} = 30$ times (with solution exchanges between the searches after the completion of each evolution except the last one). Algorithm parameters such as the number of no improvements before halting the optimization, $N_{\mathrm{stop}}$, the number of generations, $N_{\mathrm{gen}}$, and the number of iterations, $N_{\mathrm{iter}}$, are indicated in the table. The hybrid strategy (eight parallel searches using seven global algorithms) is described in the text.}
\begin{ruledtabular}
\begin{tabular}{lrrrrr}
{} & \multicolumn{5}{c}{$\omega_{\mathrm{opt}}/2\pi$}\\ 
Algorithm & $5$~MHz & $25$~MHz & $50$~MHz & $100$~MHz & $200$~MHz \\ 
\hline
Sequential Least SQuares Programming (local only) & $-4.270$ & $-4.021$ & $-3.396$ & $-2.676$ & $-1.809$\\
Compass Search (local only) & $-8.824$ & $-7.540$ & $-8.274$ & $-8.113$ & $-2.527$\\
Compass Search guided by Monotonic Basin Hopping ($N_{\mathrm{stop}} = 5$) & $-9.105$ & $-7.611$ & $-7.037$ & $-8.255$ & $-7.540$\\
Artificial Bee Colony ($N_{\mathrm{gen}} = 200$) & $-9.791$ & $-8.945$ & $-8.788$ & $-8.427$ & $-7.811$\\
Covariance Matrix Adaptation Evolution Strategy ($N_{\mathrm{gen}} = 500$) & $-9.798$ & $-8.869$ & $-8.806$ & $-8.423$ & $-7.811$\\
Differential Evolution, variant 1220 ($N_{\mathrm{gen}} = 800$) & $-9.805$ & $-8.626$ & $-8.809$ & $-8.429$ & $-7.813$\\
Differential Evolution with p-best crossover ($N_{\mathrm{gen}} = 1000$) & $-9.805$ & $-8.953$ & $-8.808$ & $-8.429$ & $-7.813$\\
Improved Harmony Search ($N_{\mathrm{iter}} = 1000$) & $-9.805$ & $-8.949$ & $-8.808$ & $-8.429$ & $-7.813$\\
Particle Swarm Optimization, variant 5 ($N_{\mathrm{gen}} = 1$) & $-9.219$ & $-8.623$ & $-7.090$ & $-8.332$ & $-7.617$\\
Particle Swarm Optimization, variant 6 ($N_{\mathrm{gen}} = 1$) & $-8.811$ & $-7.936$ & $-7.403$ & $-7.536$ & $-5.932$\\
Simple Genetic Algorithm ($N_{\mathrm{gen}} = 1000$) & $-9.805$ & $-7.665$ & $-8.809$ & $-8.429$ & $-7.813$\\
Corana's Simulated Annealing ($N_{\mathrm{iter}} = 20000$) & $-7.432$ & $-5.015$ & $-4.893$ & $-6.110$ & $-4.754$\\
Hybrid strategy (eight parallel searches using seven global algorithms) & $-9.805$ & $-8.953$ & $-8.809$ & $-8.429$ & $-7.813$\\
\end{tabular}
\end{ruledtabular}
\end{table*}

\subsection{Optimization methodology}

Preliminary optimization runs using local algorithms (e.g., Sequential Least Squares Programming) demonstrated that different choices of initial parameter values resulted in different solutions of varying quality. These results mean that the fitness landscape contains multiple local optima. In order to reach a solution of very high quality, we decided to use global search methods. Specifically, we used PyGMO, a suite of global (stochastic) algorithms~\cite{pygmo.url}. Since these global algorithms are heuristic in nature, they do not guarantee the convergence to a global optimum; in fact, as shown in Table~\ref{tab:alg-comparison}, while multiple global methods are capable of finding high-quality solutions, the performance varies between different algorithms as well as between optimizations with different values of $\omega_{\mathrm{opt}}$ for the same algorithm.

To maximize the chances of finding a globally optimal solution, we employed a hybrid strategy, where each optimization executes in parallel eight searches (using seven different global algorithms), with a fully connected topology of solution exchanges between them. These eight searches include two instances of Artificial Bee Colony and one instance of each: Covariance Matrix Adaptation Evolution Strategy, Differential Evolution variant 1220, Differential Evolution with $p$-best crossover, Improved Harmony Search, Particle Swarm Optimization variant 5, and Compass Search guided by Monotonic Basin Hopping. Each optimization uses the population size of $N_{\mathrm{pop}} = 30$ for each of the global searches, and the evolutions are repeated $N_{\mathrm{ev}} = 30$ times (with solution exchanges between the searches after the completion of each evolution except the last one); the algorithm parameters (the number of no improvements before halting the optimization, $N_{\mathrm{stop}}$, the number of generations, $N_{\mathrm{gen}}$, and the number of iterations, $N_{\mathrm{iter}}$) used in the searches are the same as those shown in Table~\ref{tab:alg-comparison} for individual algorithms. As indicated by the results in Table~\ref{tab:alg-comparison}, this hybrid strategy consistently finds the best solution, as compared to any individual algorithm. Multiple trials with larger values of $N_{\mathrm{pop}}$, $N_{\mathrm{ev}}$, $N_{\mathrm{gen}}$, and $N_{\mathrm{iter}}$ did not typically result in an improvement of the solution quality, and thus did not warrant the increased run time.

\section{Squeezing optimization results}
\label{sec:results}

\begin{figure}[htbp]
\centering
\includegraphics[width=1.0\columnwidth]{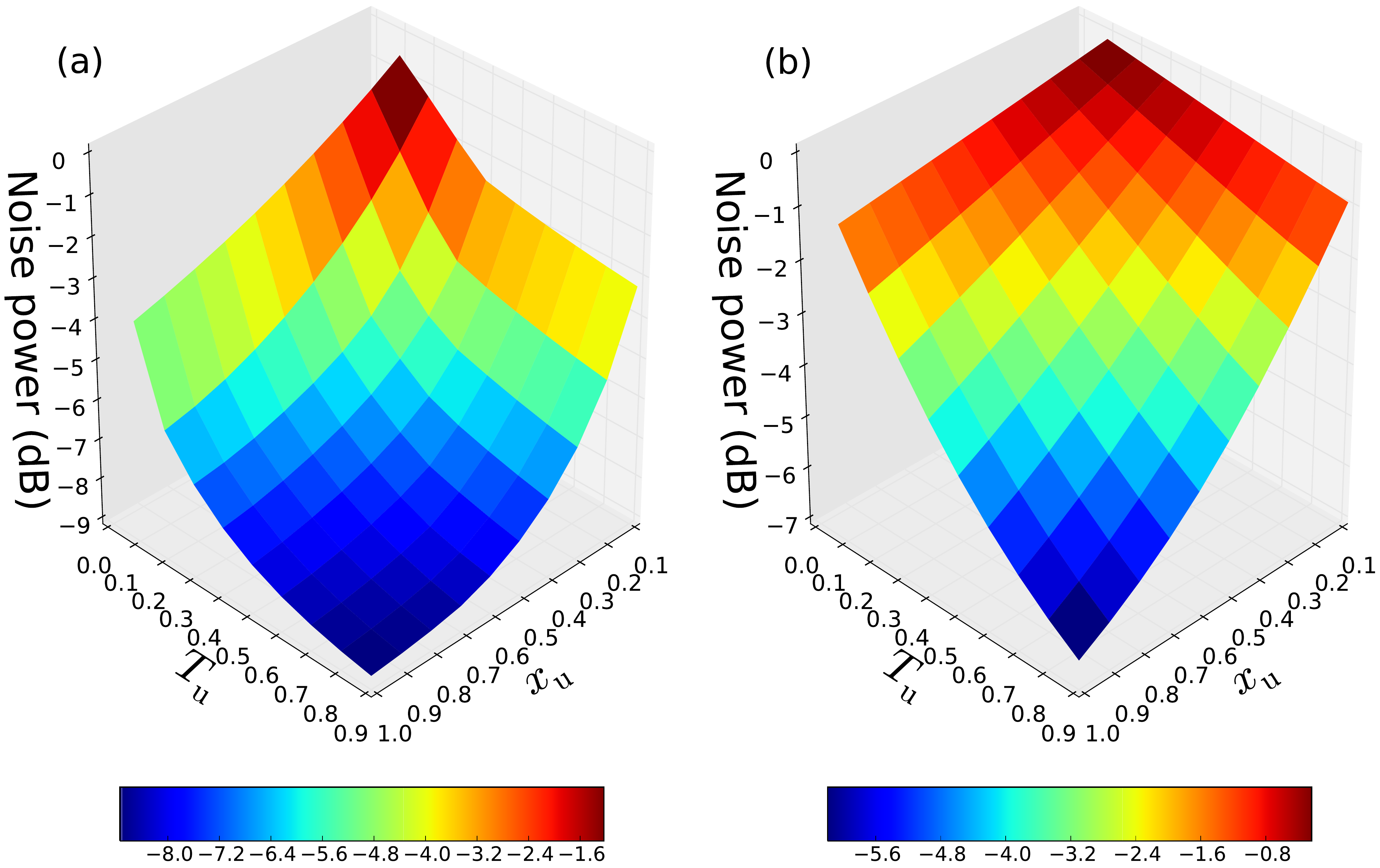}
\caption{The optimized degree of squeezing, $\mathcal{Q}^{-}(\omega_{\mathrm{opt}})$, for (a) the CQFC network of two coupled OPOs and (b) the single OPO network, versus the upper limits on the power transmittance of cavity mirrors, $T_{\mathrm{u}}$, and the scaled pump amplitude, $x_{\mathrm{u}}$. Other parameters are $\omega_{\mathrm{opt}}/2\pi = 100$~MHz, $\omega_{\mathrm{u}}/2\pi = 100$~MHz, $L = L_{\mathrm{in}} = 0.01$, $L_{\mathrm{tl}} = L_{\mathrm{out}} = 0.1$.}
\label{fig:Qmin_vs_xb_and_Rb}
\end{figure}

\begin{figure}[htbp]
\centering
\includegraphics[width=1.0\columnwidth]{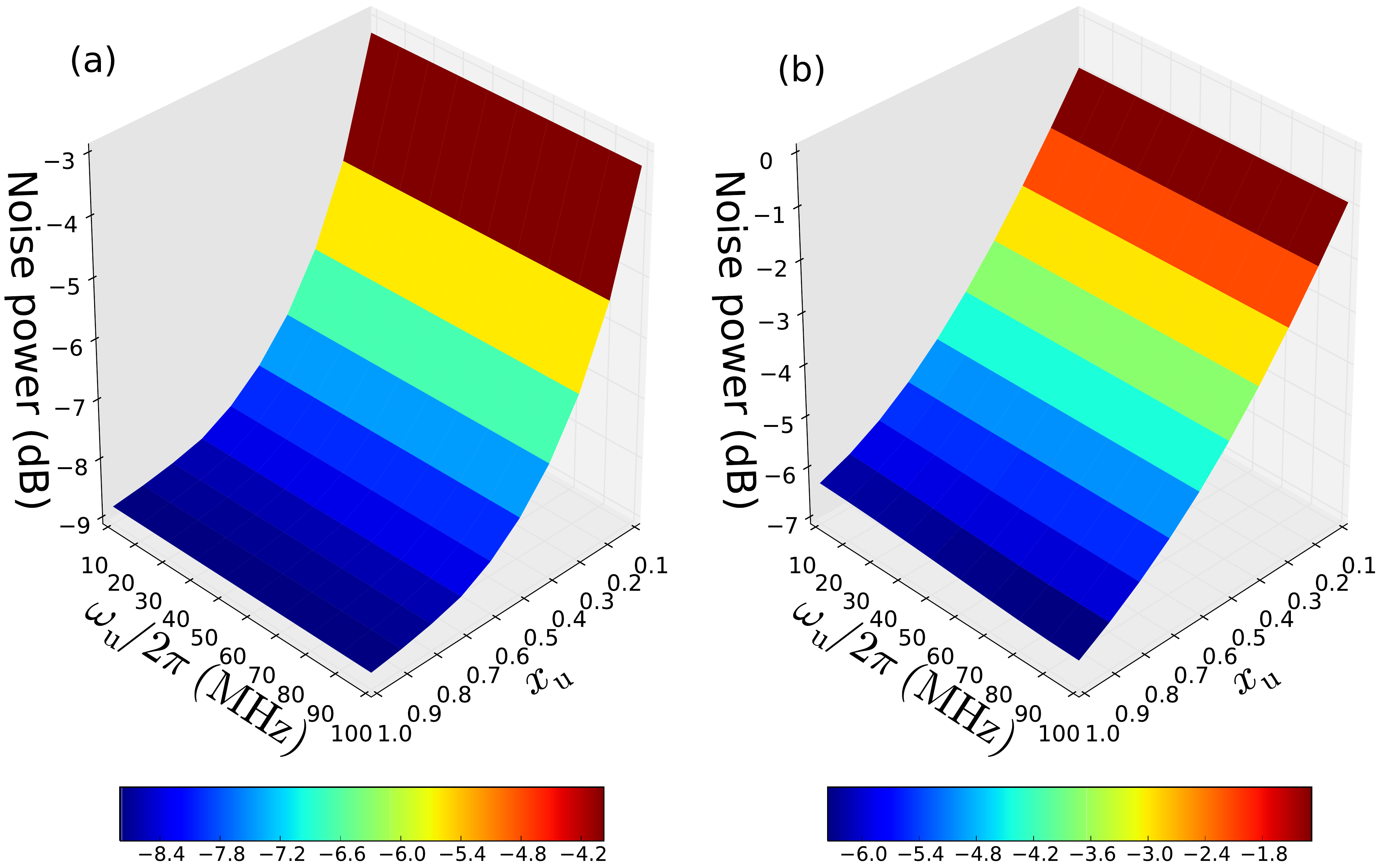}
\caption{The optimized degree of squeezing, $\mathcal{Q}^{-}(\omega_{\mathrm{opt}})$, for (a) the CQFC network of two coupled OPOs and (b) the single OPO network, versus the upper limits on the cavity detuning frequency, $\omega_{\mathrm{u}}$, and the scaled pump amplitude, $x_{\mathrm{u}}$. Other parameters are $\omega_{\mathrm{opt}}/2\pi = 100$~MHz, $T_{\mathrm{u}} = 0.9$, $L = L_{\mathrm{in}} = 0.01$, $L_{\mathrm{tl}} = L_{\mathrm{out}} = 0.1$.}
\label{fig:Qmin_vs_xb_and_wb}
\end{figure}

First of all, we would like to compare the performance of the CQFC network of two coupled OPOs versus that of the single OPO network, in terms of the maximum degree of squeezing achievable under comparable conditions. Figures~\ref{fig:Qmin_vs_xb_and_Rb}~and~\ref{fig:Qmin_vs_xb_and_wb} show the optimized degree of squeezing, $\mathcal{Q}^{-}(\omega_{\mathrm{opt}})$, at $\omega_{\mathrm{opt}}/2\pi = 100$~MHz, for both networks, versus the upper limits on various network parameters ($T_{\mathrm{u}}$ and  $x_{\mathrm{u}}$ in Fig.~\ref{fig:Qmin_vs_xb_and_Rb}, and $\omega_{\mathrm{u}}$ and  $x_{\mathrm{u}}$ in Fig.~\ref{fig:Qmin_vs_xb_and_wb}), with constant loss values: $L = L_{\mathrm{in}} = 0.01$, $L_{\mathrm{tl}} = L_{\mathrm{out}} = 0.1$. We observe that the CQFC network of two coupled OPOs generates stronger squeezing than the single OPO network, even as total losses in transmission lines in the former are three times larger than those in the latter (30\% versus 10\%). In both networks, the maximum degree of squeezing increases with both $T_{\mathrm{u}}$ (more light is allowed to leave the cavities) and $x_{\mathrm{u}}$ (higher pump power), with these increases being roughly linear for the single OPO network and faster than linear in the CQFC network of two coupled OPOs. These results demonstrate that the feedback makes it possible to more effectively utilize the available pump power.

Figure~\ref{fig:Qmin_vs_xb_and_wb} also shows that, for both networks, the maximum degree of squeezing is independent of the upper limit $\omega_{\mathrm{u}}$ on the cavity detuning frequency; furthermore, we found that in most cases the maximum degree of squeezing is actually achieved with zero detuning. In all results shown below, optimizations used the upper limit value $\omega_{\mathrm{u}}/2\pi = 100$~MHz.

\begin{figure}[htbp]
\centering
\includegraphics[width=1.0\columnwidth]{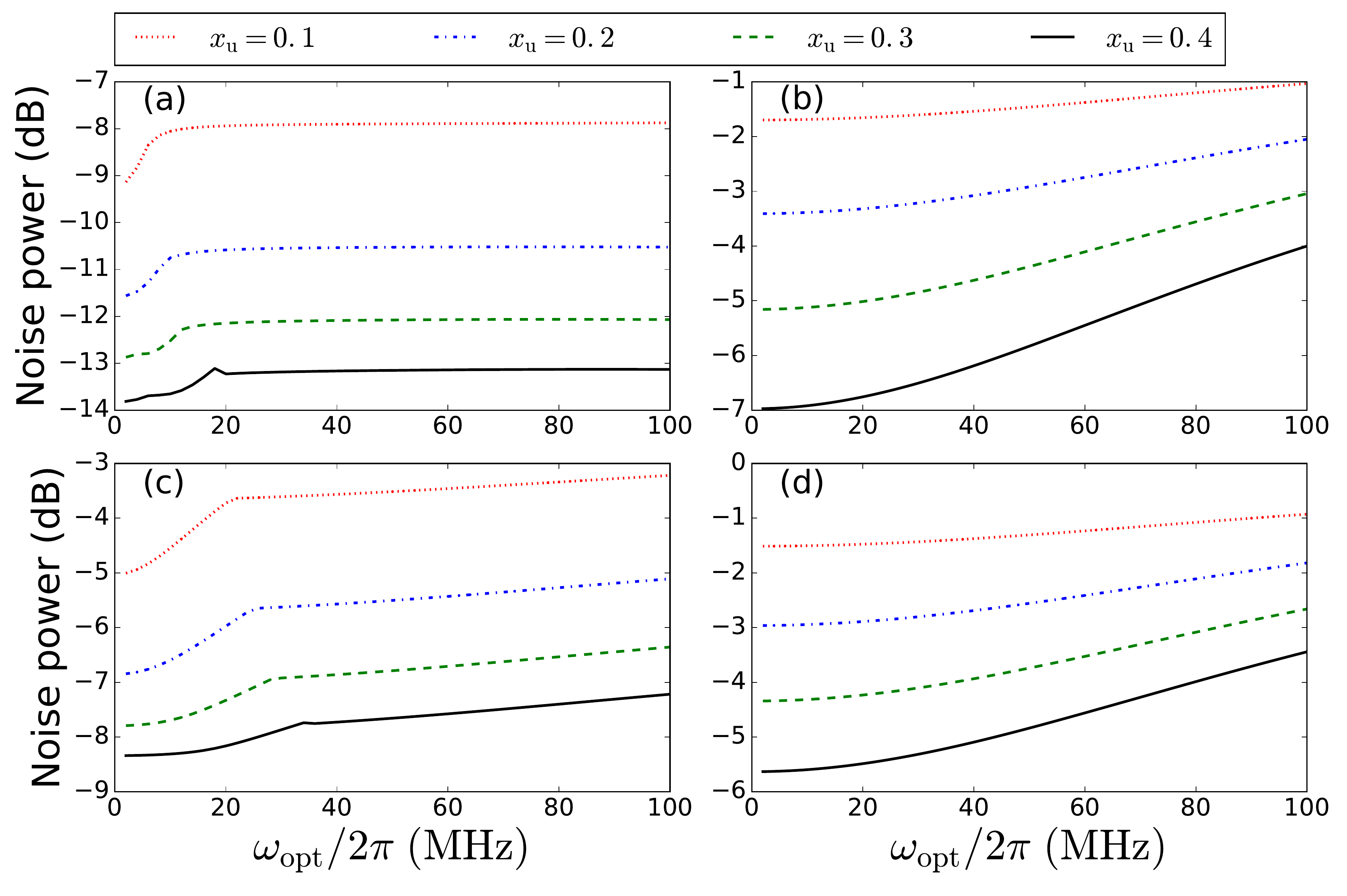}
\caption{The optimized degree of squeezing, $\mathcal{Q}^{-}(\omega_{\mathrm{opt}})$, versus $\omega_{\mathrm{opt}}$, for the CQFC network of two coupled OPOs (subplots (a) and (c)) and the single OPO network (subplots (b) and (d)). The transmission line losses are $L_{\mathrm{tl}} = L_{\mathrm{out}} = 0.01$ in subplots (a) and (b), and $L_{\mathrm{tl}} = L_{\mathrm{out}} = 0.1$ in subplots (c) and (d). Each subplot shows four curves corresponding to different values of $x_{\mathrm{u}}$ ($x_{\mathrm{u}} = \{0.1, 0.2, 0.3, 0.4\}$), as indicated in the legend. Other parameters are $T_{\mathrm{u}} = 0.9$, $L = L_{\mathrm{in}} = 0.01$.}
\label{fig:Qmin_vs_fopt_1}
\end{figure}

We also investigate the dependence of the maximum degree of squeezing, $\mathcal{Q}^{-}(\omega_{\mathrm{opt}})$, on the sideband frequency $\omega_{\mathrm{opt}}$ at which it is optimized. This dependence is shown in Fig.~\ref{fig:Qmin_vs_fopt_1}, for both networks, for different values of transmission line losses and pump amplitude bound. We observe that the CQFC network of two coupled OPOs not only generates stronger squeezing than the single OPO network, but that the degradation of squeezing associated with the increase of $\omega_{\mathrm{opt}}$ is substantially slower in the former than in the latter. The capability of the CQFC network to moderate the degradation of squeezing at higher values of $\omega_{\mathrm{opt}}$ is associated with a rather abrupt change in the regime of network operation, which is manifested by a rapid change in the slope of the curves in subplots (a) and (c) of Fig.~\ref{fig:Qmin_vs_fopt_1}. 

\begin{figure}[htbp]
\centering
\includegraphics[width=1.0\columnwidth]{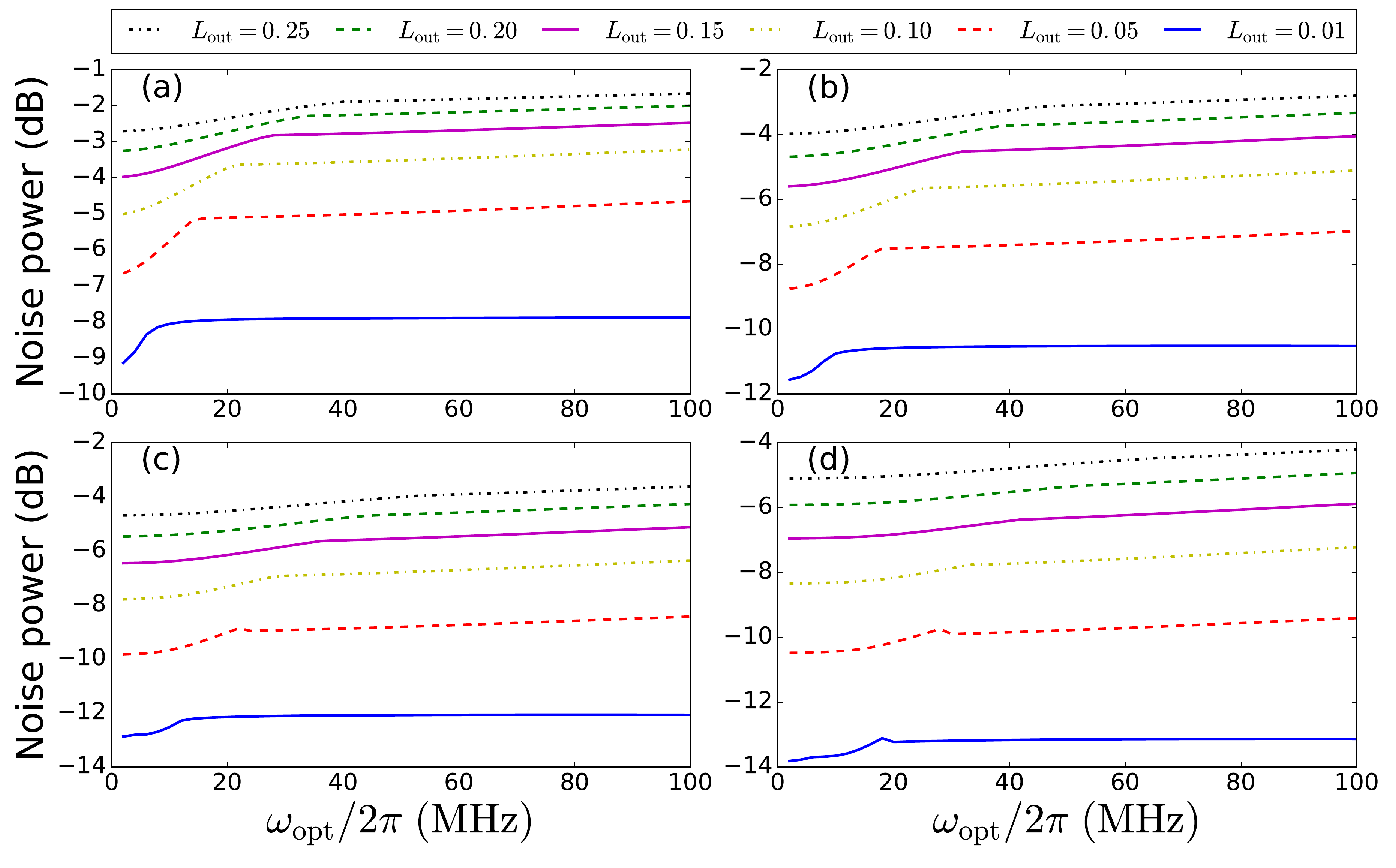}
\caption{The optimized degree of squeezing, $\mathcal{Q}^{-}(\omega_{\mathrm{opt}})$, versus $\omega_{\mathrm{opt}}$, for the CQFC network of two coupled OPOs. The values of $x_{\mathrm{u}}$ are: (a) $x_{\mathrm{u}} = 0.1$, (b) $x_{\mathrm{u}} = 0.2$, (c) $x_{\mathrm{u}} = 0.3$, and (d) $x_{\mathrm{u}} = 0.4$. Each subplot shows six curves corresponding to different values of transmission line losses: $L_{\mathrm{out}} = \{0.01, 0.05, 0.1, 0.15, 0.2, 0.25\}$, as indicated in the legend. Other parameters are $T_{\mathrm{u}} = 0.9$, $L_{\mathrm{in}} = 0.01$.}
\label{fig:Qmin_vs_fopt_2}
\end{figure}

To explore further the emergence of this new operation regime, we focus on the CQFC network of two coupled OPOs, with Fig.~\ref{fig:Qmin_vs_fopt_2} showing the dependence of the maximum degree of squeezing on $\omega_{\mathrm{opt}}$ for more values of transmission line losses. We observe that the value of $\omega_{\mathrm{opt}}$ at which the operation regime switches, increases with both $x_{\mathrm{u}}$ and $L_{\mathrm{out}}$. The difference between the curve slopes in the low-$\omega_{\mathrm{opt}}$ and high-$\omega_{\mathrm{opt}}$ regimes decreases as $L_{\mathrm{out}}$ increases.

\begin{figure*}[htbp]
\centering
\includegraphics[width=1.7\columnwidth]{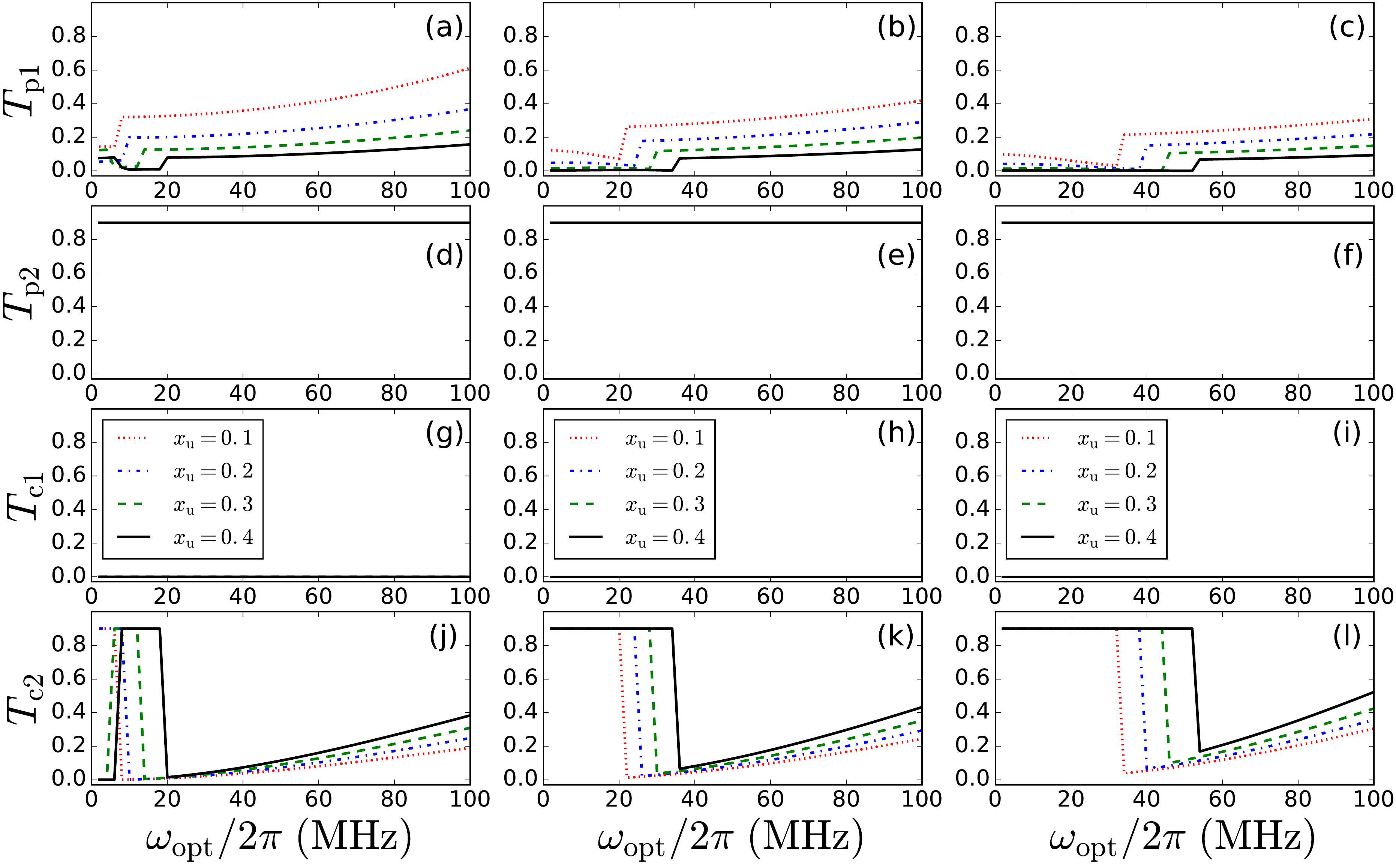}
\caption{The optimal values of power transmittances of cavity mirrors, $T_{\mathrm{p} 1}$ (subplots (a), (b), (c)), $T_{\mathrm{p} 2}$ (subplots (d), (e), (f)), $T_{\mathrm{c} 1}$ (subplots (g), (h), (i)), and $T_{\mathrm{c} 2}$ (subplots (j), (k), (l)), versus $\omega_{\mathrm{opt}}$, for the CQFC network of two coupled OPOs. The transmission line losses are $L_{\mathrm{out}} = 0.01$ (subplots (a), (d), (g), (j)), $L_{\mathrm{out}} = 0.1$ (subplots (b), (e), (h), (k)), and $L_{\mathrm{out}} = 0.2$ (subplots (c), (f), (i), (l)). Each subplot shows four curves corresponding to different values of $x_{\mathrm{u}}$ ($x_{\mathrm{u}} = \{0.1, 0.2, 0.3, 0.4\}$), as indicated in the legend. Other parameters are $T_{\mathrm{u}} = 0.9$, $L_{\mathrm{in}} = 0.01$.}
\label{fig:T_vs_fopt}
\end{figure*}

\begin{table}[b]
\caption{\label{tab:f_crit}The sideband frequency $\omega_{\mathrm{opt}}^{\star}/2\pi$~(in MHz), at which the high-$\omega_{\mathrm{opt}}$ regime commences, for the CQFC network of two coupled OPOs with $T_{\mathrm{u}} = 0.9$, $L_{\mathrm{in}} = 0.01$, and various values of $L_{\mathrm{out}}$ and $x_{\mathrm{u}}$. The accuracy of the reported values is limited by the sampling interval of $2$~MHz.}
\begin{ruledtabular}
\begin{tabular}{lrrrrrrr}
{} & \multicolumn{7}{c}{$L_{\mathrm{out}}$}\\ 
$x_{\mathrm{u}}$ & 0.01 & 0.05 & 0.10 & 0.15 & 0.20 & 0.25 & 0.30 \\ \hline
0.1 & 8 &  16 &  22 &  28 &  34 &  42 &  48 \\
0.2 & 10 &  18 &  26 &  32 &  40 &  48 &  56\\
0.3 & 14 &  24 &  30 &  38 &  46 &  56 &  66\\
0.4 & 20 &  30 &  36 &  44 &  54 &  68 &  90\\
\end{tabular}
\end{ruledtabular}
\end{table}

\begin{figure}[b]
\centering
\includegraphics[width=1.0\columnwidth]{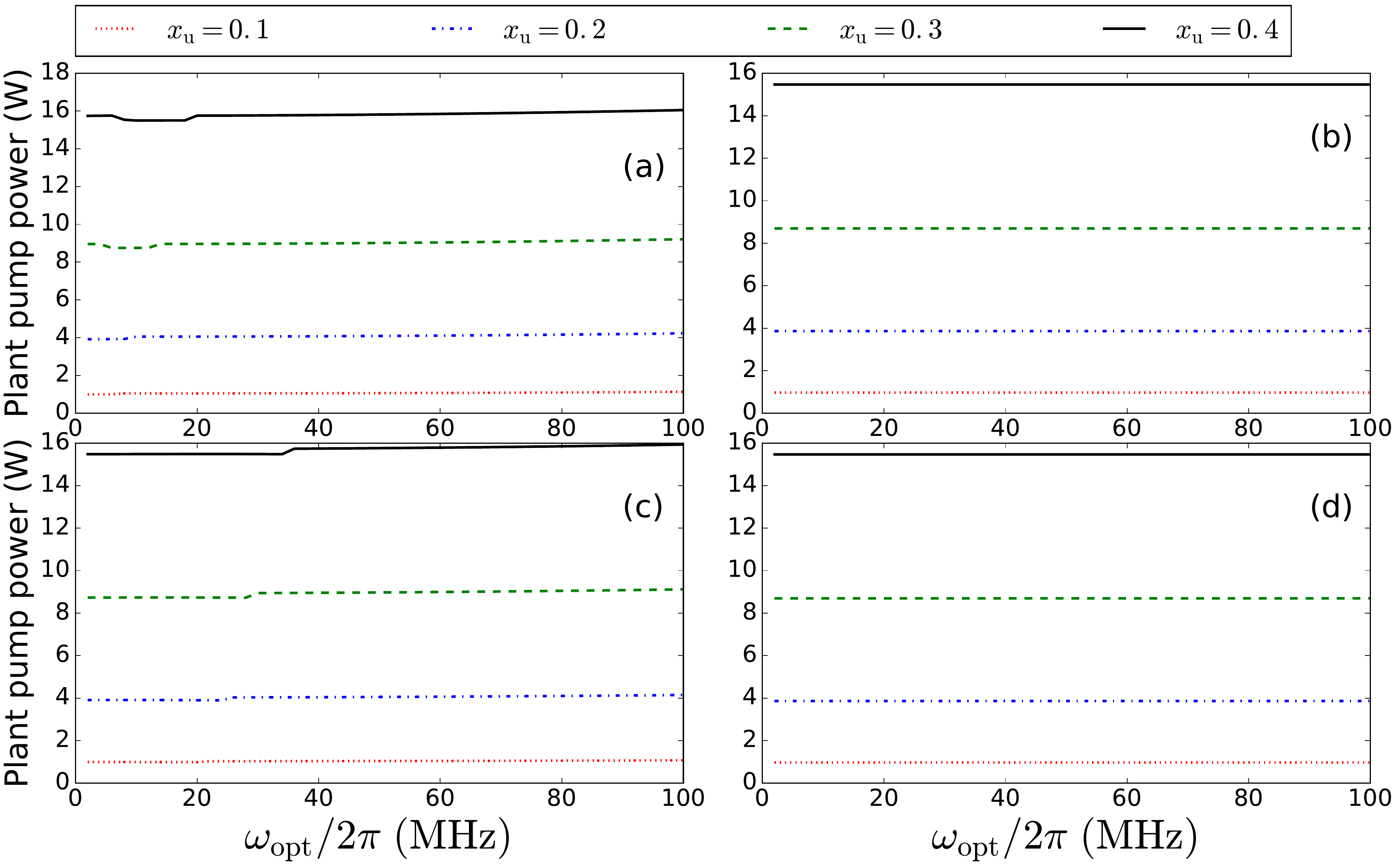}
\caption{The optimal values of the pump power for the plant OPO in the CQFC network of two coupled OPOs (subplots (a) and (c)) and for the single OPO (subplots (b) and (d)), versus $\omega_{\mathrm{opt}}$. The transmission line losses are $L_{\mathrm{tl}} = L_{\mathrm{out}} = 0.01$ in subplots (a) and (b), and $L_{\mathrm{tl}} = L_{\mathrm{out}} = 0.1$ in subplots (c) and (d). Each subplot shows four curves corresponding to different values of $x_{\mathrm{u}}$ ($x_{\mathrm{u}} = \{0.1, 0.2, 0.3, 0.4\}$), as indicated in the legend. Other parameters are $T_{\mathrm{u}} = 0.9$, $L = L_{\mathrm{in}} = 0.01$.}
\label{fig:Pp_vs_fopt}
\end{figure}

\begin{figure*}[htbp]
\centering
\includegraphics[width=1.7\columnwidth]{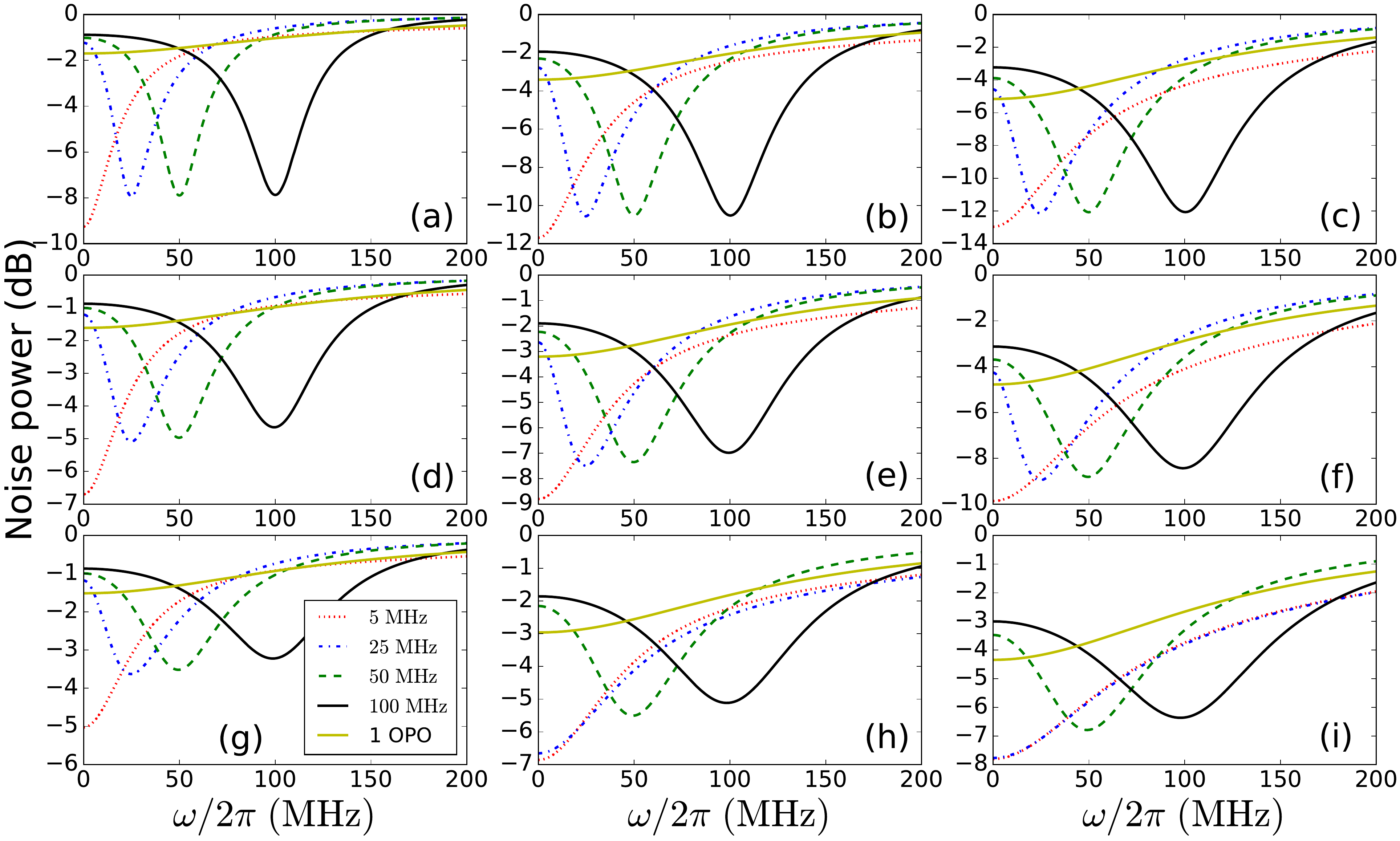}
\caption{The squeezing spectrum $\mathcal{Q}^{-}(\omega)$ for the optimal operation of both networks. Each subplot shows four curves corresponding to the optimally operated CQFC network of two coupled OPOs for different values of $\omega_{\mathrm{opt}}$ ($\omega_{\mathrm{opt}}/2\pi = \{5, 25, 50, 100\}$~MHz), along with a curve corresponding to the optimally operated single OPO network for any value of $\omega_{\mathrm{opt}}$, as indicated in the legend. The transmission line losses are $L_{\mathrm{tl}} = L_{\mathrm{out}} = 0.01$ (subplots (a), (b), (c)), $L_{\mathrm{tl}} = L_{\mathrm{out}} = 0.05$ (subplots (d), (e), (f)), and $L_{\mathrm{tl}} = L_{\mathrm{out}} = 0.1$ (subplots (g), (h), (i)). The upper limit on the scaled pump amplitude is $x_{\mathrm{u}} = 0.1$ (subplots (a), (d), (g)), $x_{\mathrm{u}} = 0.2$ (subplots (b), (e), (h)), and $x_{\mathrm{u}} = 0.3$ (subplots (c), (f), (i)). Other parameters are $T_{\mathrm{u}} = 0.9$, $L = L_{\mathrm{in}} = 0.01$.}
\label{fig:Qmin_vs_f}
\end{figure*}

To understand the physical differences between operations of the CQFC network in the low-$\omega_{\mathrm{opt}}$ and high-$\omega_{\mathrm{opt}}$ regimes, we consider the dependence of the optimal values of power transmittances of cavity mirrors, $T_{\mathrm{p} 1}$, $T_{\mathrm{p} 2}$, $T_{\mathrm{c} 1}$, and $T_{\mathrm{c} 2}$, on $\omega_{\mathrm{opt}}$. This dependence is shown in Fig.~\ref{fig:T_vs_fopt} for optimizations with $T_{\mathrm{u}} = 0.9$, $L_{\mathrm{in}} = 0.01$, and various values of $L_{\mathrm{out}}$ and $x_{\mathrm{u}}$. First, we see that the optimal values of $T_{\mathrm{p} 2}$ and $T_{\mathrm{c} 1}$ are constant over the entire range of $\omega_{\mathrm{opt}}$ values; specifically, $T_{\mathrm{p} 2} = 0.9$ is at the upper bound, which corresponds to the maximum flow from the plant cavity to the 1st output field (the one whose squeezing properties are measured), and $T_{\mathrm{c} 1} = 0$ is at the lower bound, which corresponds to the minimum flow from the controller cavity to the 5th output field (the one which is not used for either squeezing measurement or feedback). In contrast to this simple behavior of the optimal values of $T_{\mathrm{p} 2}$ and $T_{\mathrm{c} 1}$, the optimal values of $T_{\mathrm{p} 1}$ and $T_{\mathrm{c} 2}$, which regulate the feedback between the plant and controller OPOs, demonstrate much more intricate dependence on $\omega_{\mathrm{opt}}$. The optimal value of $T_{\mathrm{p} 1}$ and especially that of $T_{\mathrm{c} 2}$ undergo a substantial and rather abrupt change at the critical $\omega_{\mathrm{opt}}$ value at which the network's operation switches between the low-$\omega_{\mathrm{opt}}$ and high-$\omega_{\mathrm{opt}}$ regimes. As $\omega_{\mathrm{opt}}$ increases through the critical point, $T_{\mathrm{p} 1}$ changes from a lower to a higher value, while $T_{\mathrm{c} 2}$ decreases from the upper bound $T_{\mathrm{c} 2} = 0.9$ to a much lower value. In other words, the low-$\omega_{\mathrm{opt}}$ optimal regime is characterized by \emph{the maximum flow of light from the controller to the plant and a much lower flow in the opposite direction}, while the high-$\omega_{\mathrm{opt}}$ optimal regime is characterized by \emph{roughly similar flows of light in both directions}. These patterns characterizing the regimes of optimal network operation, their dependencies on pump and loss parameters, and the rapid switch between the regimes, are quite non-intuitive, and finding them would be rather unlikely without the use of a stochastic global search that explores vast areas of the fitness landscape.

The sharp change of the optimal value of $T_{\mathrm{c} 2}$ associated with the regime switch makes it easy to identify the sideband frequency $\omega_{\mathrm{opt}}^{\star}$, at which the high-$\omega_{\mathrm{opt}}$ regime commences (the precision of determining the $\omega_{\mathrm{opt}}^{\star}/2\pi$ values is limited by the sampling interval, which is $2$~MHz in our data). The values of $\omega_{\mathrm{opt}}^{\star}/2\pi$ are shown in Table~\ref{tab:f_crit} for $T_{\mathrm{u}} = 0.9$, $L_{\mathrm{in}} = 0.01$, and various values of $L_{\mathrm{out}}$ and $x_{\mathrm{u}}$. We see that $\omega_{\mathrm{opt}}^{\star}$ increases monotonously with both $L_{\mathrm{out}}$ and $x_{\mathrm{u}}$.

We also note that the optimal values of the scaled pump amplitudes, $x_{\mathrm{p}}$ and $x_{\mathrm{c}}$, are almost always at (or very close to) the upper bound $x_{\mathrm{u}}$, i.e., in either regime the optimally operated CQFC network usually uses all the pump power it can get. The maximum use of the pump power is also observed for the optimal operation of the single OPO network. Indeed, as seen in Fig.~\ref{fig:Pp_vs_fopt}, for both networks, the optimal values of the pump power are virtually independent of $\omega_{\mathrm{opt}}$ and losses, while they scale quadratically with $x_{\mathrm{u}}$. Due to the rapid growth of the optimal pump power with $x_{\mathrm{u}}$, only values $x_{\mathrm{u}} \leq 3$ should be considered realistic for the optimal operation with a typical tabletop experimental setup considered in this paper.

Next, we investigate the squeezing spectrum $\mathcal{Q}^{-}(\omega)$ generated under the optimal operation of either network for various values of $\omega_{\mathrm{opt}}$, $x_{\mathrm{u}}$, and transmission line losses. Figure~\ref{fig:Qmin_vs_f} shows $\mathcal{Q}^{-}(\omega)$ for both networks for various values of $\omega_{\mathrm{opt}}$, $L_{\mathrm{tl}} = L_{\mathrm{out}}$, and $x_{\mathrm{u}}$. We see that the optimally operated single OPO network generates exactly the same Lorentzian squeezing spectrum for any choice of $\omega_{\mathrm{opt}}$. In contrast, the CQFC network of two coupled OPOs is capable of generating diverse squeezing spectra, with the specific spectral shape varying to fit the selected value of $\omega_{\mathrm{opt}}$, and overall generates much stronger squeezing over a major portion of the spectrum (especially, at frequencies around $\omega_{\mathrm{opt}}$). Interestingly, the capability of the CQFC network to generate a squeezing spectrum $\mathcal{Q}^{-}(\omega)$ that has the minimum at $\omega = \omega_{\mathrm{opt}}$ is attained only if the selected value of $\omega_{\mathrm{opt}}$ is within the high-$\omega_{\mathrm{opt}}$ regime of optimal network operation, i.e., $\omega_{\mathrm{opt}} \geq \omega_{\mathrm{opt}}^{\star}$ (for a given set of bound and loss values). Conversely, as seen for $\omega_{\mathrm{opt}}/2\pi = 5$~MHz in all subplots of Fig.~\ref{fig:Qmin_vs_f} and for $\omega_{\mathrm{opt}}/2\pi = 25$~MHz in subplots (h) and (i) of Fig.~\ref{fig:Qmin_vs_f}, the squeezing spectrum has the minimum at $\omega = 0$ if $\omega_{\mathrm{opt}}$ is within the low-$\omega_{\mathrm{opt}}$ regime of optimal network operation.

\begin{figure}[htbp]
\centering
\includegraphics[width=1.0\columnwidth]{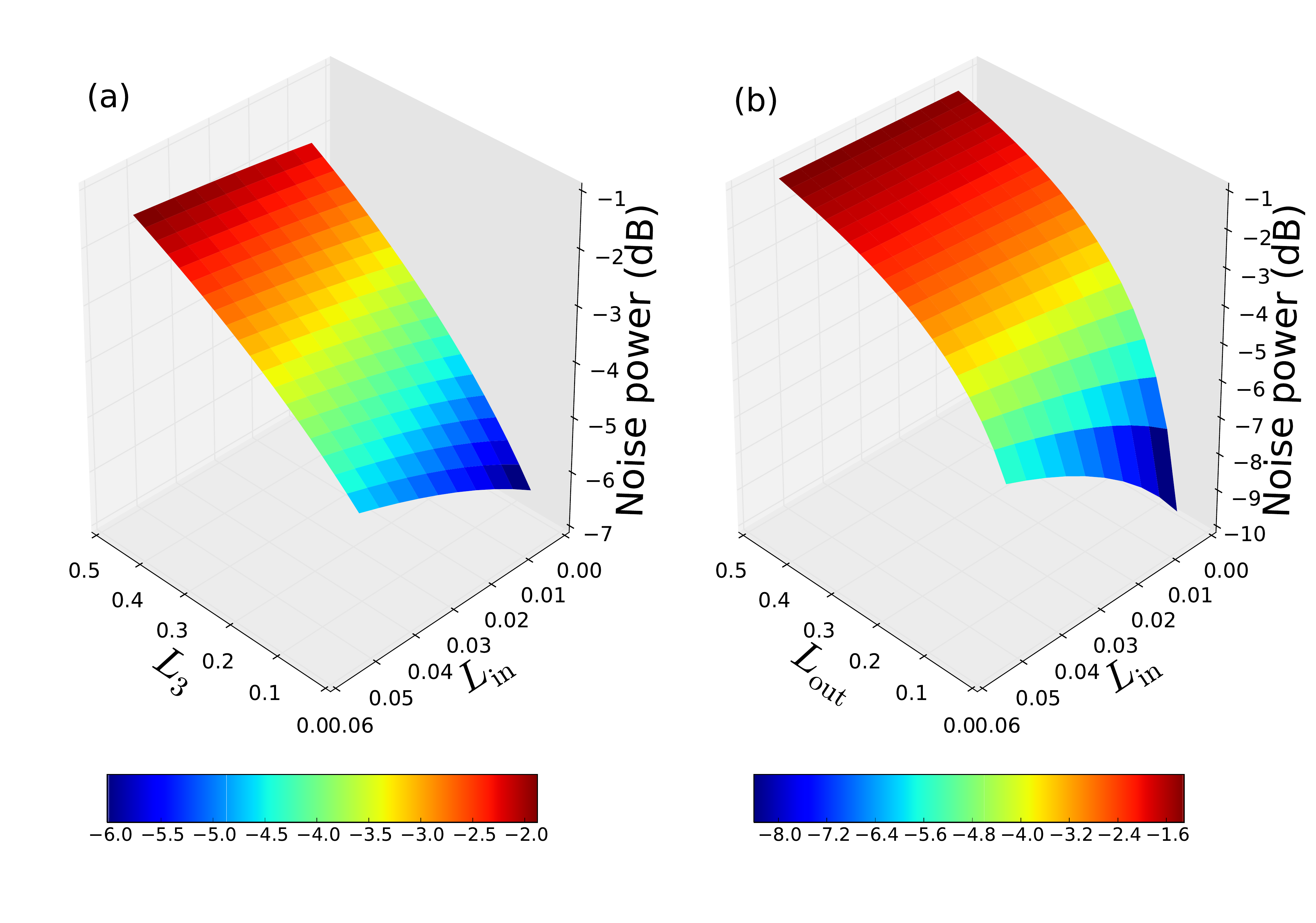}
\caption{The optimized degree of squeezing, $\mathcal{Q}^{-}(\omega_{\mathrm{opt}})$, for the CQFC network of two coupled OPOs, versus (a) $L_{\mathrm{in}}$ and $L_3$ (with $L_1 = L_2 = 0.1$), and (b) $L_{\mathrm{in}}$ and $L_{\mathrm{out}}$. Other parameters are $\omega_{\mathrm{opt}}/2\pi = 100$~MHz, $x_{\mathrm{u}} = 0.2$, $T_{\mathrm{u}} = 0.9$.}
\label{fig:Qmin_vs_Lin_and_L3}
\end{figure}

Finally, we explore further the dependence of the optimized degree of squeezing, $\mathcal{Q}^{-}(\omega_{\mathrm{opt}})$, on the intracavity and transmission line losses for the CQFC network of two coupled OPOs. Figure~\ref{fig:Qmin_vs_Lin_and_L3} shows $\mathcal{Q}^{-}(\omega_{\mathrm{opt}})$ versus (a) $L_{\mathrm{in}}$ and $L_3$ (with $L_1 = L_2 = 0.1$), and (b) $L_{\mathrm{in}}$ and $L_{\mathrm{out}}$. The situation when $L_3 \neq L_1 = L_2$ is practically relevant since $L_3$ includes, in addition to losses in the output transmission line, inefficiencies in the homodyne detector used to measure the squeezing spectrum of the output field. The results shown in Fig.~\ref{fig:Qmin_vs_Lin_and_L3} confirm that any increase in losses is detrimental to squeezing and quantify this relationship.

\section{Correlations between optimal values of phase variables}
\label{sec:corr}

Since the phase parameters play a significant role in tuning the quantum interference that governs the CQFC network's performance, an interesting question is whether their optimal values are correlated. Optimal values of a parameter can be cast as a vector each element of which corresponds to a distinct value of $\omega_{\mathrm{opt}}$, and correlations can be computed between pairs of such vectors. Specifically, we computed the Pearson correlation coefficient for all six pairs of four phase variables ($\theta_{\mathrm{p}}$, $\theta_{\mathrm{c}}$, $\phi_1$, $\phi_2$), and found that substantial correlations only exist between $\phi_1$ and $\phi_2$. Table~\ref{tab:corr_1} shows the values of the Pearson correlation coefficient $r(\phi_1, \phi_2)$, computed for $T_{\mathrm{u}} = 0.9$, $L_{\mathrm{in}} = 0.01$, and various values of $L_{\mathrm{out}}$ and $x_{\mathrm{u}}$.

\begin{table}[htbp]
\caption{\label{tab:corr_1}The Pearson correlation coefficient $r(\phi_1, \phi_2)$, for the CQFC network of two coupled OPOs with $T_{\mathrm{u}} = 0.9$, $L_{\mathrm{in}} = 0.01$, and various values of $L_{\mathrm{out}}$ and $x_{\mathrm{u}}$.}
\begin{ruledtabular}
\begin{tabular}{lrrrrrr}
{} & \multicolumn{6}{c}{$L_{\mathrm{out}}$}\\ 
$x_{\mathrm{u}}$ & 0.01 & 0.05 & 0.10 & 0.15 & 0.20 & 0.25 \\ \hline
0.1 & 0.575  & 0.438  & 0.436  & 0.354  & 0.275   & 0.223 \\
0.2 & 0.381  & 0.474  & 0.471  & 0.244  & 0.169 & -0.041 \\
0.3 & 0.588  & 0.322  & 0.316  & 0.165  & 0.132 & -0.121 \\
0.4 & 0.423  & 0.373  & 0.189  & 0.172 & -0.057 & -0.002 \\
\end{tabular}
\end{ruledtabular}
\end{table}

The correlation in Table~\ref{tab:corr_1} generally decreases as $L_{\mathrm{out}}$ and $x_{\mathrm{u}}$ increase. This trend can be compared to the one observed in Table~\ref{tab:f_crit} where the sideband frequency $\omega_{\mathrm{opt}}^{\star}$ at which the high-$\omega_{\mathrm{opt}}$ regime commences increases as $L_{\mathrm{out}}$ and $x_{\mathrm{u}}$ increase. Since the vectors $\phi_1 (\omega_{\mathrm{opt}})$ and $\phi_2 (\omega_{\mathrm{opt}})$ contain elements corresponding to both operation regimes, the trend observed in Table~\ref{tab:corr_1} implies that the correlation $r(\phi_1, \phi_2)$ generally decreases as the number of vector components corresponding to the high-$\omega_{\mathrm{opt}}$ regime decreases. A plausible explanation of this behavior is that the correlation between the two phase variables is higher in the high-$\omega_{\mathrm{opt}}$ regime. To test this hypothesis, we computed the Pearson correlation coefficient $r(\phi'_1, \phi'_2)$ for the pair of vectors $\phi'_1 = \phi_1 (\omega_{\mathrm{opt}} \geq \omega_{\mathrm{opt}}^{\star})$ and $\phi'_2 = \phi_2 (\omega_{\mathrm{opt}} \geq \omega_{\mathrm{opt}}^{\star})$ that include only elements corresponding to the high-$\omega_{\mathrm{opt}}$ regime. The values of $r(\phi'_1, \phi'_2)$ are shown in Table~\ref{tab:corr_2} for $T_{\mathrm{u}} = 0.9$, $L_{\mathrm{in}} = 0.01$, and various values of $L_{\mathrm{out}}$ and $x_{\mathrm{u}}$. The correlations in Table~\ref{tab:corr_2} are consistently larger than $0.5$, and, furthermore, we find that $r(\sin\phi'_1, \sin\phi'_2) = 1.0$ and $r(\cos\phi'_1, \cos\phi'_2) = -1.0$ (up to numerical precision) for all considered values of $L_{\mathrm{out}}$ and $x_{\mathrm{u}}$. These findings indicate a significant degree of concerted action in how the CQFC network of two coupled OPOs operates in the high-$\omega_{\mathrm{opt}}$ regime.

\begin{table}[htbp]
\caption{\label{tab:corr_2}The Pearson correlation coefficient $r(\phi'_1, \phi'_2)$, for the CQFC network of two coupled OPOs with $T_{\mathrm{u}} = 0.9$, $L_{\mathrm{in}} = 0.01$, and various values of $L_{\mathrm{out}}$ and $x_{\mathrm{u}}$.}
\begin{ruledtabular}
\begin{tabular}{lrrrrrr}
{} & \multicolumn{6}{c}{$L_{\mathrm{out}}$}\\ 
$x_{\mathrm{u}}$ & 0.01 & 0.05 & 0.10 & 0.15 & 0.20 & 0.25 \\ \hline
0.1 & 0.620  & 0.546  & 0.592  & 0.629  & 0.532  & 0.599 \\
0.2 & 0.610  & 0.604  & 0.597  & 0.578  & 0.650  & 0.679 \\
0.3 & 0.649  & 0.671  & 0.559  & 0.644  & 0.602  & 0.606 \\
0.4 & 0.555  & 0.646  & 0.540  & 0.687  & 0.604  & 0.582 \\
\end{tabular}
\end{ruledtabular}
\end{table}

\section{Robustness of optimal solutions}
\label{sec:robust}

Any practical implementation of a quantum optical network inevitably involves imprecisions and imperfections, which may affect the desired performance. This issue is of especial importance in a CQFC network, which relies on a precise quantum interference between the pump and controller fields to manipulate the properties of the output field (see, for example, the superposition of $a_{\mathrm{p}}$ and $a_{\mathrm{c}}$ in the first element of the $\mathbf{L}$ vector in Eq.~\eqref{eq:L-2OPOs}). This interference depends on the values of phase variables, and a key question is how robust is an optimal solution to small variations in these values. To analyze this robustness, we computed the Hessian of the objective function $J$ with respect to the phase variables, for a variety of optimal sets of network parameters.

For the single OPO network, $J$ depends on one phase variable $\theta_{\xi}$, and the Hessian $\mathsf{H}$ has one element $\partial^2 J/\partial \theta_{\xi}^2$. $\mathsf{H}$ was computed for 3500 optimal solutions (all combinations of $\omega_{\mathrm{opt}}/2\pi = \{ 2, 4, \ldots, 100 \}$~MHz, $x_{\mathrm{u}} = \{0.1, 0.2, \ldots, 0.5\}$, $T_{\mathrm{u}} = \{0.5, 0.9\}$, and $L_{\mathrm{tl}} = \{0.01, 0.05, 0.1, \ldots, 0.3\}$, with $L = 0.01$). The numerical analysis shows that the Hessian is zero (up to numerical precision) for all of these optimal solutions. Therefore, small fluctuations in the value of the pump phase $\theta_{\xi}$ should have no effect on the optimized degree of squeezing.

\begin{figure}[htbp]
\centering
\includegraphics[width=1.0\columnwidth]{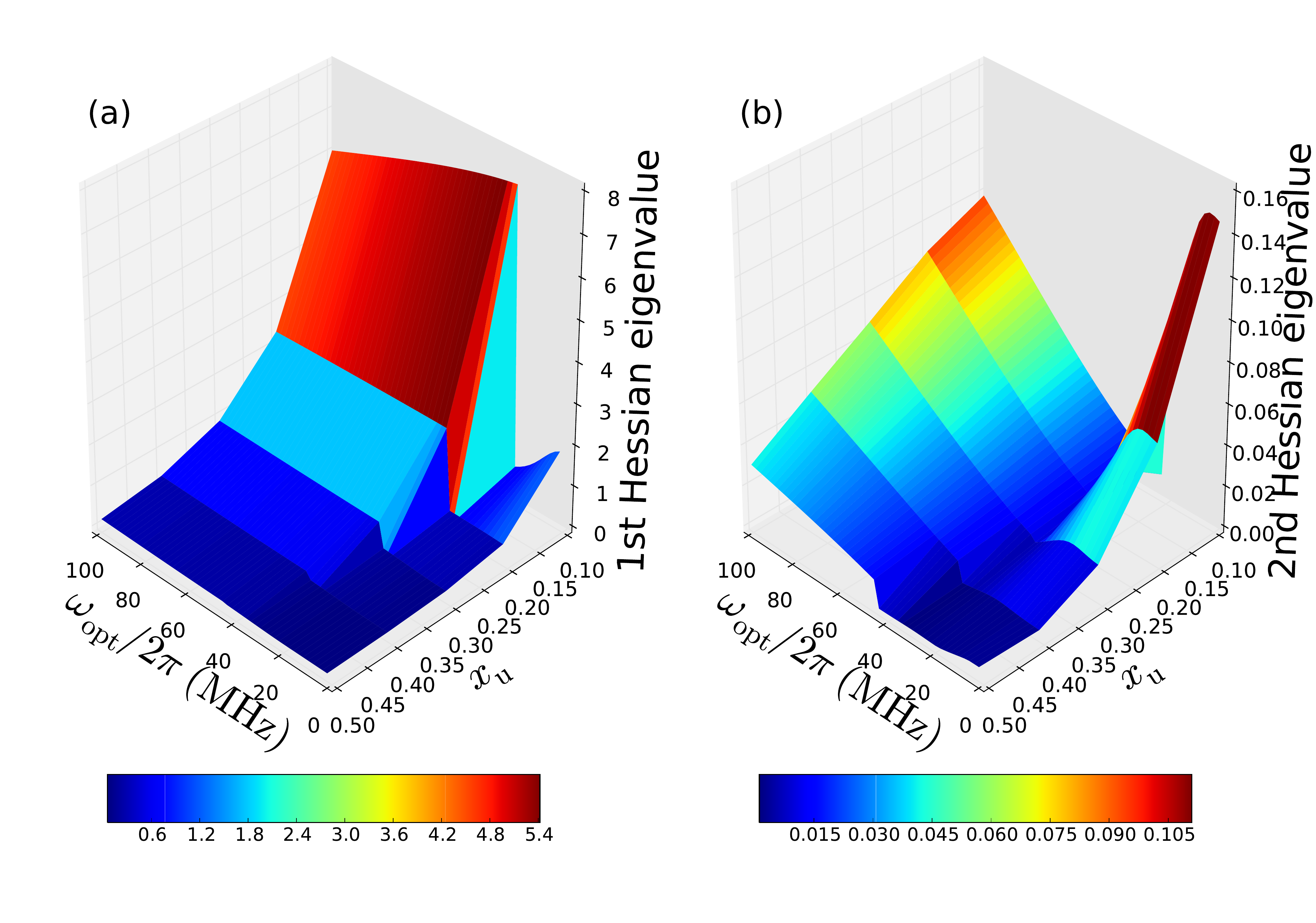}
\caption{The first (a) and second (b) Hessian eigenvalues for the CQFC network of two coupled OPOs, versus $\omega_{\mathrm{opt}}$ and $x_{\mathrm{u}}$. Other parameters are $L_{\mathrm{out}} = 0.1$, $T_{\mathrm{u}} = 0.9$, $L_{\mathrm{in}} = 0.01$.} 
\label{fig:HE_vs_fopt_and_xb}
\end{figure}

\begin{figure}[htbp]
\centering
\includegraphics[width=1.0\columnwidth]{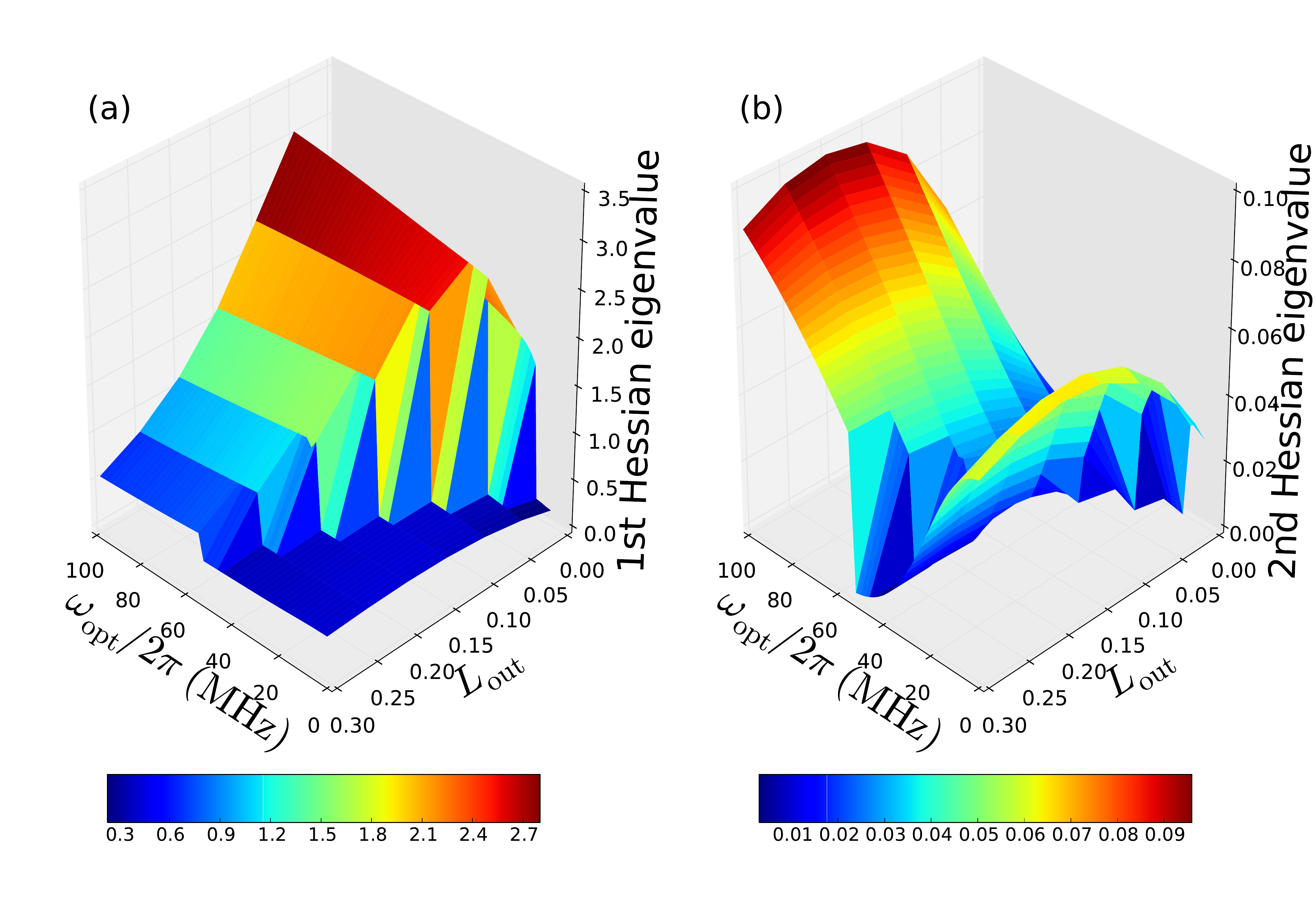}
\caption{The first (a) and second (b) Hessian eigenvalues for the CQFC network of two coupled OPOs, versus $\omega_{\mathrm{opt}}$ and $L_{\mathrm{out}}$. Other parameters are $x_{\mathrm{u}} = 0.2$, $T_{\mathrm{u}} = 0.9$, $L_{\mathrm{in}} = 0.01$.}
\label{fig:HE_vs_fopt_and_Lout}
\end{figure}

\begin{figure}[htbp]
\centering
\includegraphics[width=1.0\columnwidth]{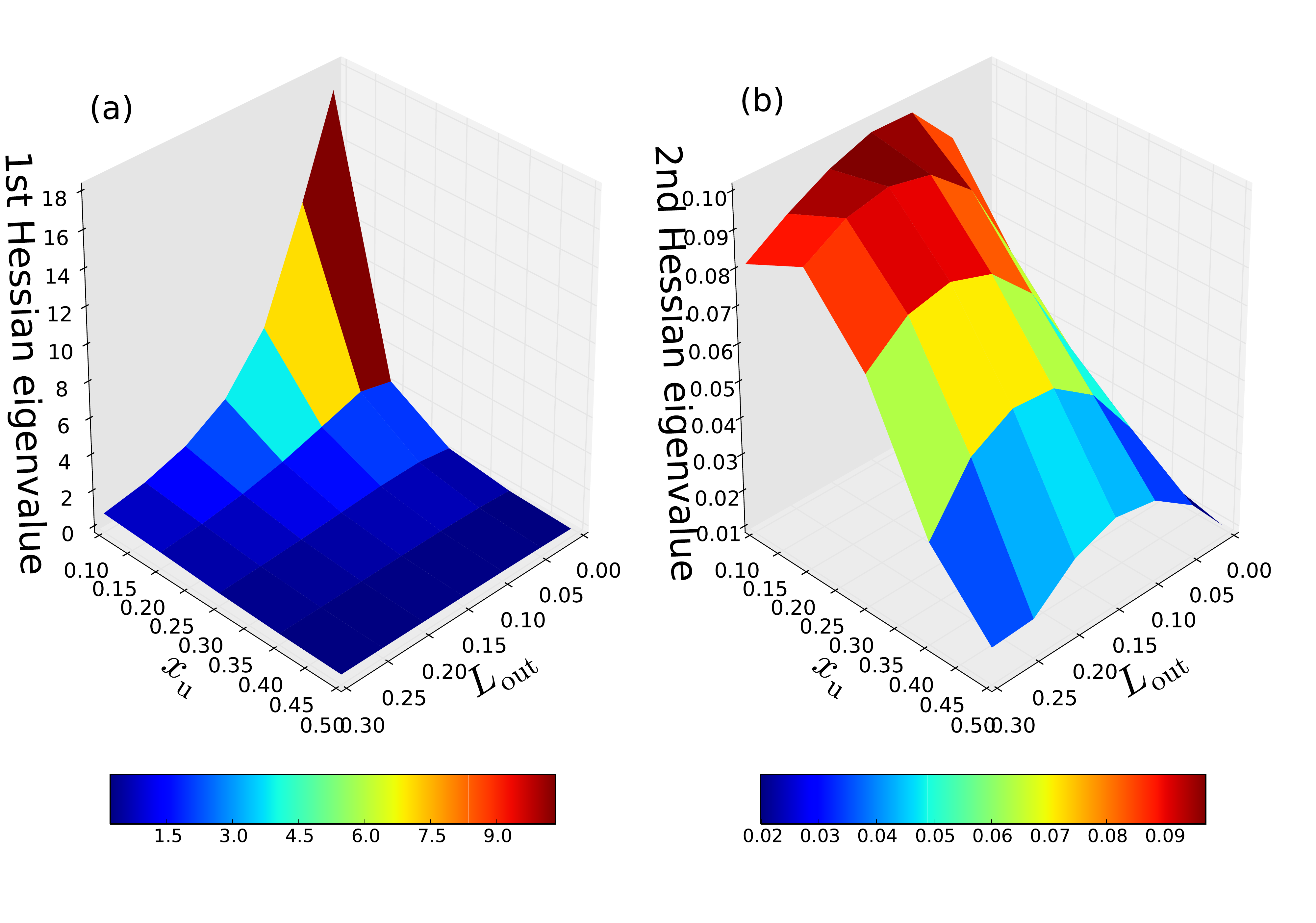}
\caption{The first (a) and second (b) Hessian eigenvalues for the CQFC network of two coupled OPOs, versus $x_{\mathrm{u}}$ and $L_{\mathrm{out}}$. Other parameters are $\omega_{\mathrm{opt}}/2\pi = 100$~MHz, $T_{\mathrm{u}} = 0.9$, $L_{\mathrm{in}} = 0.01$.}
\label{fig:HE_vs_xb_and_Lout}
\end{figure}

For the CQFC network of two coupled OPOs, $J$ depends on four phase variables ($\theta_{\mathrm{p}}$, $\theta_{\mathrm{c}}$, $\phi_1$, $\phi_2$), and the Hessian $\mathsf{H}$ is a $4 \times 4$ matrix of second-order derivatives. We computed the eigenvalues $\{h_1, \ldots, h_4\}$ and eigenvectors $\{\mathbf{e}_1, \ldots, \mathbf{e}_4\}$ of the Hessian $\mathsf{H}$ for 3500 optimal solutions (all combinations of $\omega_{\mathrm{opt}}/2\pi = \{ 2, 4, \ldots, 100 \}$~MHz, $x_{\mathrm{u}} = \{0.1, 0.2, \ldots, 0.5\}$, $T_{\mathrm{u}} = \{0.5, 0.9\}$, and $L_{\mathrm{out}} = \{0.01, 0.05, 0.1, \ldots, 0.3\}$, with $L_{\mathrm{in}} = 0.01$). The numerical analysis shows that two of the Hessian eigenvalues ($h_3$ and $h_4$) are zero (up to numerical precision) for all of these optimal solutions. Therefore, robustness to small phase variations is determined by two nonzero Hessian eigenvalues ($h_1$ and $h_2$). Figures~\ref{fig:HE_vs_fopt_and_xb}--\ref{fig:HE_vs_xb_and_Lout} show these nonzero Hessian eigenvalues as functions of $\omega_{\mathrm{opt}}$ and $x_{\mathrm{u}}$ (Fig.~\ref{fig:HE_vs_fopt_and_xb}), $\omega_{\mathrm{opt}}$ and $L_{\mathrm{out}}$ (Fig.~\ref{fig:HE_vs_fopt_and_Lout}), and $x_{\mathrm{u}}$ and $L_{\mathrm{out}}$ (Fig.~\ref{fig:HE_vs_xb_and_Lout}). We see that $h_1$ is typically much larger than $h_2$, and hence the magnitude of $h_1$ is the main factor determining the robustness properties of the optimal solutions. 

The dependence of $h_1$ on $\omega_{\mathrm{opt}}$, seen in Figs.~\ref{fig:HE_vs_fopt_and_xb}~and~\ref{fig:HE_vs_fopt_and_Lout}, demonstrates a significant difference in robustness properties between the low-$\omega_{\mathrm{opt}}$ and high-$\omega_{\mathrm{opt}}$ regimes. The low-$\omega_{\mathrm{opt}}$ regime is intrinsically robust for a broad range of parameter values. In the high-$\omega_{\mathrm{opt}}$ regime, a reasonable degree of robustness is achieved for $x_{\mathrm{u}} \geq 0.2$ (i.e., for pump powers above 4~W for the OPO parameters considered here). Larger losses in transmission lines ($L_{\mathrm{out}} \geq 0.1$) also enhance robustness.

\begin{figure}[htbp]
\centering
\includegraphics[width=1.0\columnwidth]{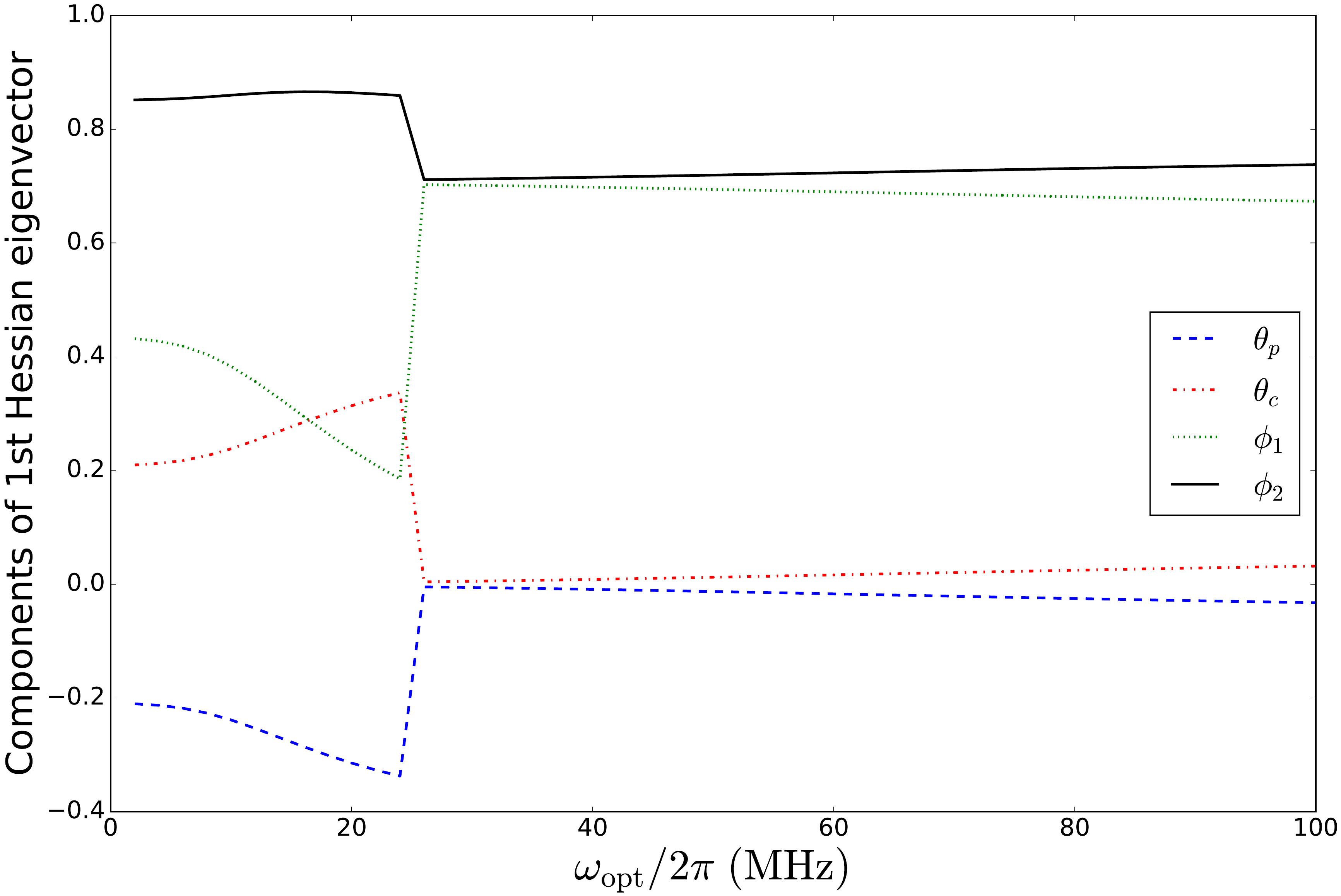}
\caption{Components of the first Hessian eigenvector for the CQFC network of two coupled OPOs, versus $\omega_{\mathrm{opt}}$. The four curves show components corresponding to the phase variables ($\theta_{\mathrm{p}}$, $\theta_{\mathrm{c}}$, $\phi_1$, $\phi_2$), as indicated in the legend. The parameters are $x_{\mathrm{u}} = 0.2$, $T_{\mathrm{u}} = 0.9$, $L_{\mathrm{in}} = 0.01$, $L_{\mathrm{out}} = 0.1$.}
\label{fig:HEV_vs_wopt}
\end{figure}

The four components of the Hessian eigenvector $\mathbf{e}_1$ (which corresponds to the largest eigenvalue $h_1$) are shown in Fig.~\ref{fig:HEV_vs_wopt} versus $\omega_{\mathrm{opt}}$. They also exhibit an abrupt change associated with the switch of the optimal operation regime at $\omega_{\mathrm{opt}}^{\star}$. In the low-$\omega_{\mathrm{opt}}$ regime, the eigenvector component corresponding to $\phi_2$ has the largest value and the rest of the components have smaller absolute values, but none is negligible. In the high-$\omega_{\mathrm{opt}}$ regime, the components corresponding to $\phi_1$ and $\phi_2$ have similar values, while the components corresponding to $\theta_{\mathrm{p}}$ and $\theta_{\mathrm{c}}$ are close to zero. These results are consistent with the findings that the low-$\omega_{\mathrm{opt}}$ regime is characterized by the maximum flow of light passing through the phase shifter P2 (from the controller to the plant), while the high-$\omega_{\mathrm{opt}}$ regime is characterized by roughly similar flows of light passing through the phase shifters P1 and P2 (in both directions).

\begin{figure}[htbp]
\centering
\includegraphics[width=1.0\columnwidth]{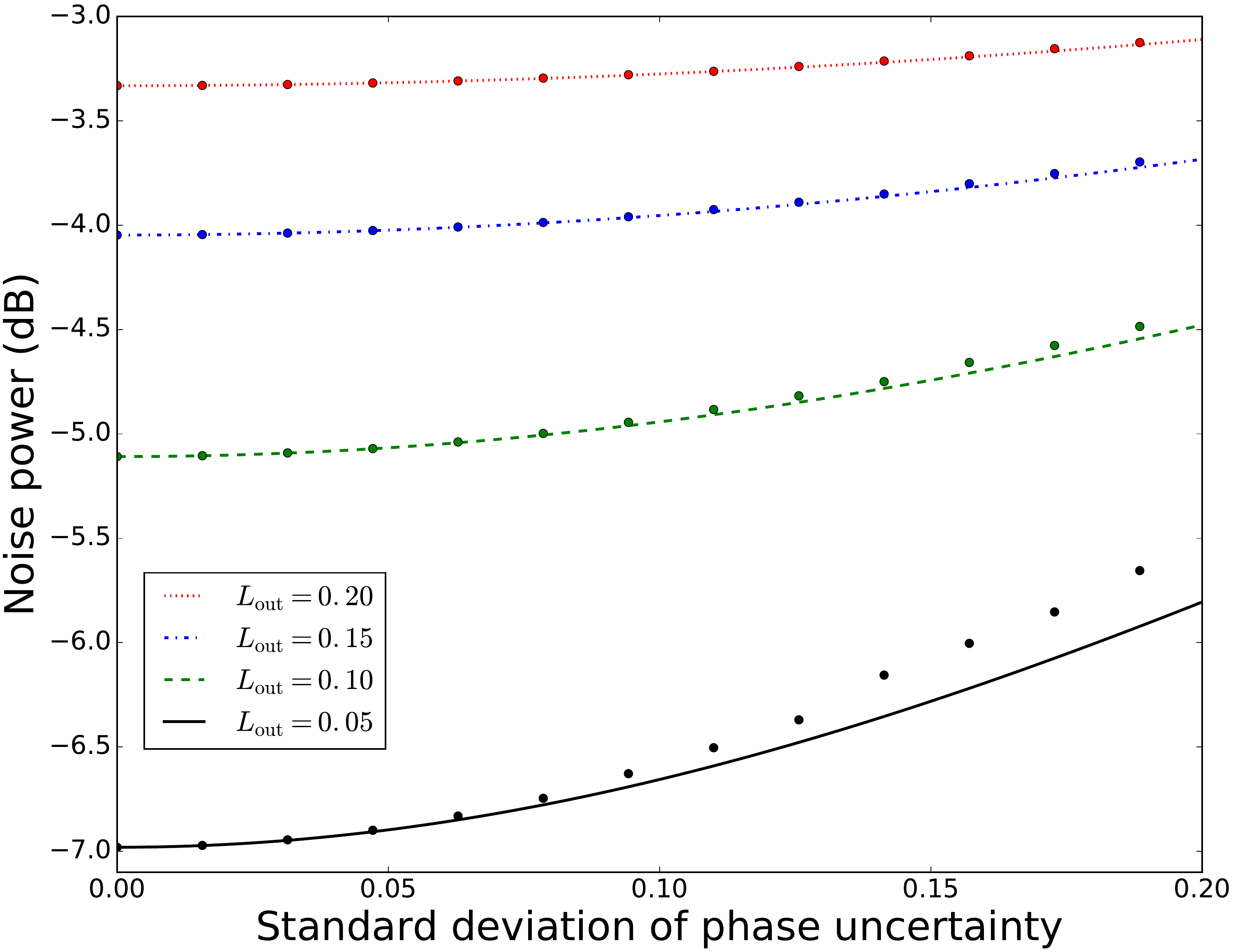}
\caption{The optimized degree of squeezing, $\mathcal{Q}^{-}(\omega_{\mathrm{opt}})$, for the CQFC network of two coupled OPOs, versus the standard deviation of phase uncertainty, $\sigma_{\mathrm{phase}}$. The four curves correspond to different values of $L_{\mathrm{out}}$ ($L_{\mathrm{out}} = \{0.05, 0.10, 0.15, 0.20\}$), as indicated in the legend. Other parameters are $\omega_{\mathrm{opt}}/2\pi = 100$~MHz, $x_{\mathrm{u}} = 0.2$, $T_{\mathrm{u}} = 0.9$, $L_{\mathrm{in}} = 0.01$. For each value of $L_{\mathrm{out}}$, the plot shows the results computed using the Hessian eigenvalues (lines) along with the data computed via Monte Carlo averaging over a random distribution of phase values (circles).}
\label{fig:Qmin_vs_dphase}
\end{figure}

The decrease of the optimized degree of squeezing due to small variations of phase parameters can be quantified using the computed Hessian eigenvalues or, alternatively, via direct Monte Carlo averaging over a random distribution of phase variable values. Figure~\ref{fig:Qmin_vs_dphase} shows the optimized degree of squeezing, $\mathcal{Q}^{-}(\omega_{\mathrm{opt}})$, for the CQFC network of two coupled OPOs (with $\omega_{\mathrm{opt}}/2\pi = 100$~MHz, $x_{\mathrm{u}} = 0.2$, $T_{\mathrm{u}} = 0.9$, and various values of $L_{\mathrm{out}}$), versus the standard deviation of phase uncertainty, $\sigma_{\mathrm{phase}}$ (for simplicity, we assume a normal distribution with zero mean and the same value of $\sigma_{\mathrm{phase}}$ for uncertainty in each of the four phase variables). We see a good agreement between the Hessian-based and Monte Carlo computations for $\sigma_{\mathrm{phase}} \leq 0.1$ (and even for $\sigma_{\mathrm{phase}} \leq 0.2$ for $L_{\mathrm{out}} \geq 0.1$). We also see that the deterioration of squeezing induced by phase variations is quite tolerable for $\sigma_{\mathrm{phase}} \leq 0.1$ (especially, for $L_{\mathrm{out}} \geq 0.1$). Note that our squeezing optimization procedure does not explicitly include a robustness requirement, and hence the observed high level of robustness might be surprising, but it is likely related to the natural tendency of stochastic optimization algorithms to eliminate solutions that are very sensitive to small parameter variations.

\section{Squeezing bandwidth optimization}
\label{sec:bw}

In CV-QKD with squeezed states, the secure key rate is proportional to the bandwidth of squeezing. Therefore, we also explored the capability of the CQFC network of two coupled OPOs to generate output states with high squeezing bandwidth, by optimizing the average degree of squeezing over a frequency interval $[0, \omega_{\mathrm{B}}]$, for various values of $\omega_{\mathrm{B}}$. Specifically, the objective function for these optimizations is
\begin{equation}
\label{eq:Pav}
J_{\mathrm{B}} = \overline{\mathcal{P}^{-}}(\omega_{\mathrm{B}}) 
\equiv \langle \mathcal{P}^{-}(\omega) \rangle
= \frac{1}{N_{\mathrm{B}}+1} \sum_{k = 0}^{N_{\mathrm{B}}} \mathcal{P}^{-}(\omega_k) .
\end{equation}
Here, $N_{\mathrm{B}} = \omega_{\mathrm{B}} / h_{\mathrm{B}}$ (i.e., $N_{\mathrm{B}} + 1$ is the number of sampling points), $\omega_k = k h_{\mathrm{B}}$, and $h_{\mathrm{B}}$ is the sampling interval. Except for the different choice of the objective function, the rest of the optimization procedure is the same as that described in Sec.~\ref{sec:optim}. In  optimization runs that minimized $J_{\mathrm{B}}$, we considered four bandwidth values $\omega_{\mathrm{B}}/2 \pi = \{ 25, 50, 75, 100 \}$~MHz and used the fixed sampling interval $h_{\mathrm{B}}/2 \pi = 1$~MHz.

For illustration purposes, we use a logarithmic measure of average squeezing,
\begin{equation}
\label{eq:Qav}
\overline{\mathcal{Q}^{-}}(\omega_{\mathrm{B}}) = 10 \log_{10} \overline{\mathcal{P}^{-}}(\omega_{\mathrm{B}}), 
\end{equation}
however, note that $\overline{\mathcal{Q}^{-}}(\omega_{\mathrm{B}}) \neq \langle \mathcal{Q}^{-}(\omega) \rangle$. Table~\ref{tab:bw} shows the best values of $\overline{\mathcal{Q}^{-}}(\omega_{\mathrm{B}})$ for $\omega_{\mathrm{B}}/2 \pi = 100$~MHz, for both the CQFC network of two coupled OPOs and the single OPO network, obtained in optimizations with $T_{\mathrm{u}} = 0.9$, $L = L_{\mathrm{in}} = 0.01$, and various values of $L_{\mathrm{tl}} = L_{\mathrm{out}}$ and $x_{\mathrm{u}}$. We see that the CQFC network of two coupled OPOs significantly outperforms the single OPO network in terms of the average squeezing generated over the $100$~MHz bandwidth, especially for lower values of transmission line losses.

\begin{table}[htbp]
\caption{\label{tab:bw}The best values of $\overline{\mathcal{Q}^{-}}(\omega_{\mathrm{B}})$ for $\omega_{\mathrm{B}}/2 \pi = 100$~MHz, for the CQFC network of two coupled OPOs and the single OPO network, obtained in optimizations with $T_{\mathrm{u}} = 0.9$, $L = L_{\mathrm{in}} = 0.01$, and various values of $L_{\mathrm{tl}} = L_{\mathrm{out}}$ and $x_{\mathrm{u}}$.}
\begin{ruledtabular}
\begin{tabular}{lrrrrrr}
\multicolumn{7}{c}{CQFC network of two coupled OPOs}\\
{} & \multicolumn{6}{c}{$L_{\mathrm{out}}$}\\ 
$x_{\mathrm{u}}$ & 0.01 & 0.05 & 0.10 & 0.15 & 0.20 & 0.25 \\ \hline
0.1 & -3.382 & -2.688 & -2.408 & -2.157 & -1.930 & -1.724 \\
0.2 & -5.773 & -4.937 & -4.339 & -3.829 & -3.385 & -2.994 \\
0.3 & -7.850 & -6.857 & -5.886 & -5.109 & -4.463 & -3.913 \\
0.4 & -9.994 & -8.441 & -7.073 & -6.049 & -5.234 & -4.559 \\ \hline
\multicolumn{7}{c}{Single OPO}\\
{} & \multicolumn{6}{c}{$L_{\mathrm{tl}}$}\\ 
$x_{\mathrm{u}}$ & 0.01 & 0.05 & 0.10 & 0.15 & 0.20 & 0.25 \\ \hline
0.1 & -1.428 & -1.361 & -1.277 & -1.196 & -1.115 & -1.037 \\
0.2 & -2.843 & -2.684 & -2.493 & -2.310 & -2.134 & -1.965 \\
0.3 & -4.248 & -3.966 & -3.638 & -3.332 & -3.047 & -2.779 \\
0.4 & -5.637 & -5.193 & -4.696 & -4.249 & -3.845 & -3.475 \\
\end{tabular}
\end{ruledtabular}
\end{table}

\begin{figure*}[htbp]
\centering
\includegraphics[width=1.82\columnwidth]{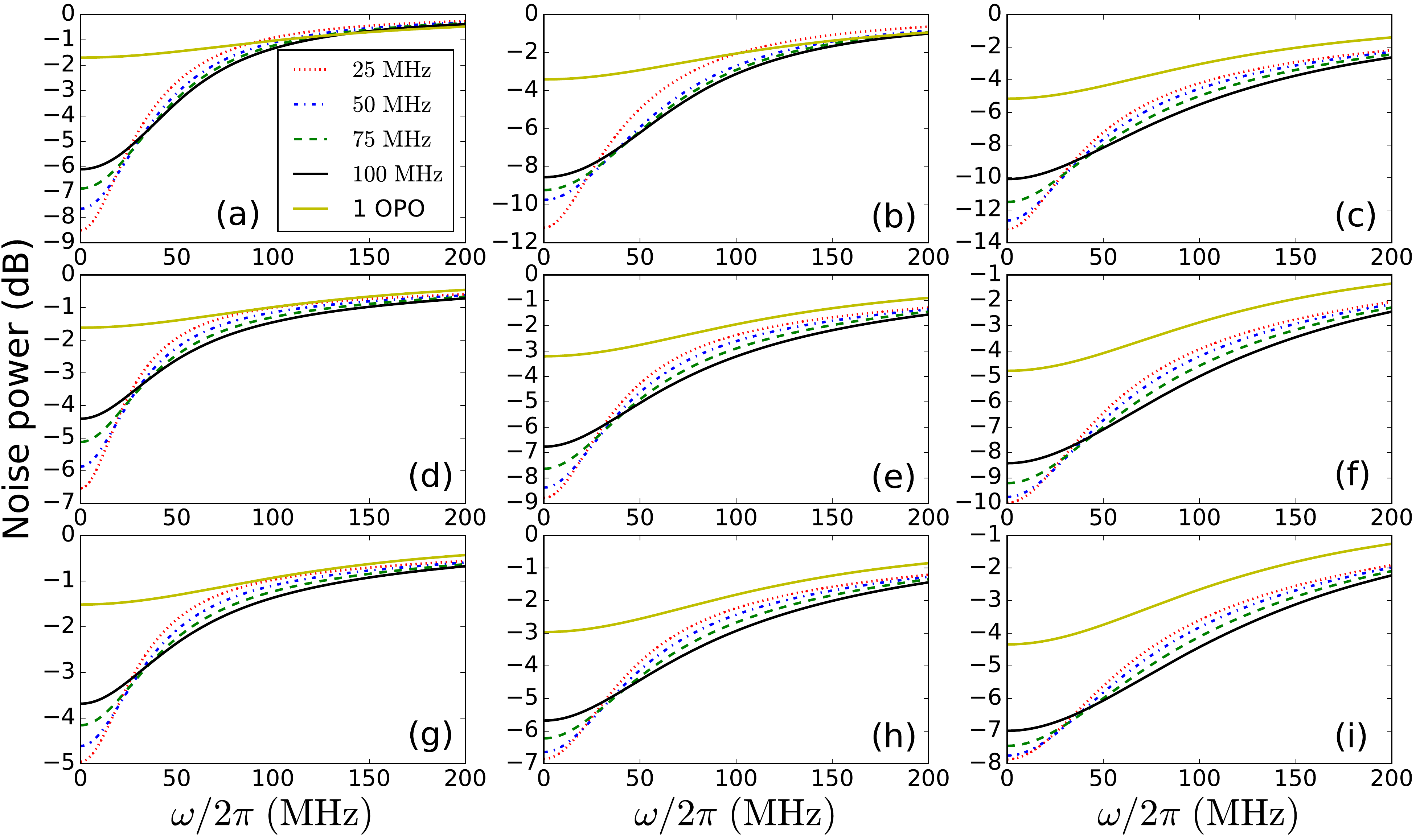}
\caption{The squeezing spectrum $\mathcal{Q}^{-}(\omega)$ for the optimal operation of both networks under the minimization of $J_{\mathrm{B}} = \overline{\mathcal{P}^{-}}(\omega_{\mathrm{B}})$ of Eq.~\eqref{eq:Pav}. Each subplot shows four curves corresponding to the optimally operated CQFC network of two coupled OPOs for different values of $\omega_{\mathrm{B}}$ ($\omega_{\mathrm{B}}/2\pi = \{25, 75, 50, 100\}$~MHz), along with a curve corresponding to the optimally operated single OPO network for any value of $\omega_{\mathrm{B}}$, as indicated in the legend. The transmission line losses are $L_{\mathrm{tl}} = L_{\mathrm{out}} = 0.01$ (subplots (a), (b), (c)), $L_{\mathrm{tl}} = L_{\mathrm{out}} = 0.05$ (subplots (d), (e), (f)), and $L_{\mathrm{tl}} = L_{\mathrm{out}} = 0.1$ (subplots (g), (h), (i)). The upper limit on the scaled pump amplitude is $x_{\mathrm{u}} = 0.1$ (subplots (a), (d), (g)), $x_{\mathrm{u}} = 0.2$ (subplots (b), (e), (h)), and $x_{\mathrm{u}} = 0.3$ (subplots (c), (f), (i)). Other parameters are $T_{\mathrm{u}} = 0.9$, $L = L_{\mathrm{in}} = 0.01$.}
\label{fig:Qmin_vs_f_bw}
\end{figure*}

It is also interesting to examine the squeezing spectrum $\mathcal{Q}^{-}(\omega)$ generated under the optimal operation of either network when we minimize $J_{\mathrm{B}} = \overline{\mathcal{P}^{-}}(\omega_{\mathrm{B}})$. Figure~\ref{fig:Qmin_vs_f_bw} shows $\mathcal{Q}^{-}(\omega)$ for both networks for various values of $\omega_{\mathrm{B}}$, $L_{\mathrm{tl}} = L_{\mathrm{out}}$, and $x_{\mathrm{u}}$. Similarly to the results shown in Sec.~\ref{sec:results} (cf.~Fig.~\ref{fig:Qmin_vs_f}), we find that the optimally operated single OPO network generates exactly the same Lorentzian squeezing spectrum for any choice of $\omega_{\mathrm{B}}$. In contrast, the CQFC network of two coupled OPOs is capable of adapting the generated squeezing spectrum depending on the selected value of $\omega_{\mathrm{B}}$ and overall produces much higher squeezing bandwidth.

\section{Conclusions}
\label{sec:conclusions}

We modeled the squeezing spectrum of the output field of the CQFC network of two coupled OPOs and used a suite of global optimization methods to examine the limits to which this spectrum can be varied under conditions typical for tabletop experiments. We found that, in contrast to a single OPO, the CQFC network can utilize the interference between the fields in the plant OPO and the controller OPO to significantly modify the squeezing spectrum of the output field in response to the selected optimization objective. In particular, when the objective is to maximize the degree of squeezing at a high-frequency sideband $\omega_{\mathrm{opt}}$, the CQFC network can operate in an optimal regime characterized by a high degree of cooperativity between the plant OPO and the controller OPO, as quantified by the flows of light between them and the correlation between the phase shifts $\phi_1$ and $\phi_2$. In this operation regime, the optimized squeezing spectrum $\mathcal{Q}^{-}(\omega)$ of the CQFC network of two coupled OPOs has the minimum at $\omega = \omega_{\mathrm{opt}}$, while the minimum of the optimized spectrum of the single OPO network is always at zero sideband frequency.

For both types of optimization objectives considered in this work (maximizing the degree of squeezing at a selected sideband frequency and maximizing the average degree of squeezing over a selected bandwidth), the CQFC network of two coupled OPOs significantly outperforms a single OPO in terms of squeezing achieved under similar conditions, even with higher losses in the CQFC network due to additional components and transmission lines. Also, the CQFC network is more effective in terms of converting a higher pump power into a stronger squeezing. While this superior performance of the CQFC network of two coupled OPOs relies on a phase-sensitive interference between multiple fields, we discovered, perhaps surprisingly, that squeezing generated by the optimally operated CQFC network is rather robust to small variations of phase parameters. This robustness can be attributed to the tendency of global optimization algorithms to avoid solutions that are overly sensitive to small parameter variations, but the fact that such robust network configurations do actually exist is quite remarkable.

Overall, our results strongly indicate that CQFC networks provide a very effective tool for engineering quantum optical systems with new properties and unprecedented levels of performance. This work also demonstrates the usefulness of advanced optimization methods for analyzing and improving the performance of such networks.

\acknowledgments
This work was supported by the Laboratory Directed Research and Development program at Sandia National Laboratories. Sandia is a multi-mission laboratory managed and operated  by Sandia Corporation, a wholly owned subsidiary of Lockheed Martin Corporation, for the United States Department of Energy's National Nuclear Security Administration under contract DE-AC04-94AL85000.

\bibliography{OPO_CQF_optimization}

\begin{thebibliography}{76}%
\makeatletter
\providecommand \@ifxundefined [1]{%
 \@ifx{#1\undefined}
}%
\providecommand \@ifnum [1]{%
 \ifnum #1\expandafter \@firstoftwo
 \else \expandafter \@secondoftwo
 \fi
}%
\providecommand \@ifx [1]{%
 \ifx #1\expandafter \@firstoftwo
 \else \expandafter \@secondoftwo
 \fi
}%
\providecommand \natexlab [1]{#1}%
\providecommand \enquote  [1]{``#1''}%
\providecommand \bibnamefont  [1]{#1}%
\providecommand \bibfnamefont [1]{#1}%
\providecommand \citenamefont [1]{#1}%
\providecommand \href@noop [0]{\@secondoftwo}%
\providecommand \href [0]{\begingroup \@sanitize@url \@href}%
\providecommand \@href[1]{\@@startlink{#1}\@@href}%
\providecommand \@@href[1]{\endgroup#1\@@endlink}%
\providecommand \@sanitize@url [0]{\catcode `\\12\catcode `\$12\catcode
  `\&12\catcode `\#12\catcode `\^12\catcode `\_12\catcode `\%12\relax}%
\providecommand \@@startlink[1]{}%
\providecommand \@@endlink[0]{}%
\providecommand \url  [0]{\begingroup\@sanitize@url \@url }%
\providecommand \@url [1]{\endgroup\@href {#1}{\urlprefix }}%
\providecommand \urlprefix  [0]{URL }%
\providecommand \Eprint [0]{\href }%
\providecommand \doibase [0]{http://dx.doi.org/}%
\providecommand \selectlanguage [0]{\@gobble}%
\providecommand \bibinfo  [0]{\@secondoftwo}%
\providecommand \bibfield  [0]{\@secondoftwo}%
\providecommand \translation [1]{[#1]}%
\providecommand \BibitemOpen [0]{}%
\providecommand \bibitemStop [0]{}%
\providecommand \bibitemNoStop [0]{.\EOS\space}%
\providecommand \EOS [0]{\spacefactor3000\relax}%
\providecommand \BibitemShut  [1]{\csname bibitem#1\endcsname}%
\let\auto@bib@innerbib\@empty
\bibitem [{\citenamefont {Brif}\ \emph {et~al.}(2010)\citenamefont {Brif},
  \citenamefont {Chakrabarti},\ and\ \citenamefont
  {Rabitz}}]{Brif.NJP.12.075008.2010}%
  \BibitemOpen
  \bibfield  {author} {\bibinfo {author} {\bibfnamefont {Constantin}\
  \bibnamefont {Brif}}, \bibinfo {author} {\bibfnamefont {Raj}\ \bibnamefont
  {Chakrabarti}}, \ and\ \bibinfo {author} {\bibfnamefont {Herschel}\
  \bibnamefont {Rabitz}},\ }\bibfield  {title} {\enquote {\bibinfo {title}
  {Control of quantum phenomena: past, present and future},}\ }\href {\doibase
  10.1088/1367-2630/12/7/075008} {\bibfield  {journal} {\bibinfo  {journal}
  {New J. Phys.}\ }\textbf {\bibinfo {volume} {12}},\ \bibinfo {pages} {075008}
  (\bibinfo {year} {2010})}\BibitemShut {NoStop}%
\bibitem [{\citenamefont {Wiseman}\ and\ \citenamefont
  {Milburn}(2014)}]{Wiseman.Milburn.book.2014}%
  \BibitemOpen
  \bibfield  {author} {\bibinfo {author} {\bibfnamefont {Howard~M.}\
  \bibnamefont {Wiseman}}\ and\ \bibinfo {author} {\bibfnamefont {Gerard~J.}\
  \bibnamefont {Milburn}},\ }\href@noop {} {\emph {\bibinfo {title} {Quantum
  Measurement and Control}}}\ (\bibinfo  {publisher} {Cambridge University
  Press},\ \bibinfo {address} {Cambridge, UK},\ \bibinfo {year}
  {2014})\BibitemShut {NoStop}%
\bibitem [{\citenamefont {Zhang}\ and\ \citenamefont
  {James}(2012)}]{Zhang.James.CSB.57.2200.2012}%
  \BibitemOpen
  \bibfield  {author} {\bibinfo {author} {\bibfnamefont {Guofeng}\ \bibnamefont
  {Zhang}}\ and\ \bibinfo {author} {\bibfnamefont {Matthew~R.}\ \bibnamefont
  {James}},\ }\bibfield  {title} {\enquote {\bibinfo {title} {{Quantum feedback
  networks and control: A brief survey}},}\ }\href {\doibase
  10.1007/s11434-012-5199-7} {\bibfield  {journal} {\bibinfo  {journal} {Chin.
  Sci. Bull.}\ }\textbf {\bibinfo {volume} {57}},\ \bibinfo {pages}
  {2200--2214} (\bibinfo {year} {2012})}\BibitemShut {NoStop}%
\bibitem [{\citenamefont {Gough}(2012)}]{Gough.PTRSA.370.5241.2012}%
  \BibitemOpen
  \bibfield  {author} {\bibinfo {author} {\bibfnamefont {John~E.}\ \bibnamefont
  {Gough}},\ }\bibfield  {title} {\enquote {\bibinfo {title} {Principles and
  applications of quantum control engineering},}\ }\href {\doibase
  10.1098/rsta.2012.0370} {\bibfield  {journal} {\bibinfo  {journal} {Phil.
  Trans. R. Soc. A}\ }\textbf {\bibinfo {volume} {370}},\ \bibinfo {pages}
  {5241--5258} (\bibinfo {year} {2012})}\BibitemShut {NoStop}%
\bibitem [{\citenamefont {Combes}\ \emph {et~al.}(2016)\citenamefont {Combes},
  \citenamefont {Kerckhoff},\ and\ \citenamefont
  {Sarovar}}]{Combes.arXiv.1611.00375.2016}%
  \BibitemOpen
  \bibfield  {author} {\bibinfo {author} {\bibfnamefont {Joshua}\ \bibnamefont
  {Combes}}, \bibinfo {author} {\bibfnamefont {Joseph}\ \bibnamefont
  {Kerckhoff}}, \ and\ \bibinfo {author} {\bibfnamefont {Mohan}\ \bibnamefont
  {Sarovar}},\ }\href@noop {} {\enquote {\bibinfo {title} {{The SLH framework
  for modeling quantum input-output networks}},}\ } (\bibinfo {year} {2016}),\
  \Eprint {http://arxiv.org/abs/arXiv:1611.00375} {arXiv:1611.00375 [quant-ph]}
  \BibitemShut {NoStop}%
\bibitem [{\citenamefont {Jacobs}\ \emph {et~al.}(2014)\citenamefont {Jacobs},
  \citenamefont {Wang},\ and\ \citenamefont
  {Wiseman}}]{Jacobs.NJP.16.073036.2014}%
  \BibitemOpen
  \bibfield  {author} {\bibinfo {author} {\bibfnamefont {Kurt}\ \bibnamefont
  {Jacobs}}, \bibinfo {author} {\bibfnamefont {Xiaoting}\ \bibnamefont {Wang}},
  \ and\ \bibinfo {author} {\bibfnamefont {Howard~M.}\ \bibnamefont
  {Wiseman}},\ }\bibfield  {title} {\enquote {\bibinfo {title} {Coherent
  feedback that beats all measurement-based feedback protocols},}\ }\href
  {\doibase 10.1088/1367-2630/16/7/073036} {\bibfield  {journal} {\bibinfo
  {journal} {New J. Phys.}\ }\textbf {\bibinfo {volume} {16}},\ \bibinfo
  {pages} {073036} (\bibinfo {year} {2014})}\BibitemShut {NoStop}%
\bibitem [{\citenamefont {Yamamoto}(2014)}]{Yamamoto.PRX.4.041029.2014}%
  \BibitemOpen
  \bibfield  {author} {\bibinfo {author} {\bibfnamefont {Naoki}\ \bibnamefont
  {Yamamoto}},\ }\bibfield  {title} {\enquote {\bibinfo {title} {Coherent
  versus measurement feedback: Linear systems theory for quantum
  information},}\ }\href {\doibase 10.1103/PhysRevX.4.041029} {\bibfield
  {journal} {\bibinfo  {journal} {Phys. Rev. X}\ }\textbf {\bibinfo {volume}
  {4}},\ \bibinfo {pages} {041029} (\bibinfo {year} {2014})}\BibitemShut
  {NoStop}%
\bibitem [{\citenamefont {Hudson}\ and\ \citenamefont
  {Parthasarathy}(1984)}]{Hudson.CMP.93.301.1984}%
  \BibitemOpen
  \bibfield  {author} {\bibinfo {author} {\bibfnamefont {R.~L.}\ \bibnamefont
  {Hudson}}\ and\ \bibinfo {author} {\bibfnamefont {K.~R.}\ \bibnamefont
  {Parthasarathy}},\ }\bibfield  {title} {\enquote {\bibinfo {title} {{Quantum
  Ito's formula and stochastic evolutions}},}\ }\href {\doibase
  10.1007/BF01258530} {\bibfield  {journal} {\bibinfo  {journal} {Comm. Math.
  Phys.}\ }\textbf {\bibinfo {volume} {93}},\ \bibinfo {pages} {301--323}
  (\bibinfo {year} {1984})}\BibitemShut {NoStop}%
\bibitem [{\citenamefont {Gardiner}\ and\ \citenamefont
  {Collett}(1985)}]{Gardiner.Collett.PRA.31.3761.1985}%
  \BibitemOpen
  \bibfield  {author} {\bibinfo {author} {\bibfnamefont {C.~W.}\ \bibnamefont
  {Gardiner}}\ and\ \bibinfo {author} {\bibfnamefont {M.~J.}\ \bibnamefont
  {Collett}},\ }\bibfield  {title} {\enquote {\bibinfo {title} {{Input and
  output in damped quantum systems: Quantum stochastic differential equations
  and the master equation}},}\ }\href {\doibase 10.1103/PhysRevA.31.3761}
  {\bibfield  {journal} {\bibinfo  {journal} {Phys. Rev. A}\ }\textbf {\bibinfo
  {volume} {31}},\ \bibinfo {pages} {3761--3774} (\bibinfo {year}
  {1985})}\BibitemShut {NoStop}%
\bibitem [{\citenamefont {Gardiner}(1993)}]{Gardiner.PRL.70.2269.1993}%
  \BibitemOpen
  \bibfield  {author} {\bibinfo {author} {\bibfnamefont {C.~W.}\ \bibnamefont
  {Gardiner}},\ }\bibfield  {title} {\enquote {\bibinfo {title} {Driving a
  quantum system with the output field from another driven quantum system},}\
  }\href {\doibase 10.1103/PhysRevLett.70.2269} {\bibfield  {journal} {\bibinfo
   {journal} {Phys. Rev. Lett.}\ }\textbf {\bibinfo {volume} {70}},\ \bibinfo
  {pages} {2269--2272} (\bibinfo {year} {1993})}\BibitemShut {NoStop}%
\bibitem [{\citenamefont {Wiseman}\ and\ \citenamefont
  {Milburn}(1994)}]{Wiseman.Milburn.PRA.49.4110.1994}%
  \BibitemOpen
  \bibfield  {author} {\bibinfo {author} {\bibfnamefont {H.~M.}\ \bibnamefont
  {Wiseman}}\ and\ \bibinfo {author} {\bibfnamefont {G.~J.}\ \bibnamefont
  {Milburn}},\ }\bibfield  {title} {\enquote {\bibinfo {title} {All-optical
  versus electro-optical quantum-limited feedback},}\ }\href {\doibase
  10.1103/PhysRevA.49.4110} {\bibfield  {journal} {\bibinfo  {journal} {Phys.
  Rev. A}\ }\textbf {\bibinfo {volume} {49}},\ \bibinfo {pages} {4110--4125}
  (\bibinfo {year} {1994})}\BibitemShut {NoStop}%
\bibitem [{\citenamefont {Gough}\ and\ \citenamefont
  {James}(2009{\natexlab{a}})}]{Gough.James.IEEE-TAC.54.2530.2007}%
  \BibitemOpen
  \bibfield  {author} {\bibinfo {author} {\bibfnamefont {John}\ \bibnamefont
  {Gough}}\ and\ \bibinfo {author} {\bibfnamefont {Matthew~R.}\ \bibnamefont
  {James}},\ }\bibfield  {title} {\enquote {\bibinfo {title} {The series
  product and its application to quantum feedforward and feedback networks},}\
  }\href {\doibase 10.1109/TAC.2009.2031205} {\bibfield  {journal} {\bibinfo
  {journal} {IEEE Trans. Autom. Control}\ }\textbf {\bibinfo {volume} {54}},\
  \bibinfo {pages} {2530--2544} (\bibinfo {year}
  {2009}{\natexlab{a}})}\BibitemShut {NoStop}%
\bibitem [{\citenamefont {Gough}\ and\ \citenamefont
  {James}(2009{\natexlab{b}})}]{Gough.James.CMP.287.1109.2009}%
  \BibitemOpen
  \bibfield  {author} {\bibinfo {author} {\bibfnamefont {J.}~\bibnamefont
  {Gough}}\ and\ \bibinfo {author} {\bibfnamefont {M.~R.}\ \bibnamefont
  {James}},\ }\bibfield  {title} {\enquote {\bibinfo {title} {{Quantum feedback
  networks: Hamiltonian formulation}},}\ }\href {\doibase
  10.1007/s00220-008-0698-8} {\bibfield  {journal} {\bibinfo  {journal}
  {Commun. Math. Phys.}\ }\textbf {\bibinfo {volume} {287}},\ \bibinfo {pages}
  {1109--1132} (\bibinfo {year} {2009}{\natexlab{b}})}\BibitemShut {NoStop}%
\bibitem [{\citenamefont {Gough}\ \emph {et~al.}(2010)\citenamefont {Gough},
  \citenamefont {James},\ and\ \citenamefont
  {Nurdin}}]{Gough.PRA.81.023804.2010}%
  \BibitemOpen
  \bibfield  {author} {\bibinfo {author} {\bibfnamefont {J.~E.}\ \bibnamefont
  {Gough}}, \bibinfo {author} {\bibfnamefont {M.~R.}\ \bibnamefont {James}}, \
  and\ \bibinfo {author} {\bibfnamefont {H.~I.}\ \bibnamefont {Nurdin}},\
  }\bibfield  {title} {\enquote {\bibinfo {title} {Squeezing components in
  linear quantum feedback networks},}\ }\href {\doibase
  10.1103/PhysRevA.81.023804} {\bibfield  {journal} {\bibinfo  {journal} {Phys.
  Rev. A}\ }\textbf {\bibinfo {volume} {81}},\ \bibinfo {pages} {023804}
  (\bibinfo {year} {2010})}\BibitemShut {NoStop}%
\bibitem [{\citenamefont {Tezak}\ \emph {et~al.}(2012)\citenamefont {Tezak},
  \citenamefont {Niederberger}, \citenamefont {Pavlichin}, \citenamefont
  {Sarma},\ and\ \citenamefont {Mabuchi}}]{Tezak.PTRSA.370.5270.2012}%
  \BibitemOpen
  \bibfield  {author} {\bibinfo {author} {\bibfnamefont {Nikolas}\ \bibnamefont
  {Tezak}}, \bibinfo {author} {\bibfnamefont {Armand}\ \bibnamefont
  {Niederberger}}, \bibinfo {author} {\bibfnamefont {Dmitri~S.}\ \bibnamefont
  {Pavlichin}}, \bibinfo {author} {\bibfnamefont {Gopal}\ \bibnamefont
  {Sarma}}, \ and\ \bibinfo {author} {\bibfnamefont {Hideo}\ \bibnamefont
  {Mabuchi}},\ }\bibfield  {title} {\enquote {\bibinfo {title} {Specification
  of photonic circuits using quantum hardware description language},}\ }\href
  {\doibase 10.1098/rsta.2011.0526} {\bibfield  {journal} {\bibinfo  {journal}
  {Phil. Trans. R. Soc. A}\ }\textbf {\bibinfo {volume} {370}},\ \bibinfo
  {pages} {5270--5290} (\bibinfo {year} {2012})}\BibitemShut {NoStop}%
\bibitem [{QNE()}]{QNET.url}%
  \BibitemOpen
  \href {http://mabuchilab.github.io/QNET} {}\bibinfo {note} {QNET software
  package, http://mabuchilab.github.io/QNET}\BibitemShut {NoStop}%
\bibitem [{\citenamefont {Kerckhoff}\ \emph {et~al.}(2010)\citenamefont
  {Kerckhoff}, \citenamefont {Nurdin}, \citenamefont {Pavlichin},\ and\
  \citenamefont {Mabuchi}}]{Kerckhoff.PRL.105.040502.2010}%
  \BibitemOpen
  \bibfield  {author} {\bibinfo {author} {\bibfnamefont {Joseph}\ \bibnamefont
  {Kerckhoff}}, \bibinfo {author} {\bibfnamefont {Hendra~I.}\ \bibnamefont
  {Nurdin}}, \bibinfo {author} {\bibfnamefont {Dmitri~S.}\ \bibnamefont
  {Pavlichin}}, \ and\ \bibinfo {author} {\bibfnamefont {Hideo}\ \bibnamefont
  {Mabuchi}},\ }\bibfield  {title} {\enquote {\bibinfo {title} {Designing
  quantum memories with embedded control: Photonic circuits for autonomous
  quantum error correction},}\ }\href {\doibase 10.1103/PhysRevLett.105.040502}
  {\bibfield  {journal} {\bibinfo  {journal} {Phys. Rev. Lett.}\ }\textbf
  {\bibinfo {volume} {105}},\ \bibinfo {pages} {040502} (\bibinfo {year}
  {2010})}\BibitemShut {NoStop}%
\bibitem [{\citenamefont {Kerckhoff}\ \emph {et~al.}(2011)\citenamefont
  {Kerckhoff}, \citenamefont {Pavlichin}, \citenamefont {Chalabi},\ and\
  \citenamefont {Mabuchi}}]{Kerckhoff.NJP.13.055022.2011}%
  \BibitemOpen
  \bibfield  {author} {\bibinfo {author} {\bibfnamefont {J.}~\bibnamefont
  {Kerckhoff}}, \bibinfo {author} {\bibfnamefont {D.~S.}\ \bibnamefont
  {Pavlichin}}, \bibinfo {author} {\bibfnamefont {H.}~\bibnamefont {Chalabi}},
  \ and\ \bibinfo {author} {\bibfnamefont {H.}~\bibnamefont {Mabuchi}},\
  }\bibfield  {title} {\enquote {\bibinfo {title} {Design of nanophotonic
  circuits for autonomous subsystem quantum error correction},}\ }\href
  {\doibase 10.1088/1367-2630/13/5/055022} {\bibfield  {journal} {\bibinfo
  {journal} {New J. Phys.}\ }\textbf {\bibinfo {volume} {13}},\ \bibinfo
  {pages} {055022} (\bibinfo {year} {2011})}\BibitemShut {NoStop}%
\bibitem [{\citenamefont {Hamerly}\ and\ \citenamefont
  {Mabuchi}(2012)}]{Hamerly.PRL.109.173602.2012}%
  \BibitemOpen
  \bibfield  {author} {\bibinfo {author} {\bibfnamefont {Ryan}\ \bibnamefont
  {Hamerly}}\ and\ \bibinfo {author} {\bibfnamefont {Hideo}\ \bibnamefont
  {Mabuchi}},\ }\bibfield  {title} {\enquote {\bibinfo {title} {Advantages of
  coherent feedback for cooling quantum oscillators},}\ }\href {\doibase
  10.1103/PhysRevLett.109.173602} {\bibfield  {journal} {\bibinfo  {journal}
  {Phys. Rev. Lett.}\ }\textbf {\bibinfo {volume} {109}},\ \bibinfo {pages}
  {173602} (\bibinfo {year} {2012})}\BibitemShut {NoStop}%
\bibitem [{\citenamefont {Hamerly}\ and\ \citenamefont
  {Mabuchi}(2013)}]{Hamerly.PRA.87.013815.2013}%
  \BibitemOpen
  \bibfield  {author} {\bibinfo {author} {\bibfnamefont {Ryan}\ \bibnamefont
  {Hamerly}}\ and\ \bibinfo {author} {\bibfnamefont {Hideo}\ \bibnamefont
  {Mabuchi}},\ }\bibfield  {title} {\enquote {\bibinfo {title} {Coherent
  controllers for optical-feedback cooling of quantum oscillators},}\ }\href
  {\doibase 10.1103/PhysRevA.87.013815} {\bibfield  {journal} {\bibinfo
  {journal} {Phys. Rev. A}\ }\textbf {\bibinfo {volume} {87}},\ \bibinfo
  {pages} {013815} (\bibinfo {year} {2013})}\BibitemShut {NoStop}%
\bibitem [{\citenamefont {Zhang}\ \emph {et~al.}(2012)\citenamefont {Zhang},
  \citenamefont {Wu}, \citenamefont {x.~Liu}, \citenamefont {Li},\ and\
  \citenamefont {Tarn}}]{Zhang.IEEE-TAC.57.1997.2012}%
  \BibitemOpen
  \bibfield  {author} {\bibinfo {author} {\bibfnamefont {J.}~\bibnamefont
  {Zhang}}, \bibinfo {author} {\bibfnamefont {R.~B.}\ \bibnamefont {Wu}},
  \bibinfo {author} {\bibfnamefont {Y.}~\bibnamefont {x.~Liu}}, \bibinfo
  {author} {\bibfnamefont {C.~W.}\ \bibnamefont {Li}}, \ and\ \bibinfo {author}
  {\bibfnamefont {T.~J.}\ \bibnamefont {Tarn}},\ }\bibfield  {title} {\enquote
  {\bibinfo {title} {Quantum coherent nonlinear feedback with applications to
  quantum optics on chip},}\ }\href {\doibase 10.1109/TAC.2012.2195871}
  {\bibfield  {journal} {\bibinfo  {journal} {IEEE Trans. Autom. Control}\
  }\textbf {\bibinfo {volume} {57}},\ \bibinfo {pages} {1997--2008} (\bibinfo
  {year} {2012})}\BibitemShut {NoStop}%
\bibitem [{\citenamefont {Liu}\ \emph {et~al.}(2015)\citenamefont {Liu},
  \citenamefont {Liu},\ and\ \citenamefont {Zhang}}]{Liu.JPB.48.105501.2015}%
  \BibitemOpen
  \bibfield  {author} {\bibinfo {author} {\bibfnamefont {Yu-Long}\ \bibnamefont
  {Liu}}, \bibinfo {author} {\bibfnamefont {Zhong-Peng}\ \bibnamefont {Liu}}, \
  and\ \bibinfo {author} {\bibfnamefont {Jing}\ \bibnamefont {Zhang}},\
  }\bibfield  {title} {\enquote {\bibinfo {title} {Coherent-feedback-induced
  controllable optical bistability and photon blockade},}\ }\href {\doibase
  10.1088/0953-4075/48/10/105501} {\bibfield  {journal} {\bibinfo  {journal}
  {J. Phys. B: At. Mol. Opt. Phys.}\ }\textbf {\bibinfo {volume} {48}},\
  \bibinfo {pages} {105501} (\bibinfo {year} {2015})}\BibitemShut {NoStop}%
\bibitem [{\citenamefont {Mabuchi}(2008)}]{Mabuchi.PRA.78.032323.2008}%
  \BibitemOpen
  \bibfield  {author} {\bibinfo {author} {\bibfnamefont {Hideo}\ \bibnamefont
  {Mabuchi}},\ }\bibfield  {title} {\enquote {\bibinfo {title}
  {Coherent-feedback quantum control with a dynamic compensator},}\ }\href
  {\doibase 10.1103/PhysRevA.78.032323} {\bibfield  {journal} {\bibinfo
  {journal} {Phys. Rev. A}\ }\textbf {\bibinfo {volume} {78}},\ \bibinfo
  {pages} {032323} (\bibinfo {year} {2008})}\BibitemShut {NoStop}%
\bibitem [{\citenamefont {Nurdin}\ and\ \citenamefont
  {Gough}(2015)}]{Nurdin.Gough.QIC.15.1017.2015}%
  \BibitemOpen
  \bibfield  {author} {\bibinfo {author} {\bibfnamefont {Hendra~I.}\
  \bibnamefont {Nurdin}}\ and\ \bibinfo {author} {\bibfnamefont {John~E.}\
  \bibnamefont {Gough}},\ }\bibfield  {title} {\enquote {\bibinfo {title}
  {Modular quantum memories using passive linear optics and coherent
  feedback},}\ }\href@noop {} {\bibfield  {journal} {\bibinfo  {journal}
  {Quantum Inf. Comput.}\ }\textbf {\bibinfo {volume} {15}},\ \bibinfo {pages}
  {1017--1040} (\bibinfo {year} {2015})},\ \Eprint
  {http://arxiv.org/abs/arXiv:1409.7473} {arXiv:1409.7473} \BibitemShut
  {NoStop}%
\bibitem [{\citenamefont {Gough}\ and\ \citenamefont
  {Wildfeuer}(2009)}]{Gough.Wildfeuer.PRA.80.042107.2009}%
  \BibitemOpen
  \bibfield  {author} {\bibinfo {author} {\bibfnamefont {J.~E.}\ \bibnamefont
  {Gough}}\ and\ \bibinfo {author} {\bibfnamefont {S.}~\bibnamefont
  {Wildfeuer}},\ }\bibfield  {title} {\enquote {\bibinfo {title} {Enhancement
  of field squeezing using coherent feedback},}\ }\href {\doibase
  10.1103/PhysRevA.80.042107} {\bibfield  {journal} {\bibinfo  {journal} {Phys.
  Rev. A}\ }\textbf {\bibinfo {volume} {80}},\ \bibinfo {pages} {042107}
  (\bibinfo {year} {2009})}\BibitemShut {NoStop}%
\bibitem [{\citenamefont {Iida}\ \emph {et~al.}(2012)\citenamefont {Iida},
  \citenamefont {Yukawa}, \citenamefont {Yonezawa}, \citenamefont {Yamamoto},\
  and\ \citenamefont {Furusawa}}]{Iida.IEEE-TAC.57.2045.2012}%
  \BibitemOpen
  \bibfield  {author} {\bibinfo {author} {\bibfnamefont {S.}~\bibnamefont
  {Iida}}, \bibinfo {author} {\bibfnamefont {M.}~\bibnamefont {Yukawa}},
  \bibinfo {author} {\bibfnamefont {H.}~\bibnamefont {Yonezawa}}, \bibinfo
  {author} {\bibfnamefont {N.}~\bibnamefont {Yamamoto}}, \ and\ \bibinfo
  {author} {\bibfnamefont {A.}~\bibnamefont {Furusawa}},\ }\bibfield  {title}
  {\enquote {\bibinfo {title} {Experimental demonstration of coherent feedback
  control on optical field squeezing},}\ }\href {\doibase
  10.1109/TAC.2012.2195831} {\bibfield  {journal} {\bibinfo  {journal} {IEEE
  Trans. Autom. Control}\ }\textbf {\bibinfo {volume} {57}},\ \bibinfo {pages}
  {2045--2050} (\bibinfo {year} {2012})}\BibitemShut {NoStop}%
\bibitem [{\citenamefont {Crisafulli}\ \emph {et~al.}(2013)\citenamefont
  {Crisafulli}, \citenamefont {Tezak}, \citenamefont {Soh}, \citenamefont
  {Armen},\ and\ \citenamefont {Mabuchi}}]{Crisafulli.OE.21.18371.2013}%
  \BibitemOpen
  \bibfield  {author} {\bibinfo {author} {\bibfnamefont {Orion}\ \bibnamefont
  {Crisafulli}}, \bibinfo {author} {\bibfnamefont {Nikolas}\ \bibnamefont
  {Tezak}}, \bibinfo {author} {\bibfnamefont {Daniel B.~S.}\ \bibnamefont
  {Soh}}, \bibinfo {author} {\bibfnamefont {Michael~A.}\ \bibnamefont {Armen}},
  \ and\ \bibinfo {author} {\bibfnamefont {Hideo}\ \bibnamefont {Mabuchi}},\
  }\bibfield  {title} {\enquote {\bibinfo {title} {Squeezed light in an optical
  parametric oscillator network with coherent feedback quantum control},}\
  }\href {\doibase 10.1364/OE.21.018371} {\bibfield  {journal} {\bibinfo
  {journal} {Opt. Expr.}\ }\textbf {\bibinfo {volume} {21}},\ \bibinfo {pages}
  {18371--18386} (\bibinfo {year} {2013})}\BibitemShut {NoStop}%
\bibitem [{\citenamefont {N\'emet}\ and\ \citenamefont
  {Parkins}(2016)}]{Nemet.Parkins.PRA.94.023809.2016}%
  \BibitemOpen
  \bibfield  {author} {\bibinfo {author} {\bibfnamefont {Nikolett}\
  \bibnamefont {N\'emet}}\ and\ \bibinfo {author} {\bibfnamefont {Scott}\
  \bibnamefont {Parkins}},\ }\bibfield  {title} {\enquote {\bibinfo {title}
  {Enhanced optical squeezing from a degenerate parametric amplifier via
  time-delayed coherent feedback},}\ }\href {\doibase
  10.1103/PhysRevA.94.023809} {\bibfield  {journal} {\bibinfo  {journal} {Phys.
  Rev. A}\ }\textbf {\bibinfo {volume} {94}},\ \bibinfo {pages} {023809}
  (\bibinfo {year} {2016})}\BibitemShut {NoStop}%
\bibitem [{\citenamefont {Yan}\ \emph {et~al.}(2011)\citenamefont {Yan},
  \citenamefont {Jia}, \citenamefont {Xie},\ and\ \citenamefont
  {Peng}}]{Yan.PRA.84.062304.2011}%
  \BibitemOpen
  \bibfield  {author} {\bibinfo {author} {\bibfnamefont {Zhihui}\ \bibnamefont
  {Yan}}, \bibinfo {author} {\bibfnamefont {Xiaojun}\ \bibnamefont {Jia}},
  \bibinfo {author} {\bibfnamefont {Changde}\ \bibnamefont {Xie}}, \ and\
  \bibinfo {author} {\bibfnamefont {Kunchi}\ \bibnamefont {Peng}},\ }\bibfield
  {title} {\enquote {\bibinfo {title} {Coherent feedback control of
  multipartite quantum entanglement for optical fields},}\ }\href {\doibase
  10.1103/PhysRevA.84.062304} {\bibfield  {journal} {\bibinfo  {journal} {Phys.
  Rev. A}\ }\textbf {\bibinfo {volume} {84}},\ \bibinfo {pages} {062304}
  (\bibinfo {year} {2011})}\BibitemShut {NoStop}%
\bibitem [{\citenamefont {Zhou}\ \emph {et~al.}(2015)\citenamefont {Zhou},
  \citenamefont {Jia}, \citenamefont {Li}, \citenamefont {Yu}, \citenamefont
  {Xie},\ and\ \citenamefont {Peng}}]{Zhou.SciRep.5.11132.2015}%
  \BibitemOpen
  \bibfield  {author} {\bibinfo {author} {\bibfnamefont {Yaoyao}\ \bibnamefont
  {Zhou}}, \bibinfo {author} {\bibfnamefont {Xiaojun}\ \bibnamefont {Jia}},
  \bibinfo {author} {\bibfnamefont {Fang}\ \bibnamefont {Li}}, \bibinfo
  {author} {\bibfnamefont {Juan}\ \bibnamefont {Yu}}, \bibinfo {author}
  {\bibfnamefont {Changde}\ \bibnamefont {Xie}}, \ and\ \bibinfo {author}
  {\bibfnamefont {Kunchi}\ \bibnamefont {Peng}},\ }\bibfield  {title} {\enquote
  {\bibinfo {title} {Quantum coherent feedback control for generation system of
  optical entangled state},}\ }\href {\doibase 10.1038/srep11132} {\bibfield
  {journal} {\bibinfo  {journal} {Sci. Rep.}\ }\textbf {\bibinfo {volume}
  {5}},\ \bibinfo {pages} {11132} (\bibinfo {year} {2015})}\BibitemShut
  {NoStop}%
\bibitem [{\citenamefont {Shi}\ and\ \citenamefont
  {Nurdin}(2015{\natexlab{a}})}]{Shi.Nurdin.QIP.14.337.2015}%
  \BibitemOpen
  \bibfield  {author} {\bibinfo {author} {\bibfnamefont {Zhan}\ \bibnamefont
  {Shi}}\ and\ \bibinfo {author} {\bibfnamefont {Hendra~I.}\ \bibnamefont
  {Nurdin}},\ }\bibfield  {title} {\enquote {\bibinfo {title} {{Coherent
  feedback enabled distributed generation of entanglement between propagating
  Gaussian fields}},}\ }\href {\doibase 10.1007/s11128-014-0845-4} {\bibfield
  {journal} {\bibinfo  {journal} {Quantum Inf. Process.}\ }\textbf {\bibinfo
  {volume} {14}},\ \bibinfo {pages} {337--359} (\bibinfo {year}
  {2015}{\natexlab{a}})}\BibitemShut {NoStop}%
\bibitem [{\citenamefont {Shi}\ and\ \citenamefont
  {Nurdin}(2015{\natexlab{b}})}]{Shi.Nurdin.arXiv.1502.01070.2015}%
  \BibitemOpen
  \bibfield  {author} {\bibinfo {author} {\bibfnamefont {Zhan}\ \bibnamefont
  {Shi}}\ and\ \bibinfo {author} {\bibfnamefont {Hendra~I.}\ \bibnamefont
  {Nurdin}},\ }\href@noop {} {\enquote {\bibinfo {title} {{Optimization of
  distributed EPR entanglement generated between two Gaussian fields by the
  modified steepest descent method}},}\ } (\bibinfo {year}
  {2015}{\natexlab{b}}),\ \Eprint {http://arxiv.org/abs/arXiv:1502.01070}
  {arXiv:1502.01070 [quant-ph]} \BibitemShut {NoStop}%
\bibitem [{\citenamefont {Shi}\ and\ \citenamefont
  {Nurdin}(2015{\natexlab{c}})}]{Shi.Nurdin.QIC.15.1141.2015}%
  \BibitemOpen
  \bibfield  {author} {\bibinfo {author} {\bibfnamefont {Zhan}\ \bibnamefont
  {Shi}}\ and\ \bibinfo {author} {\bibfnamefont {Hendra~I.}\ \bibnamefont
  {Nurdin}},\ }\bibfield  {title} {\enquote {\bibinfo {title} {Entanglement in
  a linear coherent feedback chain of nondegenerate optical parametric
  amplifiers},}\ }\href@noop {} {\bibfield  {journal} {\bibinfo  {journal}
  {Quantum Inf. Comput.}\ }\textbf {\bibinfo {volume} {15}},\ \bibinfo {pages}
  {1141--1164} (\bibinfo {year} {2015}{\natexlab{c}})},\ \Eprint
  {http://arxiv.org/abs/arXiv:1506.03152} {arXiv:1506.03152} \BibitemShut
  {NoStop}%
\bibitem [{\citenamefont {Shi}\ and\ \citenamefont
  {Nurdin}(2015{\natexlab{d}})}]{Shi.Nurdin.arXiv.1508.04584.2015}%
  \BibitemOpen
  \bibfield  {author} {\bibinfo {author} {\bibfnamefont {Zhan}\ \bibnamefont
  {Shi}}\ and\ \bibinfo {author} {\bibfnamefont {Hendra~I.}\ \bibnamefont
  {Nurdin}},\ }\href@noop {} {\enquote {\bibinfo {title} {Local optimality of a
  coherent feedback scheme for distributed entanglement generation: the
  idealized infinite bandwidth limit},}\ } (\bibinfo {year}
  {2015}{\natexlab{d}}),\ \Eprint {http://arxiv.org/abs/arXiv:1508.04584}
  {arXiv:1508.04584 [quant-ph]} \BibitemShut {NoStop}%
\bibitem [{\citenamefont {Miao}\ \emph {et~al.}(2015)\citenamefont {Miao},
  \citenamefont {Hush},\ and\ \citenamefont {James}}]{Miao.PRA.92.012115.2015}%
  \BibitemOpen
  \bibfield  {author} {\bibinfo {author} {\bibfnamefont {Zibo}\ \bibnamefont
  {Miao}}, \bibinfo {author} {\bibfnamefont {Michael~R.}\ \bibnamefont {Hush}},
  \ and\ \bibinfo {author} {\bibfnamefont {Matthew~R.}\ \bibnamefont {James}},\
  }\bibfield  {title} {\enquote {\bibinfo {title} {Coherently tracking the
  covariance matrix of an open quantum system},}\ }\href {\doibase
  10.1103/PhysRevA.92.012115} {\bibfield  {journal} {\bibinfo  {journal} {Phys.
  Rev. A}\ }\textbf {\bibinfo {volume} {92}},\ \bibinfo {pages} {012115}
  (\bibinfo {year} {2015})}\BibitemShut {NoStop}%
\bibitem [{\citenamefont {Roy}\ \emph {et~al.}(2016)\citenamefont {Roy},
  \citenamefont {Petersen},\ and\ \citenamefont
  {Huntington}}]{Roy.arXiv.1502.03729.2016}%
  \BibitemOpen
  \bibfield  {author} {\bibinfo {author} {\bibfnamefont {Shibdas}\ \bibnamefont
  {Roy}}, \bibinfo {author} {\bibfnamefont {Ian~R.}\ \bibnamefont {Petersen}},
  \ and\ \bibinfo {author} {\bibfnamefont {Elanor~H.}\ \bibnamefont
  {Huntington}},\ }\href@noop {} {\enquote {\bibinfo {title}
  {Coherent-classical estimation for linear quantum systems},}\ } (\bibinfo
  {year} {2016}),\ \Eprint {http://arxiv.org/abs/arXiv:1502.03729}
  {arXiv:1502.03729 [math.OC]} \BibitemShut {NoStop}%
\bibitem [{\citenamefont {Mabuchi}(2009)}]{Mabuchi.PRA.80.045802.2009}%
  \BibitemOpen
  \bibfield  {author} {\bibinfo {author} {\bibfnamefont {Hideo}\ \bibnamefont
  {Mabuchi}},\ }\bibfield  {title} {\enquote {\bibinfo {title} {{Cavity-QED
  models of switches for attojoule-scale nanophotonic logic}},}\ }\href
  {\doibase 10.1103/PhysRevA.80.045802} {\bibfield  {journal} {\bibinfo
  {journal} {Phys. Rev. A}\ }\textbf {\bibinfo {volume} {80}},\ \bibinfo
  {pages} {045802} (\bibinfo {year} {2009})}\BibitemShut {NoStop}%
\bibitem [{\citenamefont {Mabuchi}(2011)}]{Mabuchi.APL.98.193109.2011}%
  \BibitemOpen
  \bibfield  {author} {\bibinfo {author} {\bibfnamefont {Hideo}\ \bibnamefont
  {Mabuchi}},\ }\bibfield  {title} {\enquote {\bibinfo {title}
  {Coherent-feedback control strategy to suppress spontaneous switching in
  ultralow power optical bistability},}\ }\href {\doibase 10.1063/1.3589994}
  {\bibfield  {journal} {\bibinfo  {journal} {Appl. Phys. Lett.}\ }\textbf
  {\bibinfo {volume} {98}},\ \bibinfo {pages} {193109} (\bibinfo {year}
  {2011})}\BibitemShut {NoStop}%
\bibitem [{\citenamefont {Santori}\ \emph {et~al.}(2014)\citenamefont
  {Santori}, \citenamefont {Pelc}, \citenamefont {Beausoleil}, \citenamefont
  {Tezak}, \citenamefont {Hamerly},\ and\ \citenamefont
  {Mabuchi}}]{Santori.PRAppl.1.054005.2014}%
  \BibitemOpen
  \bibfield  {author} {\bibinfo {author} {\bibfnamefont {Charles}\ \bibnamefont
  {Santori}}, \bibinfo {author} {\bibfnamefont {Jason~S.}\ \bibnamefont
  {Pelc}}, \bibinfo {author} {\bibfnamefont {Raymond~G.}\ \bibnamefont
  {Beausoleil}}, \bibinfo {author} {\bibfnamefont {Nikolas}\ \bibnamefont
  {Tezak}}, \bibinfo {author} {\bibfnamefont {Ryan}\ \bibnamefont {Hamerly}}, \
  and\ \bibinfo {author} {\bibfnamefont {Hideo}\ \bibnamefont {Mabuchi}},\
  }\bibfield  {title} {\enquote {\bibinfo {title} {Quantum noise in large-scale
  coherent nonlinear photonic circuits},}\ }\href {\doibase
  10.1103/PhysRevApplied.1.054005} {\bibfield  {journal} {\bibinfo  {journal}
  {Phys. Rev. Applied}\ }\textbf {\bibinfo {volume} {1}},\ \bibinfo {pages}
  {054005} (\bibinfo {year} {2014})}\BibitemShut {NoStop}%
\bibitem [{\citenamefont {Pavlichin}\ and\ \citenamefont
  {Mabuchi}(2014)}]{Pavlichin.Mabuchi.NJP.16.105017.2014}%
  \BibitemOpen
  \bibfield  {author} {\bibinfo {author} {\bibfnamefont {Dmitri~S.}\
  \bibnamefont {Pavlichin}}\ and\ \bibinfo {author} {\bibfnamefont {Hideo}\
  \bibnamefont {Mabuchi}},\ }\bibfield  {title} {\enquote {\bibinfo {title}
  {Photonic circuits for iterative decoding of a class of low-density
  parity-check codes},}\ }\href {\doibase 10.1088/1367-2630/16/10/105017}
  {\bibfield  {journal} {\bibinfo  {journal} {New J. Phys.}\ }\textbf {\bibinfo
  {volume} {16}},\ \bibinfo {pages} {105017} (\bibinfo {year}
  {2014})}\BibitemShut {NoStop}%
\bibitem [{\citenamefont {Tezak}\ and\ \citenamefont
  {Mabuchi}(2015)}]{Tezak.Mabuchi.EPJ-QT.2.10.2015}%
  \BibitemOpen
  \bibfield  {author} {\bibinfo {author} {\bibfnamefont {Nikolas}\ \bibnamefont
  {Tezak}}\ and\ \bibinfo {author} {\bibfnamefont {Hideo}\ \bibnamefont
  {Mabuchi}},\ }\bibfield  {title} {\enquote {\bibinfo {title} {A coherent
  perceptron for all-optical learning},}\ }\href {\doibase
  10.1140/epjqt/s40507-015-0023-3} {\bibfield  {journal} {\bibinfo  {journal}
  {EPJ Quantum Technology}\ }\textbf {\bibinfo {volume} {2}},\ \bibinfo {pages}
  {10} (\bibinfo {year} {2015})}\BibitemShut {NoStop}%
\bibitem [{\citenamefont {Sarovar}\ \emph {et~al.}(2016)\citenamefont
  {Sarovar}, \citenamefont {Soh}, \citenamefont {Cox}, \citenamefont {Brif},
  \citenamefont {DeRose}, \citenamefont {Camacho},\ and\ \citenamefont
  {Davids}}]{Sarovar.EPJ-QT.3.14.2016}%
  \BibitemOpen
  \bibfield  {author} {\bibinfo {author} {\bibfnamefont {Mohan}\ \bibnamefont
  {Sarovar}}, \bibinfo {author} {\bibfnamefont {Daniel B.~S.}\ \bibnamefont
  {Soh}}, \bibinfo {author} {\bibfnamefont {Jonathan}\ \bibnamefont {Cox}},
  \bibinfo {author} {\bibfnamefont {Constantin}\ \bibnamefont {Brif}}, \bibinfo
  {author} {\bibfnamefont {Christopher~T.}\ \bibnamefont {DeRose}}, \bibinfo
  {author} {\bibfnamefont {Ryan}\ \bibnamefont {Camacho}}, \ and\ \bibinfo
  {author} {\bibfnamefont {Paul}\ \bibnamefont {Davids}},\ }\bibfield  {title}
  {\enquote {\bibinfo {title} {Silicon nanophotonics for scalable quantum
  coherent feedback networks},}\ }\href {\doibase
  10.1140/epjqt/s40507-016-0052-6} {\bibfield  {journal} {\bibinfo  {journal}
  {EPJ Quantum Technology}\ }\textbf {\bibinfo {volume} {3}},\ \bibinfo {pages}
  {14} (\bibinfo {year} {2016})}\BibitemShut {NoStop}%
\bibitem [{\citenamefont {Kerckhoff}\ and\ \citenamefont
  {Lehnert}(2012)}]{Kerckhoff.PRL.109.153602.2012}%
  \BibitemOpen
  \bibfield  {author} {\bibinfo {author} {\bibfnamefont {Joseph}\ \bibnamefont
  {Kerckhoff}}\ and\ \bibinfo {author} {\bibfnamefont {K.~W.}\ \bibnamefont
  {Lehnert}},\ }\bibfield  {title} {\enquote {\bibinfo {title} {Superconducting
  microwave multivibrator produced by coherent feedback},}\ }\href {\doibase
  10.1103/PhysRevLett.109.153602} {\bibfield  {journal} {\bibinfo  {journal}
  {Phys. Rev. Lett.}\ }\textbf {\bibinfo {volume} {109}},\ \bibinfo {pages}
  {153602} (\bibinfo {year} {2012})}\BibitemShut {NoStop}%
\bibitem [{\citenamefont {Kerckhoff}\ \emph {et~al.}(2013)\citenamefont
  {Kerckhoff}, \citenamefont {Andrews}, \citenamefont {Ku}, \citenamefont
  {Kindel}, \citenamefont {Cicak}, \citenamefont {Simmonds},\ and\
  \citenamefont {Lehnert}}]{Kerckhoff.PRX.3.021013.2013}%
  \BibitemOpen
  \bibfield  {author} {\bibinfo {author} {\bibfnamefont {Joseph}\ \bibnamefont
  {Kerckhoff}}, \bibinfo {author} {\bibfnamefont {Reed~W.}\ \bibnamefont
  {Andrews}}, \bibinfo {author} {\bibfnamefont {H.~S.}\ \bibnamefont {Ku}},
  \bibinfo {author} {\bibfnamefont {William~F.}\ \bibnamefont {Kindel}},
  \bibinfo {author} {\bibfnamefont {Katarina}\ \bibnamefont {Cicak}}, \bibinfo
  {author} {\bibfnamefont {Raymond~W.}\ \bibnamefont {Simmonds}}, \ and\
  \bibinfo {author} {\bibfnamefont {K.~W.}\ \bibnamefont {Lehnert}},\
  }\bibfield  {title} {\enquote {\bibinfo {title} {Tunable coupling to a
  mechanical oscillator circuit using a coherent feedback network},}\ }\href
  {\doibase 10.1103/PhysRevX.3.021013} {\bibfield  {journal} {\bibinfo
  {journal} {Phys. Rev. X}\ }\textbf {\bibinfo {volume} {3}},\ \bibinfo {pages}
  {021013} (\bibinfo {year} {2013})}\BibitemShut {NoStop}%
\bibitem [{\citenamefont {Collett}\ and\ \citenamefont
  {Gardiner}(1984)}]{Collett.Gardiner.PRA.30.1386.1984}%
  \BibitemOpen
  \bibfield  {author} {\bibinfo {author} {\bibfnamefont {M.~J.}\ \bibnamefont
  {Collett}}\ and\ \bibinfo {author} {\bibfnamefont {C.~W.}\ \bibnamefont
  {Gardiner}},\ }\bibfield  {title} {\enquote {\bibinfo {title} {Squeezing of
  intracavity and traveling-wave light fields produced in parametric
  amplification},}\ }\href {\doibase 10.1103/PhysRevA.30.1386} {\bibfield
  {journal} {\bibinfo  {journal} {Phys. Rev. A}\ }\textbf {\bibinfo {volume}
  {30}},\ \bibinfo {pages} {1386--1391} (\bibinfo {year} {1984})}\BibitemShut
  {NoStop}%
\bibitem [{\citenamefont {Collett}\ and\ \citenamefont
  {Walls}(1985)}]{Collett.Walls.PRA.32.2887.1985}%
  \BibitemOpen
  \bibfield  {author} {\bibinfo {author} {\bibfnamefont {M.~J.}\ \bibnamefont
  {Collett}}\ and\ \bibinfo {author} {\bibfnamefont {D.~F.}\ \bibnamefont
  {Walls}},\ }\bibfield  {title} {\enquote {\bibinfo {title} {Squeezing spectra
  for nonlinear optical systems},}\ }\href {\doibase 10.1103/PhysRevA.32.2887}
  {\bibfield  {journal} {\bibinfo  {journal} {Phys. Rev. A}\ }\textbf {\bibinfo
  {volume} {32}},\ \bibinfo {pages} {2887--2892} (\bibinfo {year}
  {1985})}\BibitemShut {NoStop}%
\bibitem [{\citenamefont {Wu}\ \emph {et~al.}(1987)\citenamefont {Wu},
  \citenamefont {Xiao},\ and\ \citenamefont {Kimble}}]{Wu.JOSAB.4.1465.1987}%
  \BibitemOpen
  \bibfield  {author} {\bibinfo {author} {\bibfnamefont {Ling-An}\ \bibnamefont
  {Wu}}, \bibinfo {author} {\bibfnamefont {Min}\ \bibnamefont {Xiao}}, \ and\
  \bibinfo {author} {\bibfnamefont {H.~J.}\ \bibnamefont {Kimble}},\ }\bibfield
   {title} {\enquote {\bibinfo {title} {Squeezed states of light from an
  optical parametric oscillator},}\ }\href {\doibase 10.1364/JOSAB.4.001465}
  {\bibfield  {journal} {\bibinfo  {journal} {J. Opt. Soc. Am. B}\ }\textbf
  {\bibinfo {volume} {4}},\ \bibinfo {pages} {1465--1475} (\bibinfo {year}
  {1987})}\BibitemShut {NoStop}%
\bibitem [{\citenamefont {Lvovsky}(2015)}]{Lvovsky.chapter.2015}%
  \BibitemOpen
  \bibfield  {author} {\bibinfo {author} {\bibfnamefont {A.~I.}\ \bibnamefont
  {Lvovsky}},\ }\enquote {\bibinfo {title} {Squeezed light},}\ in\ \href@noop
  {} {\emph {\bibinfo {booktitle} {Photonics, Volume 1, Fundamentals of
  Photonics and Physics}}},\ \bibinfo {editor} {edited by\ \bibinfo {editor}
  {\bibfnamefont {David~L.}\ \bibnamefont {Andrews}}}\ (\bibinfo  {publisher}
  {Wiley},\ \bibinfo {address} {West Sussex, UK},\ \bibinfo {year} {2015})\
  Chap.~\bibinfo {chapter} {5}, pp.\ \bibinfo {pages} {121--164},\ \Eprint
  {http://arxiv.org/abs/arXiv:1401.4118} {arXiv:1401.4118} \BibitemShut
  {NoStop}%
\bibitem [{\citenamefont {Caves}(1981)}]{Caves.PRD.23.1693.1981}%
  \BibitemOpen
  \bibfield  {author} {\bibinfo {author} {\bibfnamefont {Carlton~M.}\
  \bibnamefont {Caves}},\ }\bibfield  {title} {\enquote {\bibinfo {title}
  {Quantum-mechanical noise in an interferometer},}\ }\href {\doibase
  10.1103/PhysRevD.23.1693} {\bibfield  {journal} {\bibinfo  {journal} {Phys.
  Rev. D}\ }\textbf {\bibinfo {volume} {23}},\ \bibinfo {pages} {1693--1708}
  (\bibinfo {year} {1981})}\BibitemShut {NoStop}%
\bibitem [{\citenamefont {Grote}\ \emph {et~al.}(2013)\citenamefont {Grote},
  \citenamefont {Danzmann}, \citenamefont {Dooley}, \citenamefont {Schnabel},
  \citenamefont {Slutsky},\ and\ \citenamefont
  {Vahlbruch}}]{Grote.PRL.110.181101.2013}%
  \BibitemOpen
  \bibfield  {author} {\bibinfo {author} {\bibfnamefont {H.}~\bibnamefont
  {Grote}}, \bibinfo {author} {\bibfnamefont {K.}~\bibnamefont {Danzmann}},
  \bibinfo {author} {\bibfnamefont {K.~L.}\ \bibnamefont {Dooley}}, \bibinfo
  {author} {\bibfnamefont {R.}~\bibnamefont {Schnabel}}, \bibinfo {author}
  {\bibfnamefont {J.}~\bibnamefont {Slutsky}}, \ and\ \bibinfo {author}
  {\bibfnamefont {H.}~\bibnamefont {Vahlbruch}},\ }\bibfield  {title} {\enquote
  {\bibinfo {title} {First long-term application of squeezed states of light in
  a gravitational-wave observatory},}\ }\href {\doibase
  10.1103/PhysRevLett.110.181101} {\bibfield  {journal} {\bibinfo  {journal}
  {Phys. Rev. Lett.}\ }\textbf {\bibinfo {volume} {110}},\ \bibinfo {pages}
  {181101} (\bibinfo {year} {2013})}\BibitemShut {NoStop}%
\bibitem [{\citenamefont {Furrer}\ \emph {et~al.}(2012)\citenamefont {Furrer},
  \citenamefont {Franz}, \citenamefont {Berta}, \citenamefont {Leverrier},
  \citenamefont {Scholz}, \citenamefont {Tomamichel},\ and\ \citenamefont
  {Werner}}]{Furrer.PRL.109.100502.2012}%
  \BibitemOpen
  \bibfield  {author} {\bibinfo {author} {\bibfnamefont {F.}~\bibnamefont
  {Furrer}}, \bibinfo {author} {\bibfnamefont {T.}~\bibnamefont {Franz}},
  \bibinfo {author} {\bibfnamefont {M.}~\bibnamefont {Berta}}, \bibinfo
  {author} {\bibfnamefont {A.}~\bibnamefont {Leverrier}}, \bibinfo {author}
  {\bibfnamefont {V.~B.}\ \bibnamefont {Scholz}}, \bibinfo {author}
  {\bibfnamefont {M.}~\bibnamefont {Tomamichel}}, \ and\ \bibinfo {author}
  {\bibfnamefont {R.~F.}\ \bibnamefont {Werner}},\ }\bibfield  {title}
  {\enquote {\bibinfo {title} {Continuous variable quantum key distribution:
  Finite-key analysis of composable security against coherent attacks},}\
  }\href {\doibase 10.1103/PhysRevLett.109.100502} {\bibfield  {journal}
  {\bibinfo  {journal} {Phys. Rev. Lett.}\ }\textbf {\bibinfo {volume} {109}},\
  \bibinfo {pages} {100502} (\bibinfo {year} {2012})}\BibitemShut {NoStop}%
\bibitem [{\citenamefont {Furrer}(2014)}]{Furrer.PRA.90.042325.2014}%
  \BibitemOpen
  \bibfield  {author} {\bibinfo {author} {\bibfnamefont {Fabian}\ \bibnamefont
  {Furrer}},\ }\bibfield  {title} {\enquote {\bibinfo {title}
  {Reverse-reconciliation continuous-variable quantum key distribution based on
  the uncertainty principle},}\ }\href {\doibase 10.1103/PhysRevA.90.042325}
  {\bibfield  {journal} {\bibinfo  {journal} {Phys. Rev. A}\ }\textbf {\bibinfo
  {volume} {90}},\ \bibinfo {pages} {042325} (\bibinfo {year}
  {2014})}\BibitemShut {NoStop}%
\bibitem [{\citenamefont {Madsen}\ \emph {et~al.}(2012)\citenamefont {Madsen},
  \citenamefont {Usenko}, \citenamefont {Lassen}, \citenamefont {Filip},\ and\
  \citenamefont {Andersen}}]{Madsen.NatCommun.3.1083.2012}%
  \BibitemOpen
  \bibfield  {author} {\bibinfo {author} {\bibfnamefont {Lars~S.}\ \bibnamefont
  {Madsen}}, \bibinfo {author} {\bibfnamefont {Vladyslav~C.}\ \bibnamefont
  {Usenko}}, \bibinfo {author} {\bibfnamefont {Mikael}\ \bibnamefont {Lassen}},
  \bibinfo {author} {\bibfnamefont {Radim}\ \bibnamefont {Filip}}, \ and\
  \bibinfo {author} {\bibfnamefont {Ulrik~L.}\ \bibnamefont {Andersen}},\
  }\bibfield  {title} {\enquote {\bibinfo {title} {Continuous variable quantum
  key distribution with modulated entangled states},}\ }\href {\doibase
  10.1038/ncomms2097} {\bibfield  {journal} {\bibinfo  {journal} {Nat.
  Commun.}\ }\textbf {\bibinfo {volume} {3}},\ \bibinfo {pages} {1083}
  (\bibinfo {year} {2012})}\BibitemShut {NoStop}%
\bibitem [{\citenamefont {Jacobsen}\ \emph {et~al.}(2014)\citenamefont
  {Jacobsen}, \citenamefont {Madsen}, \citenamefont {Usenko}, \citenamefont
  {Filip},\ and\ \citenamefont {Andersen}}]{Jacobsen.arXiv.1408.4566.2014}%
  \BibitemOpen
  \bibfield  {author} {\bibinfo {author} {\bibfnamefont {Christian~S.}\
  \bibnamefont {Jacobsen}}, \bibinfo {author} {\bibfnamefont {Lars~S.}\
  \bibnamefont {Madsen}}, \bibinfo {author} {\bibfnamefont {Vladyslav~C.}\
  \bibnamefont {Usenko}}, \bibinfo {author} {\bibfnamefont {Radim}\
  \bibnamefont {Filip}}, \ and\ \bibinfo {author} {\bibfnamefont {Ulrik~L.}\
  \bibnamefont {Andersen}},\ }\href@noop {} {\enquote {\bibinfo {title}
  {Elimination of information leakage in quantum information channels},}\ }
  (\bibinfo {year} {2014}),\ \Eprint {http://arxiv.org/abs/arXiv:1408.4566}
  {arXiv:1408.4566 [quant-ph]} \BibitemShut {NoStop}%
\bibitem [{\citenamefont {Eberle}\ \emph {et~al.}(2013)\citenamefont {Eberle},
  \citenamefont {H\"andchen}, \citenamefont {Duhme}, \citenamefont {Franz},
  \citenamefont {Furrer}, \citenamefont {Schnabel},\ and\ \citenamefont
  {Werner}}]{Eberle.NJP.15.053049.2013}%
  \BibitemOpen
  \bibfield  {author} {\bibinfo {author} {\bibfnamefont {Tobias}\ \bibnamefont
  {Eberle}}, \bibinfo {author} {\bibfnamefont {Vitus}\ \bibnamefont
  {H\"andchen}}, \bibinfo {author} {\bibfnamefont {J\"org}\ \bibnamefont
  {Duhme}}, \bibinfo {author} {\bibfnamefont {Torsten}\ \bibnamefont {Franz}},
  \bibinfo {author} {\bibfnamefont {Fabian}\ \bibnamefont {Furrer}}, \bibinfo
  {author} {\bibfnamefont {Roman}\ \bibnamefont {Schnabel}}, \ and\ \bibinfo
  {author} {\bibfnamefont {Reinhard~F.}\ \bibnamefont {Werner}},\ }\bibfield
  {title} {\enquote {\bibinfo {title} {Gaussian entanglement for quantum key
  distribution from a single-mode squeezing source},}\ }\href {\doibase
  10.1088/1367-2630/15/5/053049} {\bibfield  {journal} {\bibinfo  {journal}
  {New J. Phys.}\ }\textbf {\bibinfo {volume} {15}},\ \bibinfo {pages} {053049}
  (\bibinfo {year} {2013})}\BibitemShut {NoStop}%
\bibitem [{\citenamefont {Gehring}\ \emph {et~al.}(2015)\citenamefont
  {Gehring}, \citenamefont {H\"andchen}, \citenamefont {Duhme}, \citenamefont
  {Furrer}, \citenamefont {Franz}, \citenamefont {Pacher}, \citenamefont
  {Werner},\ and\ \citenamefont {Schnabel}}]{Gehring.NatCommun.6.8795.2015}%
  \BibitemOpen
  \bibfield  {author} {\bibinfo {author} {\bibfnamefont {Tobias}\ \bibnamefont
  {Gehring}}, \bibinfo {author} {\bibfnamefont {Vitus}\ \bibnamefont
  {H\"andchen}}, \bibinfo {author} {\bibfnamefont {J\"org}\ \bibnamefont
  {Duhme}}, \bibinfo {author} {\bibfnamefont {Fabian}\ \bibnamefont {Furrer}},
  \bibinfo {author} {\bibfnamefont {Torsten}\ \bibnamefont {Franz}}, \bibinfo
  {author} {\bibfnamefont {Christoph}\ \bibnamefont {Pacher}}, \bibinfo
  {author} {\bibfnamefont {Reinhard~F.}\ \bibnamefont {Werner}}, \ and\
  \bibinfo {author} {\bibfnamefont {Roman}\ \bibnamefont {Schnabel}},\
  }\bibfield  {title} {\enquote {\bibinfo {title} {Implementation of
  continuous-variable quantum key distribution with composable and
  one-sided-device-independent security against coherent attacks},}\ }\href
  {\doibase 10.1038/ncomms9795} {\bibfield  {journal} {\bibinfo  {journal}
  {Nat. Commun.}\ }\textbf {\bibinfo {volume} {6}} (\bibinfo {year} {2015}),\
  10.1038/ncomms9795}\BibitemShut {NoStop}%
\bibitem [{\citenamefont {Ast}\ \emph {et~al.}(2016)\citenamefont {Ast},
  \citenamefont {Ast}, \citenamefont {Mehmet},\ and\ \citenamefont
  {Schnabel}}]{Ast.OL.41.5094.2016}%
  \BibitemOpen
  \bibfield  {author} {\bibinfo {author} {\bibfnamefont {Stefan}\ \bibnamefont
  {Ast}}, \bibinfo {author} {\bibfnamefont {Melanie}\ \bibnamefont {Ast}},
  \bibinfo {author} {\bibfnamefont {Moritz}\ \bibnamefont {Mehmet}}, \ and\
  \bibinfo {author} {\bibfnamefont {Roman}\ \bibnamefont {Schnabel}},\
  }\bibfield  {title} {\enquote {\bibinfo {title} {Gaussian entanglement
  distribution with gigahertz bandwidth},}\ }\href {\doibase
  10.1364/OL.41.005094} {\bibfield  {journal} {\bibinfo  {journal} {Opt.
  Lett.}\ }\textbf {\bibinfo {volume} {41}},\ \bibinfo {pages} {5094--5097}
  (\bibinfo {year} {2016})}\BibitemShut {NoStop}%
\bibitem [{\citenamefont {Yukawa}\ \emph {et~al.}(2008)\citenamefont {Yukawa},
  \citenamefont {Ukai}, \citenamefont {van Loock},\ and\ \citenamefont
  {Furusawa}}]{Yukawa.PRA.78.012301.2008}%
  \BibitemOpen
  \bibfield  {author} {\bibinfo {author} {\bibfnamefont {Mitsuyoshi}\
  \bibnamefont {Yukawa}}, \bibinfo {author} {\bibfnamefont {Ryuji}\
  \bibnamefont {Ukai}}, \bibinfo {author} {\bibfnamefont {Peter}\ \bibnamefont
  {van Loock}}, \ and\ \bibinfo {author} {\bibfnamefont {Akira}\ \bibnamefont
  {Furusawa}},\ }\bibfield  {title} {\enquote {\bibinfo {title} {Experimental
  generation of four-mode continuous-variable cluster states},}\ }\href
  {\doibase 10.1103/PhysRevA.78.012301} {\bibfield  {journal} {\bibinfo
  {journal} {Phys. Rev. A}\ }\textbf {\bibinfo {volume} {78}},\ \bibinfo
  {pages} {012301} (\bibinfo {year} {2008})}\BibitemShut {NoStop}%
\bibitem [{\citenamefont {Gu}\ \emph {et~al.}(2009)\citenamefont {Gu},
  \citenamefont {Weedbrook}, \citenamefont {Menicucci}, \citenamefont {Ralph},\
  and\ \citenamefont {van Loock}}]{Gu.PRA.79.062318.2009}%
  \BibitemOpen
  \bibfield  {author} {\bibinfo {author} {\bibfnamefont {Mile}\ \bibnamefont
  {Gu}}, \bibinfo {author} {\bibfnamefont {Christian}\ \bibnamefont
  {Weedbrook}}, \bibinfo {author} {\bibfnamefont {Nicolas~C.}\ \bibnamefont
  {Menicucci}}, \bibinfo {author} {\bibfnamefont {Timothy~C.}\ \bibnamefont
  {Ralph}}, \ and\ \bibinfo {author} {\bibfnamefont {Peter}\ \bibnamefont {van
  Loock}},\ }\bibfield  {title} {\enquote {\bibinfo {title} {Quantum computing
  with continuous-variable clusters},}\ }\href {\doibase
  10.1103/PhysRevA.79.062318} {\bibfield  {journal} {\bibinfo  {journal} {Phys.
  Rev. A}\ }\textbf {\bibinfo {volume} {79}},\ \bibinfo {pages} {062318}
  (\bibinfo {year} {2009})}\BibitemShut {NoStop}%
\bibitem [{\citenamefont {Weedbrook}\ \emph {et~al.}(2012)\citenamefont
  {Weedbrook}, \citenamefont {Pirandola}, \citenamefont {Garc\'{\i}a-Patr\'on},
  \citenamefont {Cerf}, \citenamefont {Ralph}, \citenamefont {Shapiro},\ and\
  \citenamefont {Lloyd}}]{Weedbrook.RMP.84.621.2012}%
  \BibitemOpen
  \bibfield  {author} {\bibinfo {author} {\bibfnamefont {Christian}\
  \bibnamefont {Weedbrook}}, \bibinfo {author} {\bibfnamefont {Stefano}\
  \bibnamefont {Pirandola}}, \bibinfo {author} {\bibfnamefont {Ra\'ul}\
  \bibnamefont {Garc\'{\i}a-Patr\'on}}, \bibinfo {author} {\bibfnamefont
  {Nicolas~J.}\ \bibnamefont {Cerf}}, \bibinfo {author} {\bibfnamefont
  {Timothy~C.}\ \bibnamefont {Ralph}}, \bibinfo {author} {\bibfnamefont
  {Jeffrey~H.}\ \bibnamefont {Shapiro}}, \ and\ \bibinfo {author}
  {\bibfnamefont {Seth}\ \bibnamefont {Lloyd}},\ }\bibfield  {title} {\enquote
  {\bibinfo {title} {Gaussian quantum information},}\ }\href {\doibase
  10.1103/RevModPhys.84.621} {\bibfield  {journal} {\bibinfo  {journal} {Rev.
  Mod. Phys.}\ }\textbf {\bibinfo {volume} {84}},\ \bibinfo {pages} {621--669}
  (\bibinfo {year} {2012})}\BibitemShut {NoStop}%
\bibitem [{\citenamefont {Menicucci}(2014)}]{Menicucci.PRL.112.120504.2014}%
  \BibitemOpen
  \bibfield  {author} {\bibinfo {author} {\bibfnamefont {Nicolas~C.}\
  \bibnamefont {Menicucci}},\ }\bibfield  {title} {\enquote {\bibinfo {title}
  {Fault-tolerant measurement-based quantum computing with continuous-variable
  cluster states},}\ }\href {\doibase 10.1103/PhysRevLett.112.120504}
  {\bibfield  {journal} {\bibinfo  {journal} {Phys. Rev. Lett.}\ }\textbf
  {\bibinfo {volume} {112}},\ \bibinfo {pages} {120504} (\bibinfo {year}
  {2014})}\BibitemShut {NoStop}%
\bibitem [{\citenamefont {Takeno}\ \emph {et~al.}(2007)\citenamefont {Takeno},
  \citenamefont {Yukawa}, \citenamefont {Yonezawa},\ and\ \citenamefont
  {Furusawa}}]{Takeno.OE.15.4321.2007}%
  \BibitemOpen
  \bibfield  {author} {\bibinfo {author} {\bibfnamefont {Yuishi}\ \bibnamefont
  {Takeno}}, \bibinfo {author} {\bibfnamefont {Mitsuyoshi}\ \bibnamefont
  {Yukawa}}, \bibinfo {author} {\bibfnamefont {Hidehiro}\ \bibnamefont
  {Yonezawa}}, \ and\ \bibinfo {author} {\bibfnamefont {Akira}\ \bibnamefont
  {Furusawa}},\ }\bibfield  {title} {\enquote {\bibinfo {title} {{Observation
  of $-9$~dB quadrature squeezing with improvement of phase stability in
  homodyne measurement}},}\ }\href {\doibase 10.1364/OE.15.004321} {\bibfield
  {journal} {\bibinfo  {journal} {Opt. Expr.}\ }\textbf {\bibinfo {volume}
  {15}},\ \bibinfo {pages} {4321--4327} (\bibinfo {year} {2007})}\BibitemShut
  {NoStop}%
\bibitem [{\citenamefont {Vahlbruch}\ \emph {et~al.}(2006)\citenamefont
  {Vahlbruch}, \citenamefont {Chelkowski}, \citenamefont {Hage}, \citenamefont
  {Franzen}, \citenamefont {Danzmann},\ and\ \citenamefont
  {Schnabel}}]{Vahlbruch.PRL.97.011101.2006}%
  \BibitemOpen
  \bibfield  {author} {\bibinfo {author} {\bibfnamefont {Henning}\ \bibnamefont
  {Vahlbruch}}, \bibinfo {author} {\bibfnamefont {Simon}\ \bibnamefont
  {Chelkowski}}, \bibinfo {author} {\bibfnamefont {Boris}\ \bibnamefont
  {Hage}}, \bibinfo {author} {\bibfnamefont {Alexander}\ \bibnamefont
  {Franzen}}, \bibinfo {author} {\bibfnamefont {Karsten}\ \bibnamefont
  {Danzmann}}, \ and\ \bibinfo {author} {\bibfnamefont {Roman}\ \bibnamefont
  {Schnabel}},\ }\bibfield  {title} {\enquote {\bibinfo {title} {Coherent
  control of vacuum squeezing in the gravitational-wave detection band},}\
  }\href {\doibase 10.1103/PhysRevLett.97.011101} {\bibfield  {journal}
  {\bibinfo  {journal} {Phys. Rev. Lett.}\ }\textbf {\bibinfo {volume} {97}},\
  \bibinfo {pages} {011101} (\bibinfo {year} {2006})}\BibitemShut {NoStop}%
\bibitem [{\citenamefont {Eberle}\ \emph {et~al.}(2010)\citenamefont {Eberle},
  \citenamefont {Steinlechner}, \citenamefont {Bauchrowitz}, \citenamefont
  {H\"andchen}, \citenamefont {Vahlbruch}, \citenamefont {Mehmet},
  \citenamefont {M\"uller-Ebhardt},\ and\ \citenamefont
  {Schnabel}}]{Eberle.PRL.104.251102.2010}%
  \BibitemOpen
  \bibfield  {author} {\bibinfo {author} {\bibfnamefont {Tobias}\ \bibnamefont
  {Eberle}}, \bibinfo {author} {\bibfnamefont {Sebastian}\ \bibnamefont
  {Steinlechner}}, \bibinfo {author} {\bibfnamefont {J\"oran}\ \bibnamefont
  {Bauchrowitz}}, \bibinfo {author} {\bibfnamefont {Vitus}\ \bibnamefont
  {H\"andchen}}, \bibinfo {author} {\bibfnamefont {Henning}\ \bibnamefont
  {Vahlbruch}}, \bibinfo {author} {\bibfnamefont {Moritz}\ \bibnamefont
  {Mehmet}}, \bibinfo {author} {\bibfnamefont {Helge}\ \bibnamefont
  {M\"uller-Ebhardt}}, \ and\ \bibinfo {author} {\bibfnamefont {Roman}\
  \bibnamefont {Schnabel}},\ }\bibfield  {title} {\enquote {\bibinfo {title}
  {Quantum enhancement of the zero-area {Sagnac} interferometer topology for
  gravitational wave detection},}\ }\href {\doibase
  10.1103/PhysRevLett.104.251102} {\bibfield  {journal} {\bibinfo  {journal}
  {Phys. Rev. Lett.}\ }\textbf {\bibinfo {volume} {104}},\ \bibinfo {pages}
  {251102} (\bibinfo {year} {2010})}\BibitemShut {NoStop}%
\bibitem [{\citenamefont {Mehmet}\ \emph {et~al.}(2011)\citenamefont {Mehmet},
  \citenamefont {Ast}, \citenamefont {Eberle}, \citenamefont {Steinlechner},
  \citenamefont {Vahlbruch},\ and\ \citenamefont
  {Schnabel}}]{Mehmet.OE.19.25763.2011}%
  \BibitemOpen
  \bibfield  {author} {\bibinfo {author} {\bibfnamefont {Moritz}\ \bibnamefont
  {Mehmet}}, \bibinfo {author} {\bibfnamefont {Stefan}\ \bibnamefont {Ast}},
  \bibinfo {author} {\bibfnamefont {Tobias}\ \bibnamefont {Eberle}}, \bibinfo
  {author} {\bibfnamefont {Sebastian}\ \bibnamefont {Steinlechner}}, \bibinfo
  {author} {\bibfnamefont {Henning}\ \bibnamefont {Vahlbruch}}, \ and\ \bibinfo
  {author} {\bibfnamefont {Roman}\ \bibnamefont {Schnabel}},\ }\bibfield
  {title} {\enquote {\bibinfo {title} {{Squeezed light at 1550 nm with a
  quantum noise reduction of 12.3 dB}},}\ }\href {\doibase
  10.1364/OE.19.025763} {\bibfield  {journal} {\bibinfo  {journal} {Opt.
  Expr.}\ }\textbf {\bibinfo {volume} {19}},\ \bibinfo {pages} {25763--25772}
  (\bibinfo {year} {2011})}\BibitemShut {NoStop}%
\bibitem [{\citenamefont {Khalaidovski}\ \emph {et~al.}(2012)\citenamefont
  {Khalaidovski}, \citenamefont {Vahlbruch}, \citenamefont {Lastzka},
  \citenamefont {Gräf}, \citenamefont {Danzmann}, \citenamefont {Grote},\ and\
  \citenamefont {Schnabel}}]{Khalaidovski.CQG.29.075001.2012}%
  \BibitemOpen
  \bibfield  {author} {\bibinfo {author} {\bibfnamefont {Alexander}\
  \bibnamefont {Khalaidovski}}, \bibinfo {author} {\bibfnamefont {Henning}\
  \bibnamefont {Vahlbruch}}, \bibinfo {author} {\bibfnamefont {Nico}\
  \bibnamefont {Lastzka}}, \bibinfo {author} {\bibfnamefont {Christian}\
  \bibnamefont {Gräf}}, \bibinfo {author} {\bibfnamefont {Karsten}\
  \bibnamefont {Danzmann}}, \bibinfo {author} {\bibfnamefont {Hartmut}\
  \bibnamefont {Grote}}, \ and\ \bibinfo {author} {\bibfnamefont {Roman}\
  \bibnamefont {Schnabel}},\ }\bibfield  {title} {\enquote {\bibinfo {title}
  {Long-term stable squeezed vacuum state of light for gravitational wave
  detectors},}\ }\href {\doibase 10.1088/0264-9381/29/7/075001} {\bibfield
  {journal} {\bibinfo  {journal} {Class. Quantum Grav.}\ }\textbf {\bibinfo
  {volume} {29}},\ \bibinfo {pages} {075001} (\bibinfo {year}
  {2012})}\BibitemShut {NoStop}%
\bibitem [{\citenamefont {Mehmet}\ \emph {et~al.}(2010)\citenamefont {Mehmet},
  \citenamefont {Vahlbruch}, \citenamefont {Lastzka}, \citenamefont
  {Danzmann},\ and\ \citenamefont {Schnabel}}]{Mehmet.PRA.81.013814.2010}%
  \BibitemOpen
  \bibfield  {author} {\bibinfo {author} {\bibfnamefont {Moritz}\ \bibnamefont
  {Mehmet}}, \bibinfo {author} {\bibfnamefont {Henning}\ \bibnamefont
  {Vahlbruch}}, \bibinfo {author} {\bibfnamefont {Nico}\ \bibnamefont
  {Lastzka}}, \bibinfo {author} {\bibfnamefont {Karsten}\ \bibnamefont
  {Danzmann}}, \ and\ \bibinfo {author} {\bibfnamefont {Roman}\ \bibnamefont
  {Schnabel}},\ }\bibfield  {title} {\enquote {\bibinfo {title} {Observation of
  squeezed states with strong photon-number oscillations},}\ }\href {\doibase
  10.1103/PhysRevA.81.013814} {\bibfield  {journal} {\bibinfo  {journal} {Phys.
  Rev. A}\ }\textbf {\bibinfo {volume} {81}},\ \bibinfo {pages} {013814}
  (\bibinfo {year} {2010})}\BibitemShut {NoStop}%
\bibitem [{\citenamefont {Ast}\ \emph {et~al.}(2012)\citenamefont {Ast},
  \citenamefont {Samblowski}, \citenamefont {Mehmet}, \citenamefont
  {Steinlechner}, \citenamefont {Eberle},\ and\ \citenamefont
  {Schnabel}}]{Ast.OL.37.2367.2012}%
  \BibitemOpen
  \bibfield  {author} {\bibinfo {author} {\bibfnamefont {Stefan}\ \bibnamefont
  {Ast}}, \bibinfo {author} {\bibfnamefont {Aiko}\ \bibnamefont {Samblowski}},
  \bibinfo {author} {\bibfnamefont {Moritz}\ \bibnamefont {Mehmet}}, \bibinfo
  {author} {\bibfnamefont {Sebastian}\ \bibnamefont {Steinlechner}}, \bibinfo
  {author} {\bibfnamefont {Tobias}\ \bibnamefont {Eberle}}, \ and\ \bibinfo
  {author} {\bibfnamefont {Roman}\ \bibnamefont {Schnabel}},\ }\bibfield
  {title} {\enquote {\bibinfo {title} {Continuous-wave nonclassical light with
  gigahertz squeezing bandwidth},}\ }\href {\doibase 10.1364/OL.37.002367}
  {\bibfield  {journal} {\bibinfo  {journal} {Opt. Lett.}\ }\textbf {\bibinfo
  {volume} {37}},\ \bibinfo {pages} {2367--2369} (\bibinfo {year}
  {2012})}\BibitemShut {NoStop}%
\bibitem [{\citenamefont {Ast}\ \emph {et~al.}(2013)\citenamefont {Ast},
  \citenamefont {Mehmet},\ and\ \citenamefont
  {Schnabel}}]{Ast.OE.21.13572.2013}%
  \BibitemOpen
  \bibfield  {author} {\bibinfo {author} {\bibfnamefont {Stefan}\ \bibnamefont
  {Ast}}, \bibinfo {author} {\bibfnamefont {Moritz}\ \bibnamefont {Mehmet}}, \
  and\ \bibinfo {author} {\bibfnamefont {Roman}\ \bibnamefont {Schnabel}},\
  }\bibfield  {title} {\enquote {\bibinfo {title} {{High-bandwidth squeezed
  light at 1550 nm from a compact monolithic PPKTP cavity}},}\ }\href {\doibase
  10.1364/OE.21.013572} {\bibfield  {journal} {\bibinfo  {journal} {Opt.
  Expr.}\ }\textbf {\bibinfo {volume} {21}},\ \bibinfo {pages} {13572--13579}
  (\bibinfo {year} {2013})}\BibitemShut {NoStop}%
\bibitem [{\citenamefont {Baune}\ \emph {et~al.}(2015)\citenamefont {Baune},
  \citenamefont {Gniesmer}, \citenamefont {Sch\"{o}nbeck}, \citenamefont
  {Vollmer}, \citenamefont {Fiur\'{a}\v{s}ek},\ and\ \citenamefont
  {Schnabel}}]{Baune.OE.23.16035.2015}%
  \BibitemOpen
  \bibfield  {author} {\bibinfo {author} {\bibfnamefont {Christoph}\
  \bibnamefont {Baune}}, \bibinfo {author} {\bibfnamefont {Jan}\ \bibnamefont
  {Gniesmer}}, \bibinfo {author} {\bibfnamefont {Axel}\ \bibnamefont
  {Sch\"{o}nbeck}}, \bibinfo {author} {\bibfnamefont {Christina~E.}\
  \bibnamefont {Vollmer}}, \bibinfo {author} {\bibfnamefont {Jarom\'{i}r}\
  \bibnamefont {Fiur\'{a}\v{s}ek}}, \ and\ \bibinfo {author} {\bibfnamefont
  {Roman}\ \bibnamefont {Schnabel}},\ }\bibfield  {title} {\enquote {\bibinfo
  {title} {Strongly squeezed states at 532 nm based on frequency
  up-conversion},}\ }\href {\doibase 10.1364/OE.23.016035} {\bibfield
  {journal} {\bibinfo  {journal} {Opt. Expr.}\ }\textbf {\bibinfo {volume}
  {23}},\ \bibinfo {pages} {16035--16041} (\bibinfo {year} {2015})}\BibitemShut
  {NoStop}%
\bibitem [{\citenamefont {Yan}\ \emph {et~al.}(2012)\citenamefont {Yan},
  \citenamefont {Jia}, \citenamefont {Su}, \citenamefont {Duan}, \citenamefont
  {Xie},\ and\ \citenamefont {Peng}}]{Yan.PRA.85.040305.2012}%
  \BibitemOpen
  \bibfield  {author} {\bibinfo {author} {\bibfnamefont {Zhihui}\ \bibnamefont
  {Yan}}, \bibinfo {author} {\bibfnamefont {Xiaojun}\ \bibnamefont {Jia}},
  \bibinfo {author} {\bibfnamefont {Xiaolong}\ \bibnamefont {Su}}, \bibinfo
  {author} {\bibfnamefont {Zhiyuan}\ \bibnamefont {Duan}}, \bibinfo {author}
  {\bibfnamefont {Changde}\ \bibnamefont {Xie}}, \ and\ \bibinfo {author}
  {\bibfnamefont {Kunchi}\ \bibnamefont {Peng}},\ }\bibfield  {title} {\enquote
  {\bibinfo {title} {Cascaded entanglement enhancement},}\ }\href {\doibase
  10.1103/PhysRevA.85.040305} {\bibfield  {journal} {\bibinfo  {journal} {Phys.
  Rev. A}\ }\textbf {\bibinfo {volume} {85}},\ \bibinfo {pages} {040305}
  (\bibinfo {year} {2012})}\BibitemShut {NoStop}%
\bibitem [{\citenamefont {Kaiser}\ \emph {et~al.}(2016)\citenamefont {Kaiser},
  \citenamefont {Fedrici}, \citenamefont {Zavatta}, \citenamefont {D'Auria},\
  and\ \citenamefont {Tanzilli}}]{Kaiser.Optica.3.362.2016}%
  \BibitemOpen
  \bibfield  {author} {\bibinfo {author} {\bibfnamefont {F.}~\bibnamefont
  {Kaiser}}, \bibinfo {author} {\bibfnamefont {B.}~\bibnamefont {Fedrici}},
  \bibinfo {author} {\bibfnamefont {A.}~\bibnamefont {Zavatta}}, \bibinfo
  {author} {\bibfnamefont {V.}~\bibnamefont {D'Auria}}, \ and\ \bibinfo
  {author} {\bibfnamefont {S.}~\bibnamefont {Tanzilli}},\ }\bibfield  {title}
  {\enquote {\bibinfo {title} {A fully guided-wave squeezing experiment for
  fiber quantum networks},}\ }\href {\doibase 10.1364/OPTICA.3.000362}
  {\bibfield  {journal} {\bibinfo  {journal} {Optica}\ }\textbf {\bibinfo
  {volume} {3}},\ \bibinfo {pages} {362--365} (\bibinfo {year}
  {2016})}\BibitemShut {NoStop}%
\bibitem [{\citenamefont {Dutt}\ \emph {et~al.}(2015)\citenamefont {Dutt},
  \citenamefont {Luke}, \citenamefont {Manipatruni}, \citenamefont {Gaeta},
  \citenamefont {Nussenzveig},\ and\ \citenamefont
  {Lipson}}]{Dutt.PRAppl.3.044005.2015}%
  \BibitemOpen
  \bibfield  {author} {\bibinfo {author} {\bibfnamefont {Avik}\ \bibnamefont
  {Dutt}}, \bibinfo {author} {\bibfnamefont {Kevin}\ \bibnamefont {Luke}},
  \bibinfo {author} {\bibfnamefont {Sasikanth}\ \bibnamefont {Manipatruni}},
  \bibinfo {author} {\bibfnamefont {Alexander~L.}\ \bibnamefont {Gaeta}},
  \bibinfo {author} {\bibfnamefont {Paulo}\ \bibnamefont {Nussenzveig}}, \ and\
  \bibinfo {author} {\bibfnamefont {Michal}\ \bibnamefont {Lipson}},\
  }\bibfield  {title} {\enquote {\bibinfo {title} {On-chip optical
  squeezing},}\ }\href {\doibase 10.1103/PhysRevApplied.3.044005} {\bibfield
  {journal} {\bibinfo  {journal} {Phys. Rev. Applied}\ }\textbf {\bibinfo
  {volume} {3}},\ \bibinfo {pages} {044005} (\bibinfo {year}
  {2015})}\BibitemShut {NoStop}%
\bibitem [{\citenamefont {Dutt}\ \emph {et~al.}(2016)\citenamefont {Dutt},
  \citenamefont {Miller}, \citenamefont {Luke}, \citenamefont {Cardenas},
  \citenamefont {Gaeta}, \citenamefont {Nussenzveig},\ and\ \citenamefont
  {Lipson}}]{Dutt.OL.41.223.2016}%
  \BibitemOpen
  \bibfield  {author} {\bibinfo {author} {\bibfnamefont {Avik}\ \bibnamefont
  {Dutt}}, \bibinfo {author} {\bibfnamefont {Steven}\ \bibnamefont {Miller}},
  \bibinfo {author} {\bibfnamefont {Kevin}\ \bibnamefont {Luke}}, \bibinfo
  {author} {\bibfnamefont {Jaime}\ \bibnamefont {Cardenas}}, \bibinfo {author}
  {\bibfnamefont {Alexander~L.}\ \bibnamefont {Gaeta}}, \bibinfo {author}
  {\bibfnamefont {Paulo}\ \bibnamefont {Nussenzveig}}, \ and\ \bibinfo {author}
  {\bibfnamefont {Michal}\ \bibnamefont {Lipson}},\ }\bibfield  {title}
  {\enquote {\bibinfo {title} {Tunable squeezing using coupled ring resonators
  on a silicon nitride chip},}\ }\href {\doibase 10.1364/OL.41.000223}
  {\bibfield  {journal} {\bibinfo  {journal} {Opt. Lett.}\ }\textbf {\bibinfo
  {volume} {41}},\ \bibinfo {pages} {223--226} (\bibinfo {year}
  {2016})}\BibitemShut {NoStop}%
\bibitem [{pyg()}]{pygmo.url}%
  \BibitemOpen
  \href {http://esa.github.io/pygmo} {}\bibinfo {note} {PyGMO software package,
  http://esa.github.io/pygmo}\BibitemShut {NoStop}%
\bibitem [{\citenamefont {Bishop}\ and\ \citenamefont
  {Dorf}(2000)}]{Bishop.Dorf.chapter.2000}%
  \BibitemOpen
  \bibfield  {author} {\bibinfo {author} {\bibfnamefont {Robert~H.}\
  \bibnamefont {Bishop}}\ and\ \bibinfo {author} {\bibfnamefont {Richard~C.}\
  \bibnamefont {Dorf}},\ }\enquote {\bibinfo {title} {{The Routh--Hurwitz
  stability criterion}},}\ in\ \href@noop {} {\emph {\bibinfo {booktitle}
  {Control System Fundamentals}}},\ \bibinfo {editor} {edited by\ \bibinfo
  {editor} {\bibfnamefont {William~S.}\ \bibnamefont {Levine}}}\ (\bibinfo
  {publisher} {CRC Press},\ \bibinfo {address} {Boca Raton, Florida},\ \bibinfo
  {year} {2000})\ Chap.\ \bibinfo {chapter} {9.1}, pp.\ \bibinfo {pages}
  {131--135}\BibitemShut {NoStop}%
\end{thebibliography}%

\end{document}